\documentclass[12pt,a4paper,twoside,table]{report}

\usepackage{xcolor}
\usepackage[pdfborder={0 0 0}]{hyperref}  %
\usepackage[vmargin=20mm,hmargin=25mm]{geometry}  %
\usepackage{graphicx} %
\usepackage{forest}
\usepackage{parskip}  %
\usepackage{setspace} %
\usepackage{refcount} %
\usepackage{upquote}  %
\usepackage{mlmodern}
\usepackage[T1]{fontenc}
\usepackage[
    font={small,sf},
    labelfont=bf,
    format=hang,    
    format=plain,
    margin=0pt,
    width=0.8\textwidth,
]{caption}
\usepackage[list=true]{subcaption}
\usepackage{tikz}
\usetikzlibrary{fit, positioning, calc, shapes, shapes.geometric, arrows, arrows.meta,backgrounds}
\tikzset{
  big edge/.style={green, very thick,},
  big edgec/.style={big edge, -{Bar[fill=green,green,width=7,length=0,sep=0]}},
  big region/.style={draw, rectangle, rounded corners=1.5, dashed, dash pattern=on 1pt off 1pt, thin,},
  big site/.style={big region, fill=gray!60, text=black,},
  big react/.style={black, thick, -stealth, line width=3, shorten <=3, shorten >=3,},
  big react rev/.style={black, thick, stealth-stealth, line width=3, shorten <=3, shorten >=3,},
  big inst map/.style={very thick, -stealth, blue, dashed},
  lbl/.style={font=\tiny\sf, inner sep=1,},
  lbl conc/.style={font=\tiny, inner sep=1,},
  node/.style = {draw, circle},
}
\usepackage{amsmath,amssymb}
\usepackage{pgfplots}
\usepackage{pgfplotstable}
\usepackage{sansmath}
\usepackage{multirow}
\usepackage{tabularx}
\usepackage{booktabs}
\usepackage{longtable}
\usepackage{ragged2e}
\usepackage[htt]{hyphenat}
\usepackage[frozencache,cachedir=.]{minted}
\usepackage{caption}
\usepackage{dirtree}
\usepackage[Tol]{colorblind}
\usepackage{float}
\usepackage{pdflscape}
\usepackage{afterpage}
\newfloat{lstfloat}{h}{}
\floatname{lstfloat}{Listing}

\pgfplotsset{compat=1.3}
\DeclareMathOperator{\react}{\mathrel{\frac{\raisebox{0.75mm}{\begin{scriptsize}\ensuremath{\hspace*{1mm}\ \hspace*{1mm}}\end{scriptsize}}}{}} \joinrel{\!\!\vartriangleright}}

\DeclareMathOperator{\rrul}{\mathrel{\frac{\raisebox{0.75mm}{\begin{scriptsize}\ensuremath{\hspace*{1mm}\ \hspace*{1mm}}\end{scriptsize}}}{}} \joinrel{\!\!\blacktriangleright}}

\newcommand*{\defeq}{\stackrel{\text{def}}{=}}
\newcommand{\rr}[1]{\texttt{#1}}

\graphicspath{ {./} }

\title{Building Bigraphs of the real world}
\author{Kang Rong Roy Ang}
\date{2025}
\newcommand{\candidatenumber}{2320D}
\newcommand{\college}{Wolfson College}
\newcommand{\course}{Computer Science Tripos - Part II}

\begin{document}

\begin{titlepage}
\makeatletter

\Large
\hspace{\fill}
\textbf
\@author

\begin{center}
\Huge
\vspace{\fill}

\textbf
\@title

\course

\college

\@date
\vspace{\fill}

\end{center}

\makeatother
\end{titlepage}

\newpage
\section*{Declaration of originality}

I, the candidate for Part II of the Computer Science Tripos with Blind Grading Number \candidatenumber, hereby declare that this report and the work described in it are my own work, unaided except as may be specified below, and that the report does not contain material that has already been used to any substantial extent for a comparable purpose. In preparation of this report, I adhered to the Department of Computer Science and Technology AI Policy. I am content for my report to be made available to the students and staff of the University.

\bigskip 

\bigskip
\textbf{Date:} 16 May 2025
\vspace{\fill}
\makeatother

\newpage
\makeatletter
\chapter*{Proforma}
\large

Candidate Number: \candidatenumber

Project Title: \@title

Examination: \course, \@date

Word count\footnote{generated using \mintinline[breaklines, breakafter=_]{shell-session}{gs -q -dSAFER -sDEVICE=txtwrite -o - -dFirstPage=6 -dLastPage=45 2320D.pdf | egrep '[A-Za-z]{3}' | wc -w}}: 11,954

 Code Line Count: 1,888

Project Originator: The candidate

Project Supervisors: Ryan Gibb and Anil Madhavapeddy

\normalsize
\section*{Original aims of the project}
This project set out to model the physical world as a bigraph. The goal was to construct a place graph modelling the location of places marked on OpenStreetMap, and add a link graph to model the network of connected streets. The use of such bigraphs for ubiquitous systems would be demonstrated using reaction rules that allow devices to move between places and form links with other devices in its proximity.

\section*{Work completed}
This project has met all its success criteria. It developed a specification for hierarchically structuring real-world locations within a bigraph model. A tool was developed to construct a bigraph modelling all named building, streets and administrative areas contained within a specified boundary, using data extracted from OpenStreetMap. The constructed bigraphs were then transformed according to reactions rules to simulate agents moving around the world and forming connections with others. As a side-product and necessary step, algorithmic improvements were made to open-source bigraph-building tools, enabling the efficient construction and transformation of large bigraphs. 

\section*{Special difficulties}
None

\tableofcontents

\chapter{Introduction}
\pagenumbering{arabic}
\pagestyle{headings}
\label{firstcontentpage}

\textit{This dissertation proposes a formal specification for organising buildings, streets and administrative areas into a hierarchical space-partitioning tree, to model the real world as a bigraph. It delivers a tool that constructs bigraphs for any part of the world and demonstrates the use of bigraphs to model the space and motion of communicating agents.}

\section{Motivation}

For millennia, maps have been one of the most important human inventions. By encoding two-dimensional geographical information, maps of the world have served as essential tools for navigation. Geographic information systems (GIS) then enabled additional attributes and surveyed statistics such as demographic data and traffic information to be association with specific locations.

Today, our environment is saturated with computers. Ubiquitous systems observe and influence the physical world and virtual information spaces. Humans interact with these systems to effect changes in virtual domains. Ubiquitous systems are also linked to others across the world and can communicate independent of their physical proximity, using technology such as wireless networks. However, traditional maps and GIS frameworks are inadequate for modelling the mixed virtual-physical space: they represent only static physical locations and cannot capture dynamic agent behaviour or evolving virtual connections. Instead, multiple disjoint models and databases are typically used to track locality, connectivity, and effects of ubiquitous systems—leading to fragile integrations, inconsistent interpretations, and limited scalability.

Milner introduced bigraphs \cite{10.5555/1540607} with the principal aim of formally modelling the interaction of ubiquitous systems and their environment with a single unified model. The mathematical structures of bigraphs—namely, placing and linking—capture the interactions between systems that depend on both locality and connectivity. Placing also intrinsically encodes adjacency and containment, enabling efficient queries for the neighbours of an agent and the regions that enclose it.

This dissertation provides a method for representing all buildings, streets and administrative areas in the world using bigraphs. It delivers a tool implemented in OCaml that uses open geographical data from OpenStreetMap to construct a comprehensive bigraph—a digital twin of the world—that serves as a unifying foundation for bigraph models of mobile communicating agents. This work empirically demonstrates that bigraphs provide a rigorous unifying model for ubiquitous systems and their physical environment.

\section{Contributions}
The main contributions of this dissertation are:
\begin{enumerate}
\item A formal specification for organising all buildings, streets and administrative areas in the world into a hierarchical space-partitioning tree.
\item A command-line tool that, given a query region from any part of the world, extracts geographical data from OpenStreetMap and generates a bigraph modelling all named areas, streets and buildings contained within the specified boundary.
\item Algorithmic improvements to open-source bigraph-building tools that enable them to efficiently construct and transform extremely large bigraphs, achieving up to a 97× speedup among other gains.
\end{enumerate}

\section{Related work}

Milner originally introduced bigraphs to address the challenge of modelling the interactions of millions of communicating agents, which he envisioned would be everywhere. However, experiments on the use of bigraphs for geographic space have been limited to small and local built environments: Walton and Worboys \cite{10.1007/978-3-642-03832-7_22} developed a formal method to translate existing image schemas of built environments into bigraphs, and illustrated how bigraphs for indoor space \cite{10.1007/978-3-642-33024-7_17} can be used to model key indoor events and their effects. This dissertation is the first to explore modelling outdoor space using bigraphs, thereby delivering a unifying interface for previously siloed indoor bigraphs, which can be seamlessly nested into the buildings represented in the bigraphs constructed here.

Bigraph's universal formalism has also found use in modelling IoT/sensor systems \cite{wireless-home-networks,8595061}, networking protocols \cite{DBLP:journals/fac/AlbalweAS24,networking} and biological processes \cite{KRIVINE200873} amongst many others. The bigraphs of the real world constructed in this dissertation can ground the physical entities of all bigraphs of past and future work to their physical location in the world, giving insight to their spatial context to the rest of the world. This unifying capability is exemplified and evaluated in \S \ref{section:qualities}.

Ongoing work by Archibald et al. \cite{Archibald2024} seek to use bigraphs as a unifying model for adjacent heterogeneous digital twins of the various transport systems at the Port of Dover: port operations, freight shipping movement and road traffic around the port. They are interested in using the deliverables of this dissertation to construct bigraphs of geographical areas, which will provide a context which the digital twins of transport systems can be composed into. The unified model will provide better operational oversight of the whole combined transport system, which is needed to understand and explore potential solutions to challenges such as decarbonisation.

\section{Starting point}
To prepare for this dissertation, I read up on bigraphs through an introductory six-lecture series by Milner \cite{bigraph-notes}, and experimented with small example bigraphs using the BigraphER command-line tool, but not yet with its OCaml library. Besides learning OCaml in the Part IA Foundations of Computer Science course, I had no other experience with OCaml codebases. Additionally, I experimented with various tools for processing OpenStreetMap data, and briefly explored several tagging conventions that pointed to possible ways of finding the data I needed.

\chapter{Preparation}

\textit{This chapter introduces the formalism of bigraphs and features of OpenStreetMap required to build a bigraph of the real world. An analysis of the project's requirements follows, concluded by a discussion of tools and software engineering techniques used.}

\section{Bigraphs}
Milner's bigraphs \cite{10.5555/1540607} are a modelling formalism for describing the locality and connectivity of entities. This report presents the necessary features of bigraphs by means of examples, using the diagrammatic notation and its equivalent algebraic form.

\begin{figure}[h]
\centering
\begin{subfigure}[b]{0.63\linewidth}
	\centering
	\begin{forest}
	  for tree={edge = {-latex}}
  [,phantom 
      [0,big region
            [0,big site]
            [ID(William\\Gates Building),align=center,text width=2.4cm]
	  [ID(Agent A),align=center,text width=2cm]
      ]
      [1,big region 
          [ Building
              [ Agent]
              [ 1,big site]
          ]
          [ Agent]
      ]
  ]
	\end{forest}
	\caption{Place graph of bigraph $F$.}
    	\label{fig:F-place-graph}
\end{subfigure}
\begin{subfigure}[b]{0.36\linewidth}
	\centering
   	 \begin{tikzpicture}
		\node[draw,circle,minimum size=20,label={[inner sep=0.5, name=id0l,align=center]south:{ID(William\\Gates Building)}}] (id0) {};
		\node[draw,circle,minimum size=20,below=1 of id0, label={[inner sep=0.5, name=id1l,align=center]south:{ID(Agent A)}}] (id1) {};
		\node[draw,circle,minimum size=20,right=1.5 of id0, label={[inner sep=0.5, name=b0l,align=center]south:{Building}}] (b0) {};
		\node[draw,circle,minimum size=20,below=1 of b0, label={[inner sep=0.5, name=a0l,align=center]south:{Agent}}] (a0) {};
		\node[draw,circle,minimum size=20,right=0.8 of b0, label={[inner sep=0.5, name=a1l,align=center]south:{Agent}}] (a1) {};

		\draw[big edge] (id0) to[out=0,in=180] (b0);
		\draw[big edge] (id1) to[out=0,in=180] (a0);

		\node[above=0.3 of a1] (x) {$x$};
		\draw[big edge] (a1) to[out=90,in=270] (x);
	\end{tikzpicture}
  	\caption{Link graph of bigrah $F$.}
    	\label{fig:F-link-graph}
\end{subfigure}

\medskip

\begin{subfigure}[b]{\linewidth}
\centering
\begin{tikzpicture}
 	\node[big site, inner sep=10pt,] (s0) {};
	\node[below right, inner sep=0pt, shift={(0.1,-0.1)}, ] at (s0.north west) {0};
	\node[right=0.3 of s0, draw, rounded corners=2,text width = 3.05cm] (id0) {ID(William Gates Building)};
	\node[below right=0 of id0.south west, draw, rounded corners=2, shift={(0,-0.3)}] (id1) {ID(Agent A)};
	\node[big region, fit=(s0)(id0)(id1), inner sep=15pt,] (r0) {};
	\node[below right, inner sep=0pt, shift={(0.1,-0.1)}, ] at (r0.north west) {0};

   	\node[right=1.5 of r0, draw, rounded corners=2] (a0) {Agent};
	\node[right=0.3 of a0, big site, inner sep=10pt,] (s1) {};
	\node[below right, inner sep=0pt, shift={(0.1,-0.1)}, ] at (s1.north west) {1};
	\node[fit=(a0)(s1), draw, rounded corners=2,inner sep=15pt,] (b0) {};
	\node[below right, inner sep=0pt, shift={(0.1,-0.1)}, ] at (b0.north west) {Building};
   	\node[right=0.3 of b0, draw, rounded corners=2] (a1) {Agent};
	\node[big region, fit=(b0)(a1), inner sep=15pt,] (r1) {};
	\node[below right, inner sep=0pt, shift={(0.1,-0.1)}, ] at (r1.north west) {1};

	\draw[big edge] (id0)  to[out=0,in=180]  (b0.west |-,  |- id0);
	\draw[big edge] (id1)  to[out=0,in=180]  (a0);

	\node[above=0.25] at (a1|-,|-r1.north) (x) {$x$};
	\draw[big edge] (a1)  to[out=90,in=270]  (x);

	\node[above left=0.5 of r0, shift={(0,-0.2)}] (RegionLabel) {Region};
	\draw[shorten >=0pt,dashed] (RegionLabel.south) -- (r0.north west);
	\node[left=1 of s0, shift={(0,-0.1)}] (SiteLabel) {Site};
	\draw[shorten >=0pt,dashed] (SiteLabel) -- (s0);
	\node[left=2 of id0.south west, shift={(0,-0.2)}] (NodeLabel) {Node};
	\draw[shorten >=0pt,dashed] (NodeLabel) -- (id0.south west);
	\node[left=2 of id1, shift={(0,-0.5)}] (ControlLabel) {Control};
	\draw[shorten >=0pt,dashed] (ControlLabel) -- ($(id1.west)!0.15!(id1.east)$);
	\node[below=0.7 of id1, shift={(-0.3,0)}] (ParameterLabel) {String Parameter};
	\draw[shorten >=0pt,dashed] (ParameterLabel) -- ($(id1.west)!0.6!(id1.east)$);

	\node[right=0.5 of x, shift={(0,0.1)}] (OuterLabel) {Outer name};
	\draw[shorten >=0pt,dashed] (OuterLabel) -- (x);
	\node[] at (s1|-,|- ParameterLabel) (EdgeLabel) {Closed link};
	\draw[shorten >=0pt,dashed] (EdgeLabel) -- ($(id1)!0.6!(a0)$);
\end{tikzpicture}
\caption{Anatomy of Bigraph $F$.}
\label{fig:bigraph-anatomy}
\end{subfigure}
\caption{An example bigraph $F$.}
\label{fig:F-bigraph}
\end{figure}
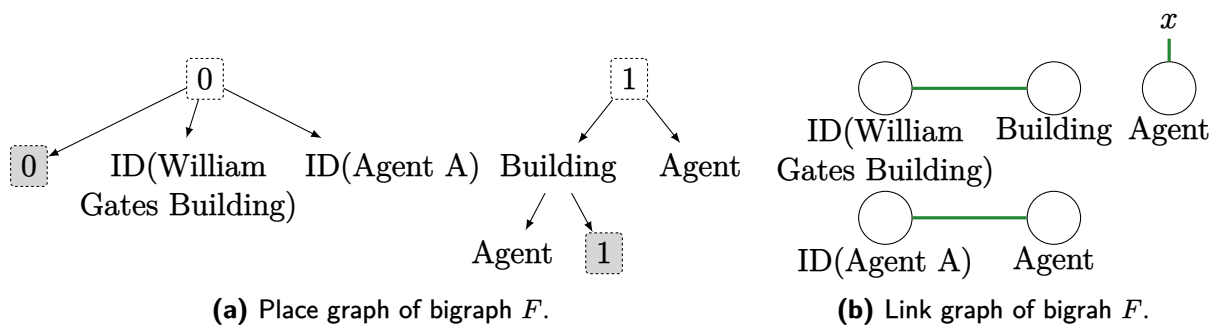
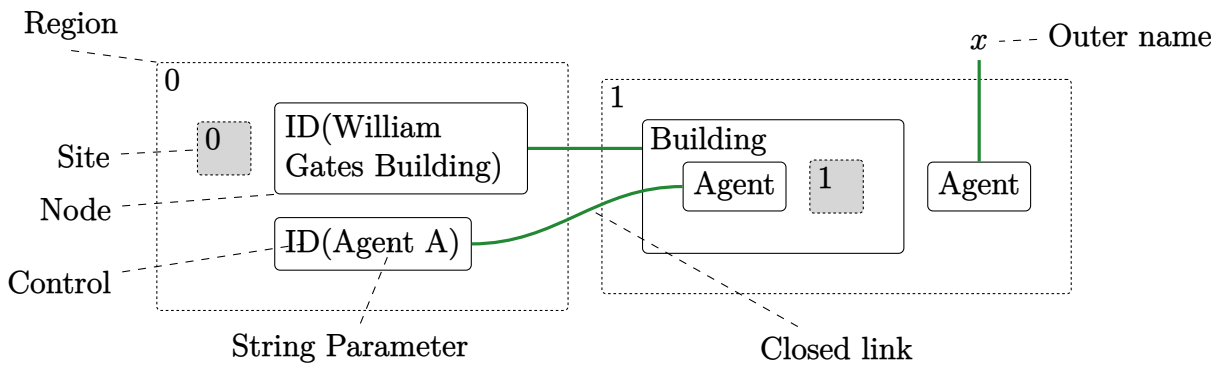

A bigraph is constituted by two orthogonal structures: a \textit{place graph} and a \textit{link graph}. Figure \ref{fig:F-bigraph} shows an example bigraph $F$, with its place graph and link graph in Figure \ref{fig:F-place-graph}  and Figure \ref{fig:F-link-graph} respectively.

The place graph is a forest that encodes the containment of a node inside its parent node. The roots of the place graph are \textit{regions} in the bigraph that denote separate components, which are illustrated in the diagrammatic notation as the outermost dashed line boundaries. \textit{Sites} specify placeholders for a component bigraph, which may be empty, and are illustrated as shaded rectangles. By convention, regions and sites are each numbered by integers starting from 0. 

The link graph is a hypergraph that specifies the non-spatial relations between nodes. It shares the set of nodes found in the place graph and contains, in addition, edges that connect one or more nodes. Besides \textit{closed links} between nodes, there are \textit{open links} that are incomplete hyperedges with \textit{outer names} (e.g. $x$ of bigraph $F$), thereby providing a way of combining with other link graphs. By convention, outer names are drawn above the bigraph.

Nodes of a bigraph are classified into different \textit{controls}, each having a fixed \textit{arity} which specifies how many links the node is connected to (e.g. ID, Building and Agent each have arity 1). Nodes can have string parameters (e.g. the top and bottom ID nodes in bigraph $F$ have parameters "William Gates Building" and "Agent A" respectively).

\subsection{Elementary bigraphs} \label{elementary-bigraphs}

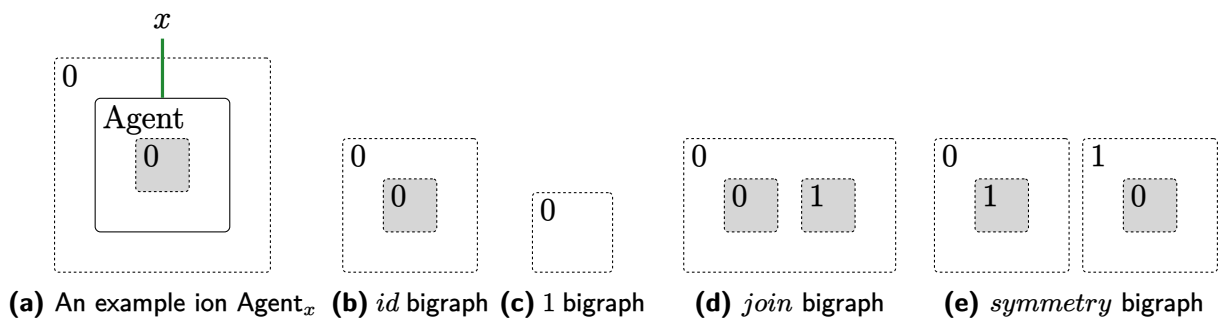
\begin{figure}[h]
\centering
\begin{subfigure}[b]{0.26\linewidth}
\centering
\begin{tikzpicture}
 	\node[big site, inner sep=10pt,] (s0) {};
	\node[below right, inner sep=0pt, shift={(0.1,-0.1)}, ] at (s0.north west) {0};
	\node[fit=(s0), draw, rounded corners=2,inner sep=15pt,] (a0) {};
	\node[below right, inner sep=0pt, shift={(0.1,-0.1)}, ] at (a0.north west) {Agent};
	\node[big region, fit=(a0), inner sep=15pt,] (r0) {};
	\node[below right, inner sep=0pt, shift={(0.1,-0.1)}, ] at (r0.north west) {0};

	\node[above=0.25 of r0, shift={(0,0)}] (x) {$x$};
	\draw[big edge] (a0)  to[out=90,in=270]  (x);
\end{tikzpicture}
\caption{An example ion Agent$_{x}$}
\label{fig:ion}
\end{subfigure}
\begin{subfigure}[b]{0.13\linewidth}
\centering
\begin{tikzpicture}
 	\node[big site, inner sep=10pt,] (s0) {};
	\node[below right, inner sep=0pt, shift={(0.1,-0.1)}, ] at (s0.north west) {0};
	\node[big region, fit=(s0), inner sep=15pt,] (r0) {};
	\node[below right, inner sep=0pt, shift={(0.1,-0.1)}, ] at (r0.north west) {0};
\end{tikzpicture}
\caption{$id$ bigraph}
\label{fig:bigraph-id}
\end{subfigure}
\begin{subfigure}[b]{0.12\linewidth}
\centering
\begin{tikzpicture}
	\node[big region, inner sep=15pt,] (r0) {};
	\node[below right, inner sep=0pt, shift={(0.1,-0.1)}, ] at (r0.north west) {0};
\end{tikzpicture}
\caption{$1$ bigraph}
\label{fig:bigraph-1}
\end{subfigure}
\begin{subfigure}[b]{0.22\linewidth}
\centering
\begin{tikzpicture}
 	\node[big site, inner sep=10pt,] (s0) {};
	\node[below right, inner sep=0pt, shift={(0.1,-0.1)}, ] at (s0.north west) {0};
	\node[right=0.3 of s0,big site, inner sep=10pt,] (s1) {};
	\node[below right, inner sep=0pt, shift={(0.1,-0.1)}, ] at (s1.north west) {1};
	\node[big region, fit=(s0)(s1), inner sep=15pt,] (r0) {};
	\node[below right, inner sep=0pt, shift={(0.1,-0.1)}, ] at (r0.north west) {0};
\end{tikzpicture}
\caption{$join$ bigraph}
\label{fig:join}
\end{subfigure}
\begin{subfigure}[b]{0.23\linewidth}
\centering
\begin{tikzpicture}
 	\node[big site, inner sep=10pt,] (s0) {};
	\node[below right, inner sep=0pt, shift={(0.1,-0.1)}, ] at (s0.north west) {1};
	\node[big region, fit=(s0), inner sep=15pt,] (r0) {};
	\node[below right, inner sep=0pt, shift={(0.1,-0.1)}, ] at (r0.north west) {0};
 	\node[right=0.7 of r0, big site, inner sep=10pt,] (s1) {};
	\node[below right, inner sep=0pt, shift={(0.1,-0.1)}, ] at (s1.north west) {0};
	\node[big region, fit=(s1), inner sep=15pt,] (r1) {};
	\node[below right, inner sep=0pt, shift={(0.1,-0.1)}, ] at (r1.north west) {1};
\end{tikzpicture}
\caption{$symmetry$ bigraph}
\label{fig:bigraph-symmetry}
\end{subfigure}
\caption{Elementary bigraphs.}
\end{figure}

An \textit{ion} is a bigraph that has a single region containing a single node, which in turn nests a single site. The ion illustrated in Figure \ref{fig:ion} has an equivalent algebraic representation Agent$_{x}$. 

The $id$ bigraph (Figure \ref{fig:bigraph-id}) has a single region which nests a single site. 

The $1$ bigraph (Figure \ref{fig:bigraph-1}) has a single region that is empty. 

The $join$ bigraph (Figure \ref{fig:join}) has a single region which nests two sites.

The $symmetry$ bigraph (Figure \ref{fig:bigraph-symmetry}) reorders the regions of a bigraph.

All bigraphs required by this dissertation can be built from ions, $id$, $1$, $join$ and $symmetry$ using the following operations.

\subsection{Operations on bigraphs} \label{bigraph-operations}
There are three methods for combining bigraphs used in this dissertation: nesting, parallel product and merge product. In addition, closure is used on links.

\subsubsection{Nesting}
Bigraph $F$ can be nested in $G$ if the number of roots in $F$ matches the number of sites in $G$. The nesting $G.F$ is constructed as follows:
\begin{itemize}
\item the place graph is formed by joining $i^{th}$ root of $F$ with the $i^{th}$ site of $G$.
\item the link graph is formed by juxtaposing the link graphs of $G$ and $F$, and combining links that share the same name.
\end{itemize}
For example, an agent that does not carry any items is Agent$_{x}.1$ (Figure \ref{fig:nesting-agent}). A building that contains only one agent is Building$_y.$Agent$_{x}.1$ (Figure \ref{fig:nesting-building}).

Nesting is associative.

\subsubsection{Parallel Product}
The parallel product $G || F$ is formed by juxtaposing the bigraphs $G$ and $F$, and combining links that share the same name. An example, $\text{ID(Agent A)}_x.1||\text{Building}_y.\text{Agent}_{x}.1$, is illustrated in Figure \ref{fig:parallel-building}.

Parallel product is associative.

\subsubsection{Merge Product}

\begin{figure}[h]
\centering
\begin{subfigure}[b]{0.25\linewidth}
\centering
\begin{tikzpicture}
   	\node[draw, rounded corners=2] (a0) {Agent};
	\node[big region, fit=(a0), inner sep=15pt,] (r0) {};
	\node[below right, inner sep=0pt, shift={(0.1,-0.1)}, ] at (r0.north west) {0};

	\node[above=0.25 of r0, shift={(0,0)}] (x) {$x$};
	\draw[big edge] (a0)  to[out=90,in=270]  (x);
\end{tikzpicture}
\caption{Agent$_{x}.1$}
\label{fig:nesting-agent}
\end{subfigure}
\begin{subfigure}[b]{0.25\linewidth}
\centering
\begin{tikzpicture}
   	\node[draw, rounded corners=2] (a0) {Agent};
	\node[fit=(a0), draw, rounded corners=2,inner sep=15pt,] (b0) {};
	\node[below right, inner sep=0pt, shift={(0.1,-0.1)}, ] at (b0.north west) {Building};
	\node[big region, fit=(b0), inner sep=15pt,] (r0) {};
	\node[below right, inner sep=0pt, shift={(0.1,-0.1)}, ] at (r0.north west) {0};

	\node[above=0.25 of r0, shift={(-0.3,0)}] (x) {$x$};
	\draw[big edge] (a0)  to[out=100,in=270]  (x);
	\node[above=0.2 of r0, shift={(0.3,0)}] (y) {$y$};
	\draw[big edge] (b0)  to[out=80,in=270]  (y);
\end{tikzpicture}
\caption{Building$_y.$Agent$_{x}.1$}
\label{fig:nesting-building}
\end{subfigure}

\begin{subfigure}[b]{0.5\linewidth}
\centering
\begin{tikzpicture}
   	\node[draw, rounded corners=2] (a0) {Agent};
	\node[fit=(a0), draw, rounded corners=2,inner sep=15pt,] (b0) {};
	\node[below right, inner sep=0pt, shift={(0.1,-0.1)}, ] at (b0.north west) {Building};
	\node[big region, fit=(b0), inner sep=15pt,] (r1) {};
	\node[below right, inner sep=0pt, shift={(0.1,-0.1)}, ] at (r1.north west) {1};

	\node[left=0.8 of r0, draw, rounded corners=2] (ida0) {ID(Agent A)};
	\node[big region, fit=(ida0), inner sep=15pt,] (r0) {};
	\node[below right, inner sep=0pt, shift={(0.1,-0.1)}, ] at (r0.north west) {0};

	\node[above=1.75] at ($(r0.east)!0.5!(r1.west)$) (x) {$x$};
	\draw[big edge] (a0)  to[out=180,in=270]  (x);
	\draw[big edge] (ida0)  to[out=0,in=270]  (x);
	\node[above=0.2 of r1, shift={(0,0)}] (y) {$y$};
	\draw[big edge] (b0)  to[out=90,in=270]  (y);
\end{tikzpicture}
\caption{$\text{ID(Agent A)}_x.1||\text{Building}_y.\text{Agent}_{x}.1$}
\label{fig:parallel-building}
\end{subfigure}
\begin{subfigure}[b]{0.3\linewidth}
\centering
\begin{tikzpicture}
   	\node[draw, rounded corners=2] (a0) {Agent};
	\node[right=0.3 of a0, big site, inner sep=10pt,] (s0) {};
	\node[below right, inner sep=0pt, shift={(0.1,-0.1)}, ] at (s0.north west) {0};
	\node[big region, fit=(a0)(s0), inner sep=15pt,] (r0) {};
	\node[below right, inner sep=0pt, shift={(0.1,-0.1)}, ] at (r0.north west) {0};

	\node[above=0.25] at (a0|-,|-r0.north) (x) {$x$};
	\draw[big edge] (a0)  to[out=90,in=270]  (x);
\end{tikzpicture}
\caption{Agent$_{x}.1|id$}
\label{fig:merge}
\end{subfigure}
\caption{Diagrammatic representation and the equivalent algebraic notation of some bigraphs.}
\end{figure}
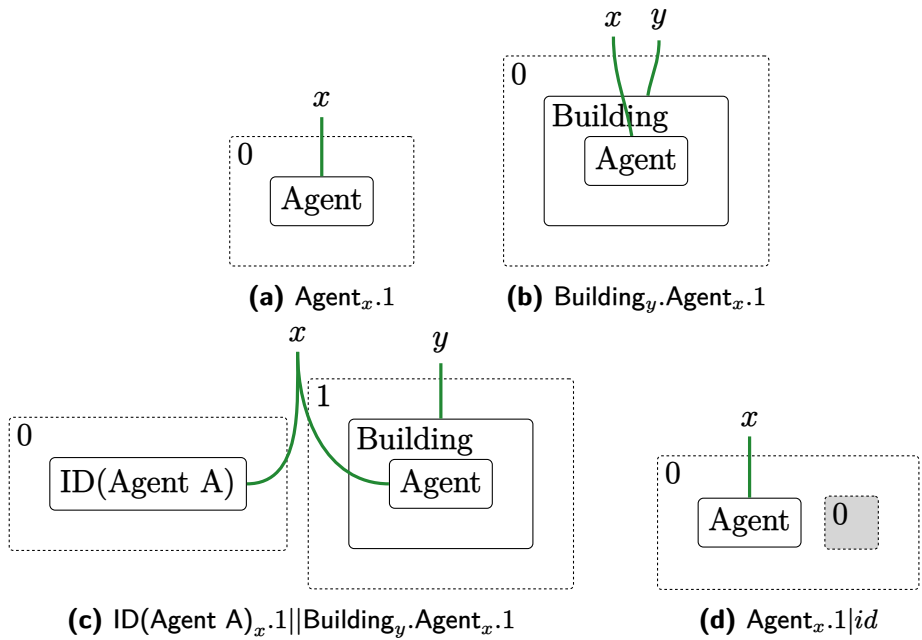

Using the elementary bigraph $join$ (Figure \ref{fig:join}), the merge product | of bigraphs $G$ and $F$ is defined as
\[G | F \defeq join.( G || F ) \]
An example, Agent$_{x}.1|id$, is illustrated in Figure \ref{fig:merge}.

Merge product is associative.

\subsubsection{Link Closure}

A closure $/x$ $F$ closes the outer name $x$ in bigraph $F$.

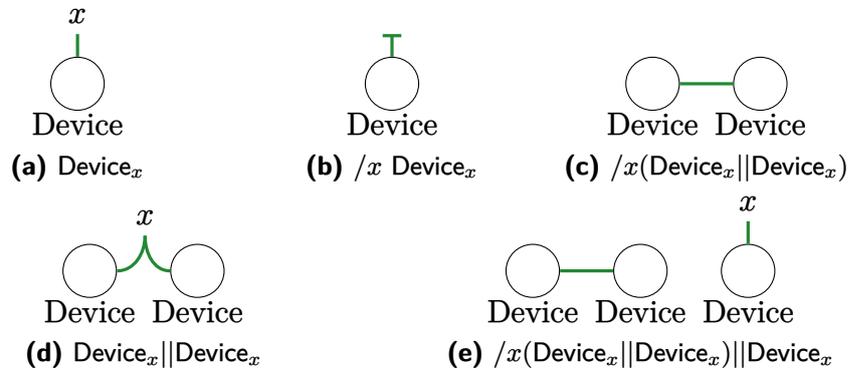
\begin{figure}[h]
\centering
\begin{subfigure}[b]{0.25\linewidth}
\centering
\begin{tikzpicture}
	\node[draw,circle,minimum size=20,label={[inner sep=0.5, name=id0l,align=center]south:{Device}}] (d0) {};
	\node[above=0.3 of d0] (x) {$x$};
	\draw[big edge] (x) to[out=270,in=90] (d0);
\end{tikzpicture}
\caption{Device$_x$}
\label{fig:device-potential}
\end{subfigure}
\begin{subfigure}[b]{0.25\linewidth}
\centering
\begin{tikzpicture}
	\node[draw,circle,minimum size=20,label={[inner sep=0.5, name=id0l,align=center]south:{Device}}] (d0) {};
	\draw[big edgec] (d0) to[out=90,in=-90] ($(d0.north) + (0,0.3)$);
\end{tikzpicture}
\caption{$/x$ Device$_x$}
\label{fig:device-disconnected}
\end{subfigure}
\begin{subfigure}[b]{0.25\linewidth}
\centering
\begin{tikzpicture}
	\node[draw,circle,minimum size=20,label={[inner sep=0.5, name=id0l,align=center]south:{Device}}] (d0) {};
	\node[right=0.7 of d0,draw,circle,minimum size=20,label={[inner sep=0.5, name=id1l,align=center]south:{Device}}] (d1) {};
	\draw[big edge] (d0) to[out=0,in=180] (d1);
\end{tikzpicture}
\caption{$/x($Device$_x||$Device$_x)$}
\label{fig:device-closed}
\end{subfigure}

\begin{subfigure}[b]{0.4\linewidth}
\centering
\begin{tikzpicture}
	\node[draw,circle,minimum size=20,label={[inner sep=0.5, name=id0l,align=center]south:{Device}}] (d0) {};
	\node[right=0.7 of d0,draw,circle,minimum size=20,label={[inner sep=0.5, name=id1l,align=center]south:{Device}}] (d1) {};
	
	\node[above right=0.3 of d0] (x) {$x$};
	\draw[big edge] (x) to[out=270,in=0] (d0);
	\draw[big edge] (x) to[out=270,in=180] (d1);
\end{tikzpicture}
\caption{Device$_x||$Device$_x$}
\label{fig:device-open}
\end{subfigure}
\begin{subfigure}[b]{0.4\linewidth}
\centering
\begin{tikzpicture}
	\node[draw,circle,minimum size=20,label={[inner sep=0.5, name=id0l,align=center]south:{Device}}] (d0) {};
	\node[right=0.7 of d0,draw,circle,minimum size=20,label={[inner sep=0.5, name=id1l,align=center]south:{Device}}] (d1) {};
	\node[right=0.7 of d1,draw,circle,minimum size=20,label={[inner sep=0.5, name=id2l,align=center]south:{Device}}] (d2) {};
	\draw[big edge] (d0) to[out=0,in=180] (d1);

	\node[above=0.3 of d2] (x) {$x$};
	\draw[big edge] (x) to[out=270,in=90] (d2);
\end{tikzpicture}
\caption{$/x($Device$_x||$Device$_x)||$Device$_x$}
\label{fig:device-bound}
\end{subfigure}
\caption{Link graphs of various devices and closures.}
\end{figure}

Let Device be a control of arity 1. The ion Device$_x$ (Figure \ref{fig:device-potential}) has one outer name $x$, which means it could potentially be linked.  

$/x$ Device$_x$ (Figure \ref{fig:device-disconnected}) is a disconnected device. 

$/x($Device$_x||$Device$_x)$ (Figure \ref{fig:device-closed}) are two connected devices with a closed link—no other connections are possible. 

Device$_x||$Device$_x$ (Figure \ref{fig:device-open}) are two connected devices with an outer name $x$—further connections are possible. 

Lastly, $/x($Device$_x||$Device$_x)||$Device$_x$ (Figure \ref{fig:device-bound}) are two connected devices with a closed link on the left and a device with an open link on the right — $x$ is bound in the former and free in the latter.

\subsection{Bigraphical reactive systems} \label{section:brs}

A \textit{Bigraphical Reactive System} (BRS) consists of bigraphs and a set of user-defined \textit{reaction rules} that specify permissible atomic changes to the bigraphs. A reaction rule consists of a pair of bigraphs: a \textit{redex} $L$ and a \textit{reactum} $R$. It is written as $L\rrul R$, specifying that an instance of $L$ can be substituted by $R$. 

Let Agent, Room and Contact be controls with arity 1. The reaction rule \rr{leave\_room} (Figure \ref{fig:react-leave-room}) allows an agent to move out of a room \cite{10.5555/1540607}. The sites numbered 0 and 1 allow the room to contain other occupants and the agent to carry any number of items respectively. The outer names in the redex allow the Agent and Room nodes to be linked to other nodes, perhaps elsewhere—any such link is retained in the reactum. Equally, these links may be disconnected—the context in which the rule is applied may close it off.

\begin{figure}[h]
\centering
\begin{tikzpicture}
 	\node[big site, inner sep=10pt,] (s0) {};
	\node[below right, inner sep=0pt, shift={(0.1,-0.1)}, ] at (s0.north west) {0};
 	\node[right=0.9 of s0, big site, inner sep=10pt,] (s1) {};
	\node[below right, inner sep=0pt, shift={(0.1,-0.1)}, ] at (s1.north west) {1};
	\node[fit=(s1), draw, rounded corners=2,inner sep=15pt,] (a0) {};
	\node[below right, inner sep=0pt, shift={(0.1,-0.1)}, ] at (a0.north west) {Agent};
	\node[fit=(s0)(a0), draw, rounded corners=2,inner sep=10pt,] (room) {};
	\node[below right, inner sep=0pt, shift={(0.1,-0.1)}, ] at (room.north west) {Room};
	\node[big region, fit=(room), inner sep=10pt,] (r0) {};
	\node[below right, inner sep=0pt, shift={(0.1,-0.1)}, ] at (r0.north west) {0};

	\node[above=0.25 of r0, shift={(0,0)}] (x) {$x$};
	\draw[big edge] (room)  to[out=90,in=270]  (x);
	\node[above=0.2] at (a0|-,|-r0.north) (y) {$y$};
	\draw[big edge] (a0)  to[out=90,in=270]  (y);
\end{tikzpicture}
\raisebox{1.7cm}{\large$\rrul$}
\begin{tikzpicture}
 	\node[big site, inner sep=10pt,] (s0) {};
	\node[below right, inner sep=0pt, shift={(0.1,-0.1)}, ] at (s0.north west) {0};
	\node[fit=(s0), draw, rounded corners=2,inner sep=15pt,] (room) {};
	\node[below right, inner sep=0pt, shift={(0.1,-0.1)}, ] at (room.north west) {Room};
 	\node[right=0.9 of room, big site, inner sep=10pt,] (s1) {};
	\node[below right, inner sep=0pt, shift={(0.1,-0.1)}, ] at (s1.north west) {1};
	\node[fit=(s1), draw, rounded corners=2,inner sep=15pt,] (a0) {};
	\node[below right, inner sep=0pt, shift={(0.1,-0.1)}, ] at (a0.north west) {Agent};
	\node[big region, fit=(room)(a0), inner sep=10pt,] (r0) {};
	\node[below right, inner sep=0pt, shift={(0.1,-0.1)}, ] at (r0.north west) {0};

	\node[above=0.25] at (room|-,|-r0.north) (x) {$x$};
	\draw[big edge] (room)  to[out=90,in=270]  (x);
	\node[above=0.2] at (a0|-,|-r0.north) (y) {$y$};
	\draw[big edge] (a0)  to[out=90,in=270]  (y);
\end{tikzpicture}
\captionsetup{justification=centering}
\caption{Reaction rule \rr{leave\_room}: Room$_x.(id|$Agent$_{y})\rrul\ $Room$_x|$Agent$_{y}$}
\label{fig:react-leave-room}
\end{figure}

If a sub-graph $L$ occurs within a larger bigraph $B$, the reaction rule can be applied to substitute $L$ with $R$, creating a new bigraph $B'$. Figure \ref{fig:built0} illustrates an example bigraph $E$ modelling an indoor space. In a BRS consisting of the bigraph $E$ and the set of reaction rules $[\rr{leave\_room}]$, \rr{leave\_room} can be applied to $E$ to move either of the two agents out of the room. Applying \rr{leave\_room} to the agent on the right results in the bigraph $E'$ (Figure \ref{fig:built1}). The relation induced by reaction rules is denoted by $\react$ i.e. $E\react E'$.

\begin{figure}[h]
\centering
\begin{subfigure}{0.45\linewidth}
\centering
\begin{tikzpicture}
   	\node[draw, rounded corners=2] (a0) {Agent};
	\node[right=0.9 of a0,draw, rounded corners=2] (k0) {Contact};
	\node[fit=(k0), draw, rounded corners=2,inner sep=15pt,] (a1) {};
	\node[below right, inner sep=0pt, shift={(0.1,-0.1)}, ] at (a1.north west) {Agent};
	\node[fit=(a0)(a1), draw, rounded corners=2,inner sep=10pt,] (ro0) {};
	\node[below right, inner sep=0pt, shift={(0.1,-0.1)}, ] at (ro0.north west) {Room};
	\node[fit=(ro0), draw, rounded corners=2,inner sep=15pt,] (b0) {};
	\node[below right, inner sep=0pt, shift={(0.1,-0.1)}, ] at (b0.north west) {Building};
	\node[big region, fit=(b0), inner sep=10pt,] (r0) {};
	\node[below right, inner sep=0pt, shift={(0.1,-0.1)}, ] at (r0.north west) {0};

	\node[above=0.25] at (a0|-,|-r0.north) (v) {$v$};
	\draw[big edge] (a0)  to[out=90,in=270]  (v);
	\node[above=0.25,shift={(-0.25,0)}] at (ro0|-,|-r0.north) (w) {$w$};
	\draw[big edge] (ro0)  to[out=95,in=270]  (w);
	\node[above=0.25,shift={(0.25,0)}] at (b0|-,|-r0.north) (x) {$x$};
	\draw[big edge] (b0)  to[out=85,in=270]  (x);
	\node[above=0.2,shift={(-0.25,0)}] at (k0|-,|-r0.north) (y) {$y$};
	\draw[big edge] (k0)  to[out=100,in=270]  (y);
	\node[above=0.25,shift={(0.25,0)}] at (a1|-,|-r0.north) (z) {$z$};
	\draw[big edge] (a1)  to[out=80,in=270]  (z);
\end{tikzpicture}
\caption{Bigraph $E$}
\label{fig:built0}
\end{subfigure}
\raisebox{2.7cm}{ \large$\react$}
\begin{subfigure}{0.45\linewidth}
\centering
\begin{tikzpicture}
	\node[draw, rounded corners=2] (a0) {Agent};
	\node[fit=(a0), draw, rounded corners=2,inner sep=15pt,] (ro0) {};
	\node[below right, inner sep=0pt, shift={(0.1,-0.1)}, ] at (ro0.north west) {Room};
	\node[right=0.9 of ro0,draw, rounded corners=2] (k0) {Contact};
	\node[fit=(k0), draw, rounded corners=2,inner sep=15pt,] (a1) {};
	\node[below right, inner sep=0pt, shift={(0.1,-0.1)}, ] at (a1.north west) {Agent};
	\node[fit=(ro0)(a1), draw, rounded corners=2,inner sep=15pt,] (b0) {};
	\node[below right, inner sep=0pt, shift={(0.1,-0.1)}, ] at (b0.north west) {Building};
	\node[big region, fit=(b0), inner sep=10pt,] (r0) {};
	\node[below right, inner sep=0pt, shift={(0.1,-0.1)}, ] at (r0.north west) {0};

	\node[above=0.25,shift={(-0.25,0)}] at (a0|-,|-r0.north) (v) {$v$};
	\draw[big edge] (a0)  to[out=95,in=270]  (v);
	\node[above=0.25,shift={(0.25,0)}] at (ro0|-,|-r0.north) (w) {$w$};
	\draw[big edge] (ro0)  to[out=85,in=270]  (w);
	\node[above=0.25] at (b0|-,|-r0.north) (x) {$x$};
	\draw[big edge] (b0)  to[out=90,in=270]  (x);
	\node[above=0.2,shift={(-0.25,0)}] at (k0|-,|-r0.north) (y) {$y$};
	\draw[big edge] (k0)  to[out=100,in=270]  (y);
	\node[above=0.25,shift={(0.25,0)}] at (a1|-,|-r0.north) (z) {$z$};
	\draw[big edge] (a1)  to[out=80,in=270]  (z);
\end{tikzpicture}
\caption{Bigraph $E'$}
\label{fig:built1}
\end{subfigure}
\caption{$E\react E'$ by applying reaction rule \rr{leave\_room}.}
\end{figure}

The inverse of the reaction rule \rr{leave\_room}—that allows agents to enter a room—can be added to the BRS to model the motion of agents in an indoor space. See \S\ref{section:brs-lock} for a BRS the models agents manipulating their environment: locking a room to prevent passage.

This concludes the features of bigraphs used in the specification of bigraphs of the real world proposed in this dissertation. For an introduction to bigraphs with sharing, which is only used in alternative methods, see \S\ref{section:sharing}.

\section{OpenStreetMap}

OpenStreetMap (OSM) is an open geographic database of stationary objects around the world, including infrastructure, points of interest and boundaries.

OSM data is made up of 3 types of elements:
\begin{itemize}
\item A \textit{node element} represents a specific point on earth, defined by its latitude and longitude.
\item A \textit{way element} is a connected line of nodes. Way elements are used to represent linear features such as roads. A closed way element, in which the first and last node are the same, can represent a polygon that describes the boundary of an area.
\item A \textit{relation element} is a collection of node, way and other relation elements, documenting a relationship between its members. It can represent a multipolygon that describes an area with holes or that consists of disjoint parts.
\end{itemize}

Each element has a unique \rr{id} attribute that identifies it. Element types have their own ID space—there could be a node element with \rr{id=100} and a way element with \rr{id=100}, but they are unlikely to be related.

Elements may have \textit{tags}: key-value pairs that give meaning to the element and describe its characteristics. While there is no fixed ontology for tags, the mapping community uses conventions and rules that are decided through common usage, voting and discussion. For example, any man-made structure with a roof is considered a building and should have a tag \rr{building=yes} or another value that classifies the type of building (e.g. \rr{house}, \rr{detached}, \rr{apartments}). Another example is the tag with the key \rr{highway}, which is used for any kind of road, street or path; its value specifies the type.

OSM is built by a community of volunteers who annotate data gathered from surveys, aerial imagery and other free geodata sources. Consequently, OSM offers the most comprehensive coverage for outdoor, street-level infrastructure such as buildings and road networks, with sufficient detail to support the development of numerous routing services. \texttt{building} and \texttt{highway} are the most commonly used keys in OSM (Figure \ref{fig:common-keys}).

\section{Requirements analysis} \label{section:requirements}
This dissertation aims to borrow the comprehensive detail of OSM data to build a bigraph of the real world. It starts by deriving a hierarchical space-partitioning tree of the world from OSM data, which is then used as the place graph of the bigraph of the real world.

\subsection*{Hierarchical space-partitioning tree}

\hangindent=2em
\hangafter=1
\textbf{All named buildings} This dissertation focuses on modelling the locations of built environments, where ubiquitous systems are typically located. OSM offers the most comprehensive building coverage, however indoor coverage is limited due to mapping challenges and privacy concerns; thus, buildings are atomic units here. 

\hangindent=2em
\hangafter=1
\textbf{Deep hierarchy and partition granularity} This dissertation aims to partition the world finely to provide more precise notions of spatial containment and adjacency.

\hangindent=2em
\hangafter=1
\textbf{Familiar partitions} The partitions must be interpretable and recognisable to humans.

\hangindent=2em
\hangafter=1
\textbf{Generalisability} The methods used to partition the world should be applicable to regions from any part of the world.

\subsection*{Bigraphs of the real world}

\hangindent=2em
\hangafter=1
\textbf{Complete road network} Road networks form the backbone of human mobility: they enable the majority of everyday travel. OSM data is very complete for street networks \cite{10.1371/journal.pone.0180698}; this dissertation aims to preserve complete street connectivity information in the bigraph, which is likely to be useful in many applications.

\hangindent=2em
\hangafter=1
\textbf{Spatial locality encoded in place graph} Placing should have strong correspondence to spatial locality in the real world, so that containment and adjacency of physical locations and objects is immediately clear from the place graph.

\hangindent=2em
\hangafter=1
\textbf{Modularity} The proposed specification should support the independent construction of bigraphs of different regions, and the capability to later combine them.

\hangindent=2em
\hangafter=1
\textbf{Unifying model} Bigraphs of the real world should unify bigraphs from past and future work.

\hangindent=2em
\hangafter=1
\textbf{Efficient computation} The time and memory required to build bigraphs of the real world should be minimised.

\hangindent=2em
\hangafter=1
\textbf{Real-time transformation} The bigraphs should support real-time transformation according to reaction rules to model the dynamic behaviour of ubiquitous systems.

While the initial scope was limited to constructing a bigraph of Cambridgeshire, a county in the UK, the extension to build bigraphs for different parts of the world was subsequently achieved.

\section{Software engineering tools and techniques}

\subsection*{Languages and libraries}
\hangindent=2em
\hangafter=1
\textbf{Overpass API and Query Language} The Overpass API's powerful query language provides functionality to filter and select elements based on their tags and location. It is distributed under the GNU AGPL v3 license.

\hangindent=2em
\hangafter=1
\textbf{osm\_xml OCaml library} osm\_xml \cite{osm-xml} is a small library that provides functionality to parse the results of an Overpass API query. It is distributed under the 3-Clause BSD License.

\hangindent=2em
\hangafter=1
\textbf{BigraphER OCaml library} BigraphER \cite{DBLP:conf/cav/SevegnaniC16} provides a library for programmatically constructing bigraphs and transforming them using reaction rules. Actively used by the bigraph research community, it supports extensions such as bigraphs with sharing (\S\ref{section:sharing}). It is distributed under the Simplified BSD License.

\hangindent=2em
\hangafter=1
\textbf{OCaml} The tool for constructing and experimenting with bigraphs of the real world was written from scratch in OCaml to interface naturally with the BigraphER library. Using OCaml also supports compatibility with other projects in the Department, such as the Spatial Name System \cite{10.1145/3626111.3628210}.

\subsection*{Licensing and open-source community}
The Overpass API was used without modifications, therefore its results can be used freely. The osm\_xml and BigraphER libraries were reused with modification; accordingly, the more restrictive 3-Clause BSD License is included with the source code. The algorithmic improvements to BigraphER described in \S \ref{section:bigraphER} were submitted as pull requests to the open-source repository and have been accepted.

\subsection*{Development methodology}

An iterative approach was used to complete the project. To manage the risks associated with attempting to construct and transform bigraphs of unprecedented size, the first iteration focused on prototyping an Overpass query to extract OSM data, parsing the results, and generating a preliminary bigraph. The bigraph produced at the end of each iteration provided a graphical visualisation that encodes adjacency and containment of geographic elements, aiding in the discovery of unexpected irregularities in OSM data.

Subsequent development progressed in two parallel tasks, each using a spiral model. The first task involved refining the specification of the hierarchical space-partitioning tree and perfecting the Overpass query. The second task focused on developing the formalism of bigraphs of the real world, programmatically constructing one from the results of the Overpass query, and defining reaction rules within BRSs to model motion and communication of agents.

The rapid feedback cycle of the iterative spiral model quickly exposed critical performance limitations of BigraphER when handling large bigraphs, prompting the extension task of implementing algorithmic improvements to enable efficient construction and real-time transformation of bigraphs of the real world.

\subsection*{Code style}
Automatic formatting with ocamlformat \cite{ocamlformat} ensured a consistent and conventional code style.

\subsection*{Hardware, version control and backup}
Development and experiments were conducted on a personal computer equipped with an AMD Ryzen 5 7640U processor and 16GB of RAM. GitHub and Bitbucket was used for version control of repositories, while OneDrive provided automatic backups of the \TeX\ source code. This robust versioning and backup strategy allowed this dissertation to survive two Windows blue screens of death.

\chapter{Implementation}

\textit{This chapter explains the specification that was developed for partitioning the real world into a hierarchical tree and modelling it with a bigraph. The structure of written source code is presented, followed by an explanation of the implementation steps taken to achieve the objectives of this dissertation.}

\section{Hierarchical space-partitioning tree of the world} \label{section:hierarchical-tree}

This dissertation proposes a specification for organising buildings, streets and administrative areas in the world into a space-partitioning tree that encodes a hierarchy of containment. The choices made here are justified by completeness and uniformity of coverage in OSM data, a result of established tagging conventions. The OSM data statistics presented refer to the global dataset (unless specified otherwise) and are based on the snapshot from 27 April 2025.

\subsection{Buildings} \label{section:building-tree}
\begin{figure}
\centering
\def\nCircle{(0,0) circle (1)}
\def\sCircle{(4.05,0.5) circle (3.9)}
\def\hCircle{(4.2,0,0) circle (4)}
\begin{tikzpicture}
\begin{scope}
	\fill[T-Q-PH5] \nCircle;
\end{scope}
\begin{scope}
          \clip \hCircle;
	\fill[T-Q-PH1] \sCircle;
\end{scope}
\begin{scope}
          \clip \nCircle;
	\fill[T-Q-PH3] \sCircle;
\end{scope}
\draw \nCircle[thick];
\draw \sCircle[thick];
\draw \hCircle[thick];

\node at (-1,1) {name};
\node at (-0.2,3) {addr:street};
\node at (8.7,-3) {addr:housenumber};

\node at (-0.35,0) {\small0.78\%};
\node at (4.2,0) {\small11.94\%};

\node at (-0.35,-2) (glabel){\small0.44\%};
\draw (0.7,0) -- (glabel.north);
\end{tikzpicture}
\caption{Venn diagram of the combination of \rr{name}, \rr{addr:street} and \rr{addr:housenumber} tags on building elements. Coloured segments are labelled with the percentage of building elements with the combination of tags. }
\label{fig:buildings-venn}
\end{figure}
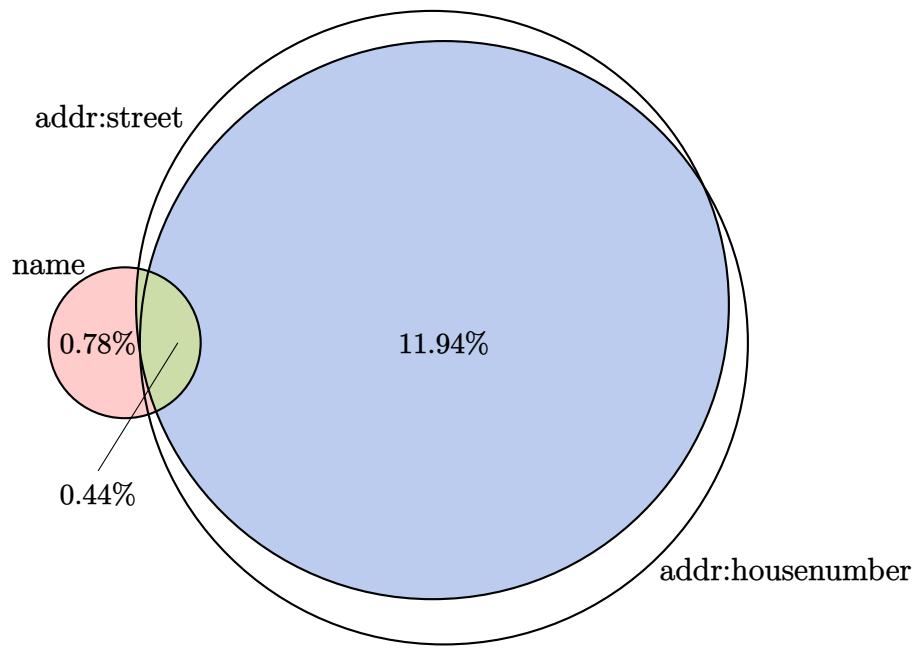

A requirement of this dissertation is that all named buildings in OSM are represented in the hierarchical tree. Buildings annotated in OSM are represented by node, way or relation elements that have a \rr{building} tag. However, only 1.22\% of buildings have a \rr{name} tag. This is primarily because most buildings have their location and outline annotated in OSM using data from aerial surveys and other geodata sources but have not yet been properly tagged; it is also because not all buildings have common names, often they are referred to by their address. OSM uses the Karlsruhe Schema for addresses, which recommends at minimum the inclusion of the \rr{addr:housenumber} and \rr{addr:street} tags. Approximately 12\% of buildings are tagged with addresses following this schema. Hence, this dissertation defines named buildings in OSM as those with a \rr{name} tag (\textcolor{red}{red}+\textcolor{green}{green} segment in Figure \ref{fig:buildings-venn}), or otherwise those with both \rr{addr:housenumber} and \rr{addr:street} tags (\textcolor{blue}{blue} segment Figure \ref{fig:buildings-venn}) for which their names shall be derived from the combination of the values of the two tags (e.g. 51 Madingley Road).

\subsection{Streets} \label{section:streets-tree}
A requirement of this dissertation is that all streets annotated in OSM are represented in the hierarchical tree. Streets are represented in OSM by way elements tagged with the key \rr{highway} and one of the following values: \rr{motorway}, \rr{trunk}, \rr{primary}, \rr{secondary}, \rr{tertiary}, \rr{unclassified}, \rr{residential}, \rr{motorway\_link}, \rr{trunk\_link}, \rr{primary\_link}, \rr{secondary\_link}, or \rr{tertiary\_link}. 41.96\% of streets have either a \rr{name} tag (e.g. \rr{name=J J Thomson Ave}) or a \rr{ref} tag (e.g. \rr{ref=M11}), the rest are often unnamed link roads that connect named streets. Streets shall be added to the hierarchical tree, each identified by the value of their \rr{name} tag or otherwise their \rr{ref} tag.

Additionally, approximately 9 in 10 named buildings have a tag with key \rr{addr:street} (\textcolor{green}{green}+\textcolor{blue}{blue} segment in Figure \ref{fig:buildings-venn}) and value of a street name. In the hierarchical tree, each buildings shall be the child of the street named in its address.

\subsection{Administrative areas}\label{section:boundaries-tree}
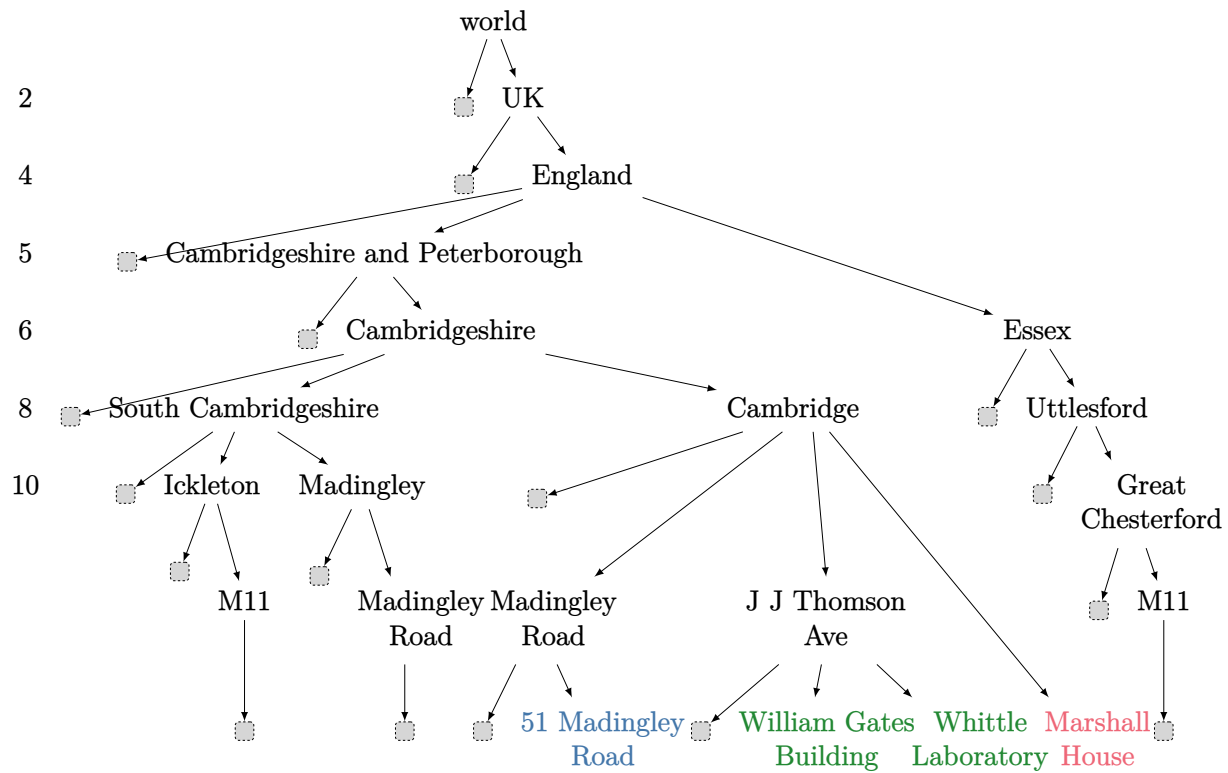
\begin{figure}[h]
\centering
\resizebox{\linewidth}{!}{
\begin{forest}
  for tree={edge = {-latex},}
[,phantom 
	[2,tier=2]
	[4,tier=4]
	[5,tier=5]
	[6,tier=6]
	[8,tier=8]
	[10,tier=10]
	[world
		[,big site]
		[UK,tier=2
			[,big site]
			[England,tier=4
				[,big site]
				[Cambridgeshire and Peterborough,tier=5
					[,big site]
					[Cambridgeshire,tier=6
						[,big site]
						[South Cambridgeshire,tier=8
							[,big site]
							[Ickleton,tier=10
								[,big site]
								[M11,tier=street
									[,big site,tier=building]
								]
							]
							[Madingley,tier=10
								[,big site]
								[Madingley\\Road,tier=street,align=center,text width=1.4cm,
									[,big site,tier=building]
								]
							]
						]
						[Cambridge,tier=8,text width=2.5cm,
							[,big site]
							[Madingley\\Road,tier=street,align=center,text width=1.4cm,
								[,big site]
								[51 Madingley\\Road,tier=building,align=center,text width=2.1cm, text=blue]
							]
							[J J Thomson\\Ave,tier=street,align=center,text width=2.5cm,
								[,big site]
								[William Gates\\Building,tier=building,align=center,text width=2cm, text=green]
								[Whittle\\Laboratory,tier=building,align=center,text width=1.4cm, text=green]
							]
							[Marshall\\House,tier=building,align=center,text width=1.2cm, text=red]
						]
					]
				]
				[Essex,tier=6
					[,big site]
					[Uttlesford,tier=8
						[,big site]
						[Great\\Chesterford,tier=10,align=center,text width=1.5cm
							[,big site]
							[M11,tier=street
								[,big site,tier=building]
							]
						]
					]
				]
			]
		]
	]
]
\end{forest}
}
\caption{Hierarchical space-partitioning tree of the world. \rr{admin\_level} values of boundaries are labelled on the left. Shaded rectangles are placeholders for other geographic locations, not included here for simplicity. Coloured buildings correspond to the coloured segments in Figure \ref{fig:buildings-venn}.}
\label{fig:tree}
\end{figure}

Administrative boundaries subdivide geographic areas into administrative units recognised by governments. They define a hierarchy: administrative areas are subdivided into administrative units (Figure \ref{fig:child-boundaries-cambridgeshire}), which in turn are subdivided into smaller units (Figure \ref{fig:descendant-boundaries-cambridgeshire}). In OSM, administrative boundaries are represented by relation elements with a \rr{boundary=adminstrative} tag in combination with a tag with the key \rr{admin\_level} and a value from 2 to 11. \rr{admin\_level=2} is used to mark the borders of the 249 countries listed on the ISO-3166-1 standard. \rr{admin\_level=3-11} can be used to mark subnational borders; the number of levels used in different countries varies. The \rr{admin\_level} values used in the UK are labelled in Figure \ref{fig:tree}.

These administrative areas shall be added to the hierarchical space-partitioning tree, identified by the value in their \rr{name} tag or \rr{short\_name} tag if present. Streets and buildings shall be made the children of the smallest enclosing administrative boundary. However, streets may cross administrative boundaries—for example, Madingley Road crosses the boundary between Madingley and Cambridge. In that case, the road is segmented into two parts: one part is contained in Madingley and the other is contained in Cambridge. Buildings are still only contained by a single boundary: 51 Madingley Road is in Cambridge and not Madingley, hence it is a child of the Madingley Road segment in Cambridge. The resulting hierarchical space-partitioning tree of the world is illustrated in Figure \ref{fig:tree}.

\subsection{Spatial names} \label{section:spatial-name}
The hierarchical space-partitioning tree constructed according to this specification also defines a system of geographic identifiers for all buildings, streets and administrative areas in the world. Spatial names, defined by appending the names of enclosing areas encountered while traversing from the root of the tree to the object, are assigned to 
{\sloppy
\begin{itemize}
\item \textbf{areas} e.g. the district of Cambridge can be addressed as "Cambridge.Cambridgeshire.Cambridgeshire and Peterborough.England.UK"
\item \textbf{buildings} e.g. William Gates Building can be addressed as "William Gates Building.J J Thomson Ave.Cambridge.Cambridgeshire.Cambridgeshire and Peterborough.England.UK"
\item \textbf{agents} e.g. a microphone in the William Gates Building can be addressed as "microphone.William Gates Building.J J Thomson Ave.Cambridge.Cambridgeshire.Cambridgeshire and Peterborough.England.UK"
\end{itemize}
}

This system of geographic identifiers fulfils the requirement for spatial names to be used in the Spatial Name System proposed by Gibb et al. \cite{10.1145/3626111.3628210}: unambiguous and unique human-readable addresses which encode a hierarchy of containment. Therefore, this dissertation delivers a specification and implementation for the assignment of spatial names to ubiquitous systems; the Spatial Name System provides possible schemes to resolve spatial names to IP, Bluetooth or Zigbee addresses.

\section{Bigraph of the real world}

The hierarchical tree in Figure \ref{fig:tree} can be used as the place graph of a bigraph that models the real world. First, however, this section discusses the methods used to encode street connectivity and add identifiers to nodes in the bigraph.

\subsection{Modelling street connectivity using linked junctions} \label{street-connectivity-link}
\begin{figure}[h]
\centering
\resizebox{\linewidth}{!}{
\begin{tikzpicture}
\coordinate (M0) at (-0.3,4.25);
\coordinate (M1) at (0,4.25);
\coordinate (M2) at (6,4.25);
\coordinate (M3) at (11,4.25);
\coordinate (M4) at (15,4.25);
\coordinate (M5) at (15.3,4.25);
\coordinate (CB1) at (1,0);
\coordinate (CB2) at (6,0);
\coordinate (CB3) at (11,0);
\coordinate (HC) at (1,3);
\coordinate (AL) at (1,-4.5);

\draw[double =white,double distance=0.5cm,line width=0.4mm, line cap=round] (M1) -- (M4)(CB1) -- (CB3)(M2)--(CB2)(M3)--(CB3)(HC)--(AL);
\draw[double =white,double distance=0.5cm,line width=0.4mm] (M0) -- (M1);
\draw[double =white,double distance=0.5cm,line width=0.4mm] (M4) -- (M5);

\node[draw,circle, fill=T-Q-PH1,inner sep=1pt] at (M1) {0};
\node[draw,circle, fill=T-Q-PH1,inner sep=1pt] at (M2) {1};
\node[draw,circle, fill=T-Q-PH1,inner sep=1pt] at (M3) {2};
\node[draw,circle, fill=T-Q-PH1,inner sep=1pt] at (M4) {3};
\node[draw,circle, fill=T-Q-PH1,inner sep=1pt] at (CB1) {4};
\node[draw,circle, fill=T-Q-PH1,inner sep=1pt] at (CB2) {5};
\node[draw,circle, fill=T-Q-PH1,inner sep=1pt] at (CB3) {6};
\node[draw,circle, fill=T-Q-PH1,inner sep=1pt] at (HC) {7};
\node[draw,circle, fill=T-Q-PH1,inner sep=1pt] at (AL) {8};

\node[] at ($(M1)!0.5!(M2)$) {Madingley Road};
\node[] at ($(M2)!0.5!(M3)$) {Madingley Road};
\node[] at ($(M3)!0.5!(M4)$) {Madingley Road};
\node[] at ($(CB1)!0.5!(CB2)$) {Charles Babbage Road};
\node[] at ($(CB2)!0.5!(CB3)$) {Charles Babbage Road};
\node[, rotate=90] at ($(CB1)!0.5!(AL)$) {Ada Lovelace Road};
\node[, rotate=90] at ($(CB1)!0.5!(HC)$) {High Cross};
\node[, rotate=90] at ($(M3)!0.5!(CB3)$) {J J Thomson Ave};
\end{tikzpicture}
}
\caption{Simple road network. Nodes are illustrated as blue circles.}
\label{fig:connected-roads}
\end{figure}
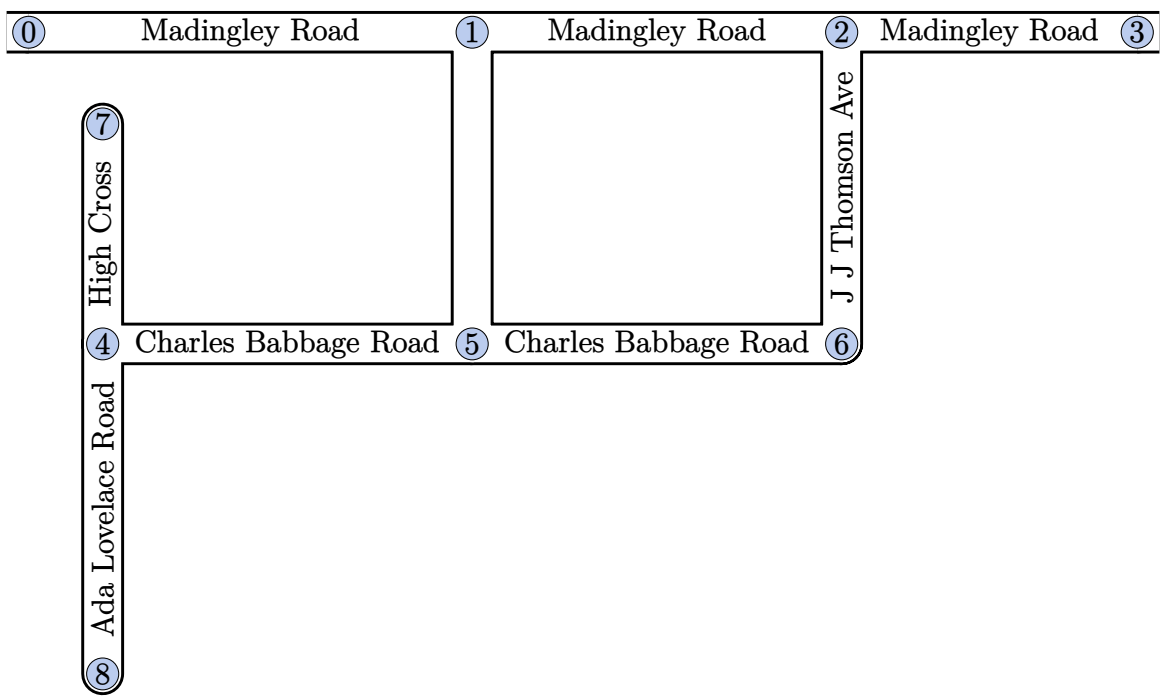

Streets are represented in OSM by way elements, which are a connected line of nodes. Streets are often split at intersections with other streets for routing purposes. In the simple road network in Figure \ref{fig:connected-roads}, there are three way elements with the \rr{name=Madingley Road} tag: one consists of the nodes 0 and 1, another consists of nodes 1 and 2, and the third consists of nodes 2 and 3. When two or more streets intersect, the road junction is mapped as a node shared by two or more way elements (e.g. nodes 1, 2, 4, 5, 6 in Figure \ref{fig:connected-roads}). 

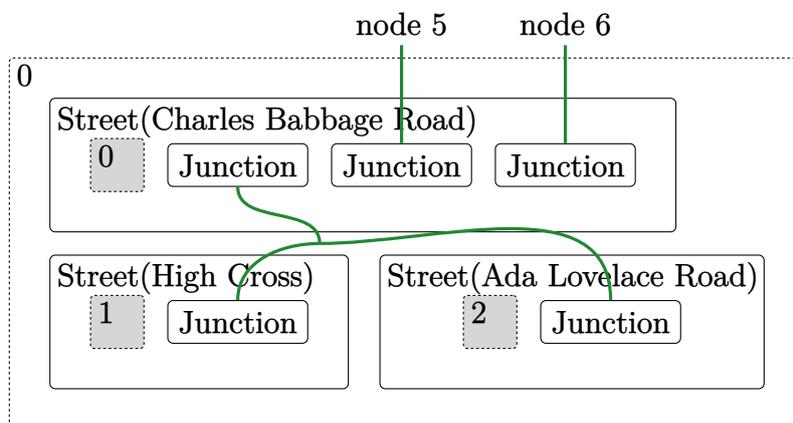
\begin{figure}[h]
\centering
\begin{tikzpicture}
 	\node[big site, inner sep=10pt,] (SiteHighCross) {};
	\node[below right, inner sep=0pt, shift={(0.1,-0.1)}, ] at (SiteHighCross.north west) {1};
	\node[right=0.3 of SiteHighCross, draw, rounded corners=2] (Junction0HighCross) {Junction};
	\node[fit=(Junction0HighCross)(SiteHighCross), draw, rounded corners=2,inner sep=15pt,] (StreetHighCross) {};
	\node[below right, inner sep=0pt, shift={(0.1,-0.1)}, ] at (StreetHighCross.north west) {Street(High Cross)};
 	\node[right=1.5 of StreetHighCross, big site, inner sep=10pt,] (SiteAda) {};
	\node[below right, inner sep=0pt, shift={(0.1,-0.1)}, ] at (SiteAda.north west) {2};
	\node[right=0.3 of SiteAda, draw, rounded corners=2] (Junction0Ada) {Junction};
	\node[fit=(Junction0Ada)(SiteAda), draw, rounded corners=2,inner sep=15pt,text width=4cm] (StreetAda) {};
	\node[below right, inner sep=0pt, shift={(0.1,-0.1)}, ] at (StreetAda.north west) {Street(Ada Lovelace Road)};
	\node[above=1.5 of Junction0HighCross, draw, rounded corners=2] (Junction0Charles) {Junction};
	\node[right=0.3 of Junction0Charles, draw, rounded corners=2] (Junction1Charles) {Junction};
	\node[right=0.3 of Junction1Charles, draw, rounded corners=2] (Junction2Charles) {Junction};
 	\node[left=0.3 of Junction0Charles, big site, inner sep=10pt,] (SiteCharlesBabbage) {};
	\node[below right, inner sep=0pt, shift={(0.1,-0.1)}, ] at (SiteCharlesBabbage.north west) {0};
	\node[fit=(Junction2Charles)(SiteCharlesBabbage), draw, rounded corners=2,inner sep=15pt] (StreetCharlesBabbage) {};
	\node[below right, inner sep=0pt, shift={(0.1,-0.1)}, ] at (StreetCharlesBabbage.north west) {Street(Charles Babbage Road)};
	\node[big region,fit=(StreetCharlesBabbage)(StreetAda),inner sep=15pt,] (r0) {};
	\node[below right, inner sep=0pt, shift={(0.1,-0.1)}, ] at (r0.north west) {0};
	\coordinate (HighCrossAdaCharles) at ($(Junction1Charles)!0.5!(Junction0HighCross)$);
	\draw[big edge] (Junction0HighCross)  to[out=90,in=180]  (HighCrossAdaCharles);
	\draw[big edge] (Junction0Ada)  to[out=90,in=0]  (HighCrossAdaCharles);
	\draw[big edge] (Junction0Charles)  to[out=270,in=90]  (HighCrossAdaCharles);
	\node[above=1.3 of Junction1Charles](5){node 5};
	\draw[big edge] (Junction1Charles)  to[out=90,in=270]  (5);
	\node[above=1.3 of Junction2Charles](6){node 6};
	\draw[big edge] (Junction2Charles)  to[out=90,in=270]  (6);
\end{tikzpicture}
\caption{Bigraph representation of intersection between Charles Babbage Road, High Cross and Ada Lovelace Road.}
\label{fig:bigraph-connected-streets}
\end{figure}

The bigraph shall have one Street node for each street name. When two or more streets intersect, their Street nodes shall each contain a Junction node, all of which shall be linked to each other. The resulting bigraph (example illustrated in Figure \ref{fig:bigraph-connected-streets}) is indeed an abstraction of OSM's representation of streets: OSM way elements with the same value in the \rr{name} tag are encoded as a single Street node in the bigraph; an OSM way element is a line of connected nodes, some of which mark intersections with other streets, the Street node in the bigraph contains a Junction node for each intersection.

Since this dissertation also requires that the complete road network is represented, unnamed highways (e.g. the way element consisting of nodes 1 and 5 in Figure \ref{fig:connected-roads}) with neither \rr{name} nor \rr{ref} tag must also be included so that no road connectivity information is lost. Therefore, all highways annotated in OSM shall be included in the bigraph: those without a \rr{name} or \rr{ref} tag are identified by their OSM way \rr{id} attribute (e.g. way 53600489).

\subsection{Separate perspectives for the physical world and IDs}
The nodes in the bigraph have the following identifiers:
\begin{itemize}
\item Boundaries are identified by the value in their \rr{name} or \rr{short\_name} tag.
\item Streets are identified by the value in their \rr{name} or \rr{ref} tag, otherwise by their way \rr{id} attribute.
\item Buildings are identified by the value in their \rr{name} tag, otherwise by a combination of the values in their \rr{addr:housenumber} and \rr{addr:street} tag.
\end{itemize}

Controls that are used in the bigraph are defined as follows:
\begin{itemize}
\item Boundary, arity 1. Single link to its identifier.
\item Street, arity 1. Single link to its identifier.
\item Building, arity 1. Single link to its identifier.
\item Junction, arity 1. Single link to other junctions that represents an intersection between streets.
\item ID, arity 1. Single link to the node it identifies.
\end{itemize}
Instead of adding string parameters to the Boundary, Street and Building nodes directly, the identifier shall be made a parameter of an ID node that is linked. This allows for the definition of general reaction rules that apply to all nodes of a control (otherwise, a parameterised reaction rule has to be created for each unique parameter for the redex to match). All ID nodes are gathered in a region (region 0 in Figure \ref{fig:bigraph-Cambridgeshire}) referred to as the ID perspective. Boundary, Street, Building and Junction nodes are contained in another region (region 1 in Figure \ref{fig:bigraph-Cambridgeshire}) referred to as the Physical perspective. This use of parallel perspectives provides a strong separation of concerns \cite{archibald2024practicalmodellingbigraphs}: the Physical perspective only contains nodes that have a physical counterpart, so placing in the bigraph directly informs of the containment and adjacency of physical locations and objects.

The bigraph of Cambridgeshire constructed according to this specification is illustrated in Figure \ref{fig:bigraph-Cambridgeshire}. The open link with outer name "node 215742" can be connected to a Junction node outside Cambridgeshire, in a neighbouring region. In fact, the complete bigraph of Cambridgeshire has 225 outer names, one for each intersection between streets inside and outside Cambridgeshire (Figure \ref{fig:cambridgeshire-outer-names}).

\begin{figure}
\centering
\begin{tikzpicture}
 	\node[big site, inner sep=10pt,] (SiteHighCross) {};
	\node[below right, inner sep=0pt, shift={(0.1,-0.1)}, ] at (SiteHighCross.north west) {11};
	\node[right=0.3 of SiteHighCross, draw, rounded corners=2] (Junction0HighCross) {Junction};
	\node[fit=(Junction0HighCross)(SiteHighCross), draw, rounded corners=2,inner sep=15pt,] (StreetHighCross) {};
	\node[below right, inner sep=0pt, shift={(0.1,-0.1)}, ] at (StreetHighCross.north west) {Street};
 	\node[right=0.9 of StreetHighCross, big site, inner sep=10pt,] (SiteAda) {};
	\node[below right, inner sep=0pt, shift={(0.1,-0.1)}, ] at (SiteAda.north west) {12};
	\node[right=0.3 of SiteAda, draw, rounded corners=2] (Junction0Ada) {Junction};
	\node[fit=(Junction0Ada)(SiteAda), draw, rounded corners=2,inner sep=15pt,] (StreetAda) {};
	\node[below right, inner sep=0pt, shift={(0.1,-0.1)}, ] at (StreetAda.north west) {Street};
	\node[below right,draw, rounded corners=2] at (-7, |- StreetHighCross.north) (IDHighCross) {ID(High Cross)};
	\draw[big edge] (IDHighCross)  to[out=0,in=180]  (StreetHighCross.west|-,|-IDHighCross);
	\node[below right,draw, rounded corners=2, shift={(0,-0.3)},text width=3.3cm] at (-7, |- IDHighCross.south) (IDAda) {ID(Ada Lovelace Road)};
	\draw[big edge] (IDAda)  to[out=0,in=180]  (StreetAda);

 	\node[above= 1.5 of SiteHighCross, big site, inner sep=10pt,] (SiteCharles) {};
	\node[below right, inner sep=0pt, shift={(0.1,-0.1)}, ] at (SiteCharles.north west) {10};
	\node[right=0.3 of SiteCharles, draw, rounded corners=2] (Junction0Charles) {Junction};
	\node[right=0.3 of Junction0Charles, draw, rounded corners=2] (Junction1Charles) {Junction};
	\coordinate (HighCrossAdaCharles) at ($(Junction1Charles)!0.5!(Junction0HighCross)$);
	\draw[big edge] (Junction0HighCross)  to[out=90,in=180]  (HighCrossAdaCharles);
	\draw[big edge] (Junction0Ada)  to[out=90,in=0]  (HighCrossAdaCharles);
	\draw[big edge] (Junction0Charles)  to[out=270,in=90]  (HighCrossAdaCharles);
	\node[right=0.3 of Junction1Charles, draw, rounded corners=2] (Junction2Charles) {Junction};
	\node[fit=(Junction2Charles)(SiteCharles), draw, rounded corners=2,inner sep=15pt,] (StreetCharles) {};
	\node[below right, inner sep=0pt, shift={(0.1,-0.1)} ] at (StreetCharles.north west) {Street};
	\node[below right,draw, rounded corners=2,text width=3.3cm] at (-7, |- StreetCharles.north) (IDCharles) {ID(Charles Babbage Road)};
	\draw[big edge] (IDCharles)  to[out=0,in=180]  (StreetCharles.west|-,|-IDCharles);

	\node[above= 1.5 of Junction0Charles, draw, rounded corners=2] (Junction0Way) {Junction};
	\draw[big edge] (Junction1Charles)  to[out=90,in=270]  (Junction0Way);
	\node[above=0.3 of Junction0Way, draw, rounded corners=2] (Junction1Way) {Junction};
	\node[fit=(Junction0Way)(Junction1Way), draw, rounded corners=2,inner sep=15pt,] (StreetWay) {};
	\node[below right, inner sep=0pt, shift={(0.1,-0.1)}, ] at (StreetWay.north west) {Street};
	\node[above=1.5 of Junction2Charles, draw, rounded corners=2] (Junction0JJ) {Junction};
	\draw[big edge] (Junction2Charles)  to[out=90,in=270]  (Junction0JJ);
   	\node[above left, draw, rounded corners=2,inner sep=12pt, text width=0.9cm, shift={(0,0.3)}] at (Junction0JJ.north east) (BuildingWhittle) {};
	\node[below right, inner sep=0pt, shift={(0.1,-0.1)}, ] at (BuildingWhittle.north west) {Building};
   	\node[above=0.3 of BuildingWhittle, draw, rounded corners=2,inner sep=12pt, text width=0.9cm] (BuildingWGB) {};
	\node[below right, inner sep=0pt, shift={(0.1,-0.1)}, ] at (BuildingWGB.north west) {Building};
	\node[above left, draw, rounded corners=2, shift={(0,0.3)}] at (BuildingWGB.north east) (Junction1JJ) {Junction};
 	\node[below right , big site, inner sep=10pt,shift={(-1,0)}] at (Junction1JJ.north west)(SiteJJ) {};
	\node[below right, inner sep=0pt, shift={(0.1,-0.1)}, ] at (SiteJJ.north west) {9};
	\node[fit=(Junction0JJ)(SiteJJ), draw, rounded corners=2,inner sep=15pt,] (StreetJJ) {};
	\node[below right, inner sep=0pt, shift={(0.1,-0.1)}, ] at (StreetJJ.north west) {Street};
	\node[below right,draw, rounded corners=2,text width=3.3cm] at (-7, |- StreetJJ.north) (IDJJ) {ID(J J Thomson Ave)};
	\draw[big edge] (IDJJ)  to[out=0,in=180]  (StreetJJ.west|-,|-IDJJ);
	\node[below right,draw, rounded corners=2, shift={(0,-0.1)},text width=3.3cm] at (-7, |- IDJJ.south) (IDWGB) {ID(William Gates Building)};
	\draw[big edge] (IDWGB)  to[out=0,in=180]  (BuildingWGB);
	\node[below right,draw, rounded corners=2, shift={(0,-0.1)},text width=3.3cm] at (-7, |- IDWGB.south) (IDWhittle) {ID(Whittle Laboratory)};
	\draw[big edge] (IDWhittle)  to[out=0,in=180]  (BuildingWhittle);
	\node[below right,draw, rounded corners=2, shift={(0,-0.3)},text width=3.3cm] at (-7, |- IDWhittle.south) (IDWay) {ID(way 53600489)};
	\draw[big edge] (IDWay)  to[out=0,in=180]  (StreetWay);

	\node[above=1.6 of Junction1JJ, draw, rounded corners=2] (Junction1Mad) {Junction};
	\draw[big edge] (Junction1JJ)  to[out=90,in=270]  (Junction1Mad);
	\node[left=0.3 of Junction1Mad, draw, rounded corners=2] (Junction0Mad) {Junction};
	\draw[big edge] (Junction1Way)  to[out=90,in=270]  (Junction0Mad);
   	\node[above left, draw, rounded corners=2,inner sep=12pt, text width=0.9cm, shift={(0,0.3)}] at (Junction1Mad.north east) (Building51Mad) {};
	\node[below right, inner sep=0pt, shift={(0.1,-0.1)}, ] at (Building51Mad.north west) {Building};
	\node[above left, draw, rounded corners=2, shift={(0,0.3)}] at (Building51Mad.north east) (Junction2Mad) {Junction};
 	\node[below right, big site, inner sep=10pt,] at (Junction0Mad.west|-,|-Junction2Mad.north) (SiteMad) {};
	\node[below right, inner sep=0pt, shift={(0.1,-0.1)}, ] at (SiteMad.north west) {8};
	\node[fit=(Junction2Mad)(Junction0Mad), draw, rounded corners=2,inner sep=15pt,] (StreetMad) {};
	\node[below right, inner sep=0pt, shift={(0.1,-0.1)}, ] at (StreetMad.north west) {Street};
	\node[above right, draw, rounded corners=2,inner sep=12pt, text width=0.9cm] at (StreetWay.west|-,|-StreetMad.south) (BuildingMarshall) {};
	\node[below right, inner sep=0pt, shift={(0.1,-0.1)}, ] at (BuildingMarshall.north west) {Building};
 	\node[below right, big site, inner sep=10pt,] at (StreetCharles.west |-, |- StreetMad.north)(SiteCambridge) {};
	\node[below right, inner sep=0pt, shift={(0.1,-0.1)}, ] at (SiteCambridge.north west) {7};
	\node[below right,draw, rounded corners=2,text width=3.3cm] at (-7, |- StreetMad.north) (IDMad) {ID(Madingley Road)};
	\draw[big edge] (IDMad)  to[out=0,in=180]  (StreetMad.west|-,|-IDMad);
	\node[below right,draw, rounded corners=2, shift={(0,-0.1)},text width=3.3cm] at (-7, |- IDMad.south) (ID51Mad) {ID(51 Madingley Road)};
	\draw[big edge] (ID51Mad)  to[out=0,in=180]  (Building51Mad);
	\node[below right,draw, rounded corners=2, shift={(0,-0.2)},text width=3.3cm] at (-7, |- ID51Mad.south) (IDMarshall) {ID(Marshall House)};
	\draw[big edge] (IDMarshall)  to[out=0,in=180]  (BuildingMarshall);

	\coordinate[shift={(0,0.3)}] (aboveStreetMad) at (SiteCambridge.north west);
	\node[fit=(aboveStreetMad)(StreetAda), draw, rounded corners=2,inner sep=15pt,] (BoundaryCambridge) {};
	\node[below right, inner sep=0pt, shift={(0.1,-0.1)}, ] at (BoundaryCambridge.north west) {Boundary};
	\node[below right,draw, rounded corners=2,text width=3.3cm] at (-7, |- BoundaryCambridge.north) (IDCambridge) {ID(Cambridge)};
	\draw[big edge] (IDCambridge)  to[out=0,in=180]  (BoundaryCambridge.west|-,|-IDCambridge);

	\node[above=3.3 of Junction2Mad, shift={(-0.55,0)}, draw, rounded corners=2] (Junction0MadMad) {Junction};
	\draw[big edge] (Junction2Mad)  to[out=90,in=270]  (Junction0MadMad);
 	\node[left=0.3 of Junction0MadMad, big site, inner sep=10pt,] (SiteMadMad) {};
	\node[below right, inner sep=0pt, shift={(0.1,-0.1)}, ] at (SiteMadMad.north west) {6};
	\node[fit=(Junction0MadMad)(SiteMadMad), draw, rounded corners=2,inner sep=15pt,] (StreetMadMad) {};
	\node[below right, inner sep=0pt, shift={(0.1,-0.1)}, ] at (StreetMadMad.north west) {Street};
 	\node[left=0.3 of StreetMadMad, big site, inner sep=10pt,] (SiteMadingley) {};
	\node[below right, inner sep=0pt, shift={(0.1,-0.1)}, ] at (SiteMadingley.north west) {5};
	\node[fit=(StreetMadMad)(SiteMadingley), draw, rounded corners=2,inner sep=15pt,] (BoundaryMad) {};
	\node[below right, inner sep=0pt, shift={(0.1,-0.1)}, ] at (BoundaryMad.north west) {Boundary};
	\node[below right,draw, rounded corners=2,text width=3.3cm] at (-7, |- BoundaryMad.north) (IDMadingley) {ID(Madingley)};
	\draw[big edge] (IDMadingley)  to[out=0,in=180]  (BoundaryMad.west|-,|-IDMadingley);
	\node[below right,draw, rounded corners=2, shift={(0,-0.1)},text width=3.3cm] at (-7, |- IDMadingley.south) (IDMadMad) {ID(Madingley Road)};
	\draw[big edge] (IDMadMad)  to[out=0,in=180]  (StreetMadMad);
 	\node[below=0.3 of IDMadMad, big site, inner sep=10pt,] (SiteID) {};
	\node[below right, inner sep=0pt, shift={(0.1,-0.1)}, ] at (SiteID.north west) {0};

	\node[above=2.5 of Junction0MadMad, draw, rounded corners=2] (Junction0M11) {Junction};
 	\node[left=0.3 of Junction0M11, big site, inner sep=10pt,] (SiteM11) {};
	\node[below right, inner sep=0pt, shift={(0.1,-0.1)}, ] at (SiteM11.north west) {4};
	\node[fit=(Junction0M11)(SiteM11), draw, rounded corners=2,inner sep=15pt,] (StreetM11) {};
	\node[below right, inner sep=0pt, shift={(0.1,-0.1)}, ] at (StreetM11.north west) {Street};
 	\node[left=0.3 of StreetM11, big site, inner sep=10pt,] (SiteIckleton) {};
	\node[below right, inner sep=0pt, shift={(0.1,-0.1)}, ] at (SiteIckleton.north west) {3};
	\node[fit=(StreetM11)(SiteIckleton), draw, rounded corners=2,inner sep=15pt,] (BoundaryIckleton) {};
	\node[below right, inner sep=0pt, shift={(0.1,-0.1)}, ] at (BoundaryIckleton.north west) {Boundary};

	\node[below left, big site, inner sep=10pt,shift={(-0.3,0)}] at (BoundaryIckleton.north west) (SiteSouthCambridgeshire) {};
	\node[below right, inner sep=0pt, shift={(0.1,-0.1)}, ] at (SiteSouthCambridgeshire.north west) {2};
	\node[fit=(BoundaryMad)(SiteSouthCambridgeshire), draw, rounded corners=2,inner sep=15pt,] (BoundarySouthCambridgeshire) {};
	\node[below right, inner sep=0pt, shift={(0.1,-0.1)}, ] at (BoundarySouthCambridgeshire.north west) {Boundary};	

 	\node[below right, big site, inner sep=10pt,] at (BoundaryCambridge.west |-, |- BoundarySouthCambridgeshire.north)(SiteCambridgeshire) {};
	\node[below right, inner sep=0pt, shift={(0.1,-0.1)}, ] at (SiteCambridgeshire.north west) {1};
	\coordinate[shift={(0,0.3)}] (aboveBoundarySouthCambridgeshire) at (BoundarySouthCambridgeshire.north west);
	\node[fit=(aboveBoundarySouthCambridgeshire)(BoundaryCambridge), draw, rounded corners=2,inner sep=15pt,] (BoundaryCambridgeshire) {};
	\node[below right, inner sep=0pt, shift={(0.1,-0.1)}, ] at (BoundaryCambridgeshire.north west) {Boundary};
	\node[below right,draw, rounded corners=2,text width=3.3cm] at (-7, |- BoundaryCambridgeshire.north) (IDCambridgeshire) {ID(Cam-bridgeshire)};
	\draw[big edge] (IDCambridgeshire)  to[out=0,in=180]  (BoundaryCambridgeshire.west|-,|-IDCambridgeshire);
	\node[below right,draw, rounded corners=2, shift={(0,-0.1)},text width=3.3cm] at (-7, |- IDCambridgeshire.south) (IDSouthCambridgeshire) {ID(South Cambridgeshire)};
	\draw[big edge] (IDSouthCambridgeshire)  to[out=0,in=180]  (BoundarySouthCambridgeshire.west|-,|-IDSouthCambridgeshire);
	\node[below right,draw, rounded corners=2, shift={(0,-0.1)},text width=3.3cm] at (-7, |- IDSouthCambridgeshire.south) (IDIckleton) {ID(Ickleton)};
	\draw[big edge] (IDIckleton)  to[out=0,in=180]  (BoundaryIckleton);
	\node[below right,draw, rounded corners=2, shift={(0,-0.1)},text width=3.3cm] at (-7, |- IDIckleton.south) (IDM11) {ID(M11)};
	\draw[big edge] (IDM11)  to[out=0,in=185]  (StreetM11);

	\node[big region,fit=(IDCambridgeshire)(IDAda),inner sep=15pt,] (r0) {};
	\node[below right, inner sep=0pt, shift={(0.1,-0.1)}, ] at (r0.north west) {0};
	\node[big region,fit=(BoundaryCambridgeshire),inner sep=15pt,] (r1) {};
	\node[below right, inner sep=0pt, shift={(0.1,-0.1)}, ] at (r1.north west) {1};
	\node[above=0.3] at(Junction0M11|-,|-r1.north) (node215742) {node 215742};
	\draw[big edge] (Junction0M11)  to[out=90,in=270]  (node215742);
\end{tikzpicture}
\caption{Bigraph of Cambridgeshire.}
\label{fig:bigraph-Cambridgeshire}
\end{figure}

\subsection{Combining bigraphs of different parts of the world} \label{composition-world}
\subsubsection{Parallel product}
The bigraph of one region can be combined by parallel product (\S \ref{bigraph-operations}) with bigraphs constructed separately for other parts of the world. Consider the parallel product of the bigraphs of Cambridgeshire (Figure \ref{fig:bigraph-Cambridgeshire}) and Essex, both of which have an open link with the outer name "node 215742". When forming the parallel product (illustrated in Figure \ref{fig:cambridgeshire_essex}), open links with the same name are combined and hence link a Junction node in Cambridgeshire to another Junction node in Essex. This corresponds to the M11 highway in the real world that runs from Cambridgeshire to Essex—the combined link represents the connection between road networks in Cambridgeshire and Essex via M11. However, the parallel product of bigraphs of different regions does not inform of their spatial relationship.

\subsubsection{Nest into contextual bigraph}
Another approach is to nest the bigraph of a region inside a contextual bigraph, to connect it to the rest of the world. Figure \ref{fig:contextual-bigraph} shows a contextual bigraph, modelling the context of Cambridgeshire described by the hierarchical tree of the world in Figure \ref{fig:tree}.
\begin{figure}[h!]
\begin{subfigure}{\linewidth}
\begin{tikzpicture}[remember picture]
 	\node[big site, inner sep=10pt,] (SiteCambridgeshireAndPeterborough0) {};
	\node[below right, inner sep=0pt, shift={(0.1,-0.1)}, ] at (SiteCambridgeshireAndPeterborough0.north west) {6};
	\node[right=0.3 of SiteCambridgeshireAndPeterborough0, big site, inner sep=10pt,] (SiteCambridgeshireAndPeterborough1) {};
	\node[below right, inner sep=0pt, shift={(0.1,-0.1)}, ] at (SiteCambridgeshireAndPeterborough1.north west) {7};
	\node[fit=(SiteCambridgeshireAndPeterborough0)(SiteCambridgeshireAndPeterborough1), draw, rounded corners=2,inner sep=15pt] (BoundaryCambridgeshireAndPeterborough) {};
	\node[below right, inner sep=0pt, shift={(0.1,-0.1)}, ] at (BoundaryCambridgeshireAndPeterborough.north west) {Boundary};
	\node[above=1.3 of SiteCambridgeshireAndPeterborough1, shift={(-0.15,0)},big site, inner sep=10pt,] (SiteEssex) {};
	\node[below right, inner sep=0pt, shift={(0.1,-0.1)}, ] at (SiteEssex.north west) {5};
	\node[fit=(SiteEssex), draw, rounded corners=2,inner sep=15pt, text width=1cm] (BoundaryEssex) {};
	\node[below right, inner sep=0pt, shift={(0.1,-0.1)}, ] at (BoundaryEssex.north west) {Boundary};
	\node[left=0.3 of BoundaryEssex, big site, inner sep=10pt,] (SiteEngland) {};
	\node[below right, inner sep=0pt, shift={(0.1,-0.1)}, ] at (SiteEngland.north west) {4};
	\node[fit=(BoundaryCambridgeshireAndPeterborough)(BoundaryEssex)(SiteEngland), draw, rounded corners=2,inner sep=15pt,] (BoundaryEngland) {};
	\node[below right, inner sep=0pt, shift={(0.1,-0.1)}, ] at (BoundaryEngland.north west) {Boundary};
	\node[left=0.3 of BoundaryEngland, big site, inner sep=10pt,] (SiteUK) {};
	\node[below right, inner sep=0pt, shift={(0.1,-0.1)}, ] at (SiteUK.north west) {3};
	\node[fit=(BoundaryEngland)(SiteUK), draw, rounded corners=2,inner sep=15pt,] (BoundaryUK) {};
	\node[below right, inner sep=0pt, shift={(0.1,-0.1)}, ] at (BoundaryUK.north west) {Boundary};
	\node[left=0.3 of BoundaryUK, big site, inner sep=10pt,] (SiteWorld) {};
	\node[below right, inner sep=0pt, shift={(0.1,-0.1)}, ] at (SiteWorld.north west) {2};
	\node[fit=(BoundaryUK)(SiteWorld), draw, rounded corners=2,inner sep=15pt,] (BoundaryWorld) {};
	\node[below right, inner sep=0pt, shift={(0.1,-0.1)}, ] at (BoundaryWorld.north west) {Boundary};
	\node[below right,draw, rounded corners=2,text width=3.4cm] at (-10, |- BoundaryWorld.north) (IDWorld) {ID(World)};
	\draw[big edge] (IDWorld)  to[out=0,in=180]  (BoundaryWorld.west|-,|-IDWorld);
	\node[below=0.1 of IDWorld,draw, rounded corners=2,text width=3.4cm] (IDUK) {ID(UK)};
	\draw[big edge] (IDUK)  to[out=0,in=180]  (BoundaryUK.west|-,|-IDUK);
	\node[below=0.1 of IDUK,draw, rounded corners=2,text width=3.4cm]  (IDEngland) {ID(England)};
	\draw[big edge] (IDEngland)  to[out=0,in=180]  (BoundaryEngland.west|-,|-IDEngland);
	\node[below=0.1 of IDEngland,draw, rounded corners=2,text width=3.4cm] (IDEssex) {ID(Essex)};
	\draw[big edge] (IDEssex)  to[out=0,in=180]  (BoundaryEssex);
	\node[below=0.1 of IDEssex,draw, rounded corners=2,text width=3.4cm] (IDCambridgeshireAndPeterborough) {ID(Cambridgeshire and Peterborough)};
	\draw[big edge] (IDCambridgeshireAndPeterborough)  to[out=0,in=180] (BoundaryCambridgeshireAndPeterborough.west|-,|-IDCambridgeshireAndPeterborough);
	\node[below=0.1 of IDCambridgeshireAndPeterborough, shift={(-0.5,0)}, big site, inner sep=10pt,] (SiteID0) {};
	\node[below right, inner sep=0pt, shift={(0.1,-0.1)}, ] at (SiteID0.north west) {0};
	\node[below=0.1 of IDCambridgeshireAndPeterborough, shift={(0.5,0)}, big site, inner sep=10pt,] (SiteID1) {};
	\node[below right, inner sep=0pt, shift={(0.1,-0.1)}, ] at (SiteID1.north west) {1};
	\node[big region,fit=(IDWorld)(SiteID0),inner sep=10pt,] (r0) {};
	\node[below right, inner sep=0pt, shift={(0.1,-0.1)}, ] at (r0.north west) {0};
	\node[big region,fit=(BoundaryWorld),inner sep=10pt,] (r1) {};
	\node[below right, inner sep=0pt, shift={(0.1,-0.1)}, ] at (r1.north west) {1};
	
	\node[above=0.3] at (SiteEssex|-,|-r1.north) (node215742) {node 215742};
	\draw[big edge] (SiteEssex)  to[out=90,in=270]  (node215742);
\end{tikzpicture}
\centering
\caption{Contextual bigraph}
\label{fig:contextual-bigraph}
\end{subfigure}

\begin{subfigure}{\linewidth}
\centering
\begin{tikzpicture}[remember picture]
	\node[draw, rounded corners=2,text width=2.1cm, shift={(0,0.5)},] (IDCambridgeshire) {ID( Cambridgeshire)};
 	\node[below=0.3 of IDCambridgeshire, big site, inner sep=10pt,] (s6) {};
	\node[below right, inner sep=0pt, shift={(0.1,-0.1)}, ] at (s6.north west) {0};
	\node[big region, fit=(IDCambridgeshire)(s6), inner sep=10pt,] (r6) {};
	\node[below right, inner sep=0pt, shift={(0.1,-0.1)}, ] at (r6.north west) {0};

	\node[right=1.2 of r6,shift={(0,-0.25)}, big site, inner sep=10pt,] (SiteCambridgeshire) {};
	\node[below right, inner sep=0pt, shift={(0.1,-0.1)}, ] at (SiteCambridgeshire.north west) {1};
	\node[fit=(SiteCambridgeshire), draw, rounded corners=2,inner sep=15pt, text width=1cm,] (BoundaryCambridgeshire) {};
	\node[below right, inner sep=0pt, shift={(0.1,-0.1)}, ] at (BoundaryCambridgeshire.north west) {Boundary};
	\node[big region, fit=(BoundaryCambridgeshire), inner sep=10pt,] (r7) {};
	\node[below right, inner sep=0pt, shift={(0.1,-0.1)}, ] at (r7.north west) {1};
	\draw[big edge] (IDCambridgeshire)  to[out=0,in=180] (BoundaryCambridgeshire);

	\node[above=0.3 of r7] (node215742) {node 215742};
	\draw[big edge] (SiteCambridgeshire)  to[out=90,in=270]  (node215742);
\end{tikzpicture}
\caption{Simplified bigraph of Cambridgeshire}
\label{fig:bigraph-Cambridgeshire-and-sites}
\end{subfigure}
\centering
\caption{Nesting the bigraph of Cambridgeshire into a contextual bigraph. The ID and Physical regions of the bigraph of Cambridgshire are nested into the sites 1 and 7 of the contextual bigraph via a placing $\phi$ that reorders the regions.}
\end{figure}
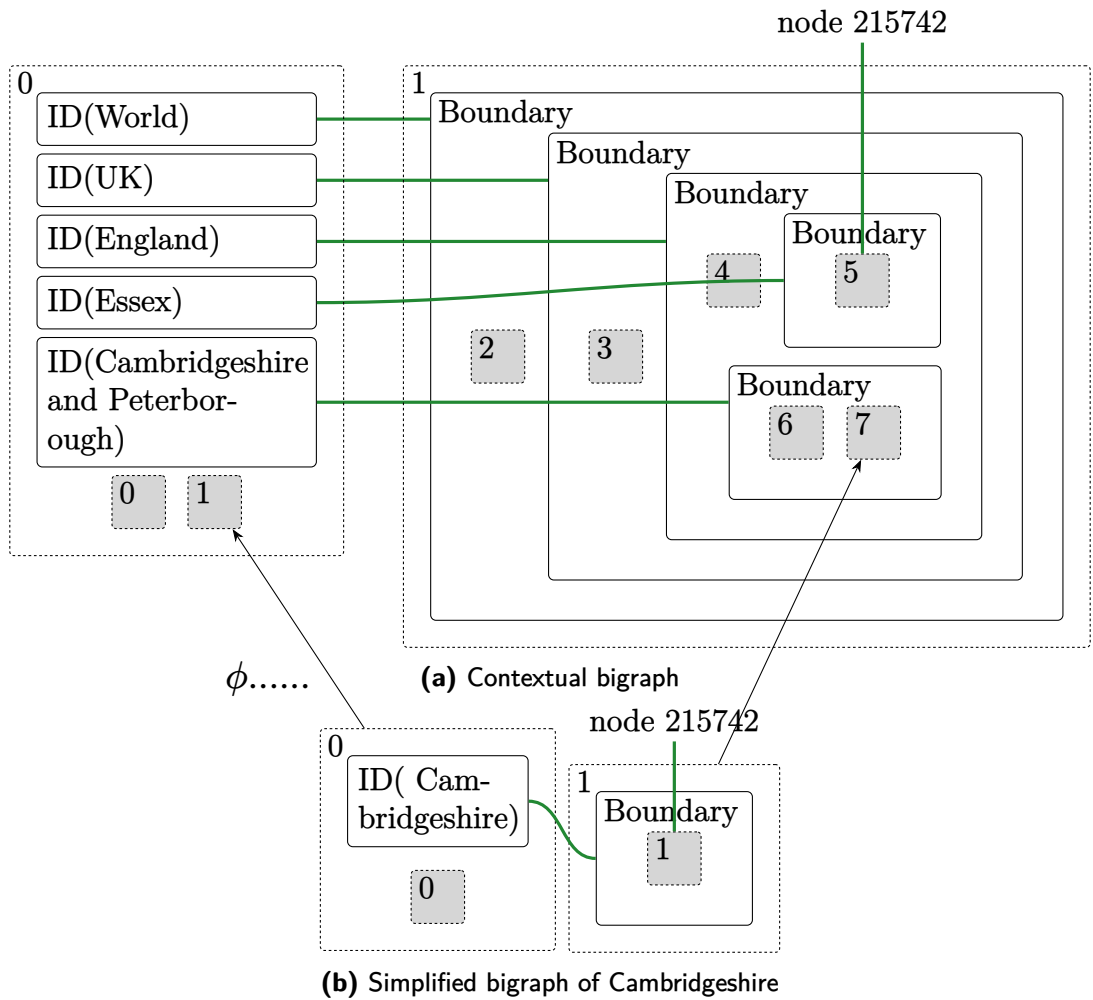
\begin{tikzpicture}[overlay, remember picture]
          \draw[-Stealth] (r6) -- (SiteID1);
          \draw[-Stealth] (r7) -- (SiteCambridgeshireAndPeterborough1);
	\node[left=0.3, inner sep=0pt, shift={(0.1,-0.1)}, ] at ($(r6)!0.5!(SiteID1)$) {\large $\phi ......$};
\end{tikzpicture}

The ID perspective of the bigraph of Cambridgeshire (region 0) can be nested into site 1, and its Physical perspective (region 1) into site 7 by using a placing $\phi$ that reorders regions, built of elementary bigraphs $symmetry$ and $id$ (\S \ref{elementary-bigraphs}). When forming the nesting, the open links with the outer name "node 215742" are combined, linking a Junction node in Cambridgeshire to another Junction node in Essex. Nesting into the contextual bigraph also informs of the spatial locality of Cambridgeshire with respect to the rest of the world, because the place graph encodes a hierarchy of containment. See \S \ref{section:results-of-combining} for details of the placing $\phi$ and a visualisation of the results of combining bigraphs of different regions.

This concludes the specification for building a bigraph that models the real world, which will guide the implementation of deliverables in \S\ref{section:OSM-implementation} and \S\ref{section:bigraph-implementation}. See \S\ref{section:shared-junctions} for an alternative method for modelling connected streets using bigraphs with sharing, accompanied by a justification for the decision to model connected streets using linking (\S\ref{street-connectivity-link}).

\section{Reactive system for movement and communication} \label{brs-implementation}

Milner introduced bigraphs in his book titled "The Space and Motion of Communicating Agents" \cite{10.5555/1540607} — it inspires this dissertation to deliver a BRS (\S \ref{section:brs}) that consists of bigraphs of the real world which model the space that ubiquitous systems exist in, and a set of reaction rules that model motion and communication.

The additional controls used are defined as follows:
\begin{itemize}
\item Agent, arity 1. Single link to its identifier.
\item Contact, arity 1. Single link to another Contact node.
\item Message, arity 1. Single link to its identifier.
\end{itemize}

\subsection{Reaction rules for motion} \label{section:react-motion}

By extending the reaction rule \rr{leave\_room} (Figure \ref{fig:react-leave-room}), the 5 reaction rules in Figure \ref{fig:reaction-rules-motion} allow Agent nodes to move around bigraphs of the real world. The full transition system of the BRS consisting of the simplified bigraph of Cambridgeshire (Figure \ref{fig:bigraph-Cambridgeshire}) and the set of reaction rules for motion is illustrated in Figure \ref{fig:transition-system-motion}.

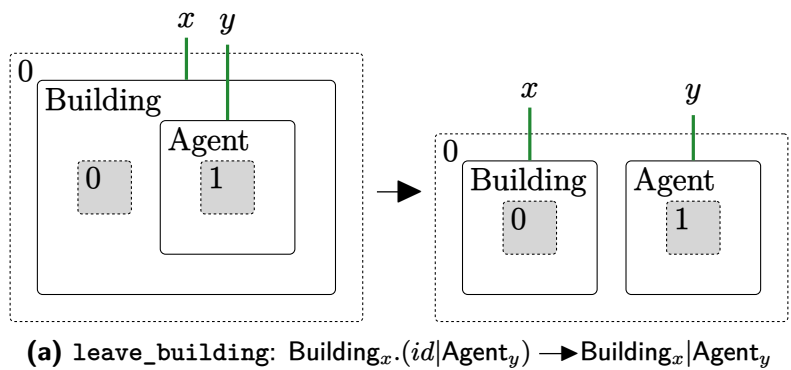
\begin{figure}[h!]
\centering
\begin{subfigure}{\linewidth}
\centering
\begin{tikzpicture}
 	\node[big site, inner sep=10pt,] (s0) {};
	\node[below right, inner sep=0pt, shift={(0.1,-0.1)}, ] at (s0.north west) {0};
 	\node[right=0.9 of s0, big site, inner sep=10pt,] (s1) {};
	\node[below right, inner sep=0pt, shift={(0.1,-0.1)}, ] at (s1.north west) {1};
	\node[fit=(s1), draw, rounded corners=2,inner sep=15pt,] (a0) {};
	\node[below right, inner sep=0pt, shift={(0.1,-0.1)}, ] at (a0.north west) {Agent};
	\node[fit=(s0)(a0), draw, rounded corners=2,inner sep=15pt,] (building) {};
	\node[below right, inner sep=0pt, shift={(0.1,-0.1)}, ] at (building.north west) {Building};
	\node[big region, fit=(building), inner sep=10pt,] (r0) {};
	\node[below right, inner sep=0pt, shift={(0.1,-0.1)}, ] at (r0.north west) {0};

	\node[above=0.2 of r0, shift={(0,0)}] (x) {$x$};
	\draw[big edge] (building)  to[out=90,in=270]  (x);
	\node[above=1.0 of a0, shift={(0,0)}] (y) {$y$};
	\draw[big edge] (a0)  to[out=90,in=270]  (y);
\end{tikzpicture}
\raisebox{1.6cm}{\large$\rrul$}
\begin{tikzpicture}
 	\node[big site, inner sep=10pt,] (s0) {};
	\node[below right, inner sep=0pt, shift={(0.1,-0.1)}, ] at (s0.north west) {0};
	\node[fit=(s0), draw, rounded corners=2,inner sep=15pt,] (building) {};
	\node[below right, inner sep=0pt, shift={(0.1,-0.1)}, ] at (building.north west) {Building};
 	\node[right=0.9 of building, big site, inner sep=10pt,] (s1) {};
	\node[below right, inner sep=0pt, shift={(0.1,-0.1)}, ] at (s1.north west) {1};
	\node[fit=(s1), draw, rounded corners=2,inner sep=15pt,] (a0) {};
	\node[below right, inner sep=0pt, shift={(0.1,-0.1)}, ] at (a0.north west) {Agent};
	\node[big region, fit=(building)(a0), inner sep=10pt,] (r0) {};
	\node[below right, inner sep=0pt, shift={(0.1,-0.1)}, ] at (r0.north west) {0};

	\node[above=0.7 of building, shift={(0,0)}] (x) {$x$};
	\draw[big edge] (building)  to[out=90,in=270]  (x);
	\node[above=0.6 of a0, shift={(0,0)}] (y) {$y$};
	\draw[big edge] (a0)  to[out=90,in=270]  (y);
\end{tikzpicture}
\captionsetup{justification=centering}
\caption{\rr{leave\_building}: Building$_x.(id|$Agent$_{y})\rrul$Building$_x|$Agent$_{y}$}
\label{fig:react-leave-building}
\end{subfigure}
\caption{Reaction rules for motion.}
\end{figure}
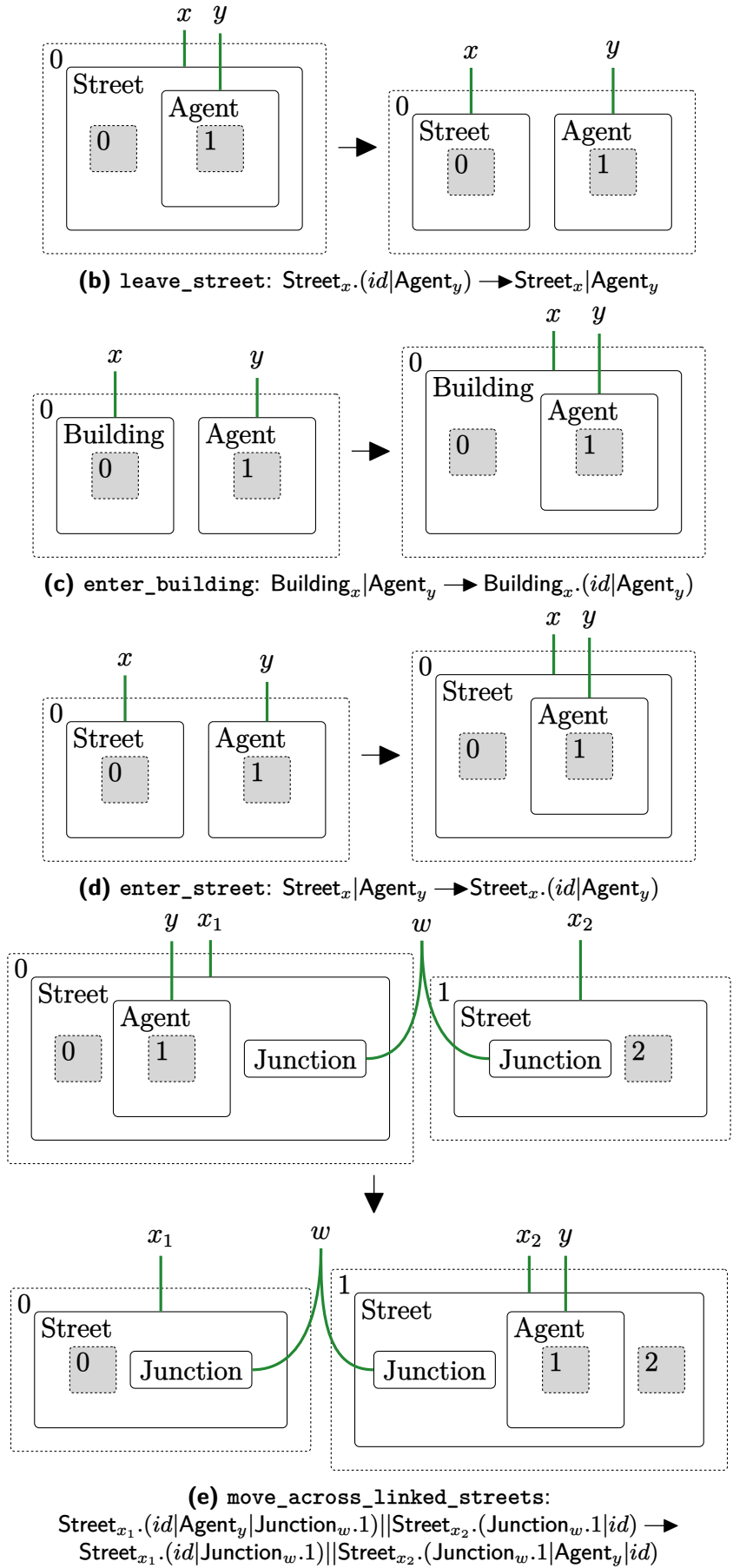
\begin{figure}\ContinuedFloat
\centering
\begin{subfigure}{\linewidth}
\centering
\begin{tikzpicture}
 	\node[big site, inner sep=10pt,] (s0) {};
	\node[below right, inner sep=0pt, shift={(0.1,-0.1)}, ] at (s0.north west) {0};
 	\node[right=0.9 of s0, big site, inner sep=10pt,] (s1) {};
	\node[below right, inner sep=0pt, shift={(0.1,-0.1)}, ] at (s1.north west) {1};
	\node[fit=(s1), draw, rounded corners=2,inner sep=15pt,] (a0) {};
	\node[below right, inner sep=0pt, shift={(0.1,-0.1)}, ] at (a0.north west) {Agent};
	\node[fit=(s0)(a0), draw, rounded corners=2,inner sep=10pt,] (street) {};
	\node[below right, inner sep=0pt, shift={(0.1,-0.1)}, ] at (street.north west) {Street};
	\node[big region, fit=(street), inner sep=10pt,] (r0) {};
	\node[below right, inner sep=0pt, shift={(0.1,-0.1)}, ] at (r0.north west) {0};

	\node[above=0.2 of r0, shift={(0,0)}] (x) {$x$};
	\draw[big edge] (street)  to[out=90,in=270]  (x);
	\node[above=0.15 of a0] at (a0|-,|-r0.north) (y) {$y$};
	\draw[big edge] (a0)  to[out=90,in=270]  (y);
\end{tikzpicture}
\raisebox{1.5cm}{\large$\rrul$}
\begin{tikzpicture}
 	\node[big site, inner sep=10pt,] (s0) {};
	\node[below right, inner sep=0pt, shift={(0.1,-0.1)}, ] at (s0.north west) {0};
	\node[fit=(s0), draw, rounded corners=2,inner sep=15pt,] (street) {};
	\node[below right, inner sep=0pt, shift={(0.1,-0.1)}, ] at (street.north west) {Street};
 	\node[right=0.9 of street, big site, inner sep=10pt,] (s1) {};
	\node[below right, inner sep=0pt, shift={(0.1,-0.1)}, ] at (s1.north west) {1};
	\node[fit=(s1), draw, rounded corners=2,inner sep=15pt,] (a0) {};
	\node[below right, inner sep=0pt, shift={(0.1,-0.1)}, ] at (a0.north west) {Agent};
	\node[big region, fit=(street)(a0), inner sep=10pt,] (r0) {};
	\node[below right, inner sep=0pt, shift={(0.1,-0.1)}, ] at (r0.north west) {0};

	\node[above=0.7 of street, shift={(0,0)}] (x) {$x$};
	\draw[big edge] (building)  to[out=90,in=270]  (x);
	\node[above=0.7 of a0, shift={(0,0)}] (y) {$y$};
	\draw[big edge] (a0)  to[out=90,in=270]  (y);
\end{tikzpicture}
\captionsetup{justification=centering}
\caption{\rr{leave\_street}: Street$_x.(id|$Agent$_{y})\rrul$Street$_x|$Agent$_{y}$}
\label{fig:react-leave-street}
\end{subfigure}

\begin{subfigure}{\linewidth}
\centering
\begin{tikzpicture}
 	\node[big site, inner sep=10pt,] (s0) {};
	\node[below right, inner sep=0pt, shift={(0.1,-0.1)}, ] at (s0.north west) {0};
	\node[fit=(s0), draw, rounded corners=2,inner sep=15pt,] (building) {};
	\node[below right, inner sep=0pt, shift={(0.1,-0.1)}, ] at (building.north west) {Building};
 	\node[right=0.9 of building, big site, inner sep=10pt,] (s1) {};
	\node[below right, inner sep=0pt, shift={(0.1,-0.1)}, ] at (s1.north west) {1};
	\node[fit=(s1), draw, rounded corners=2,inner sep=15pt,] (a0) {};
	\node[below right, inner sep=0pt, shift={(0.1,-0.1)}, ] at (a0.north west) {Agent};
	\node[big region, fit=(building)(a0), inner sep=10pt,] (r0) {};
	\node[below right, inner sep=0pt, shift={(0.1,-0.1)}, ] at (r0.north west) {0};

	\node[above=0.7 of building, shift={(0,0)}] (x) {$x$};
	\draw[big edge] (building)  to[out=90,in=270]  (x);
	\node[above=0.6 of a0, shift={(0,0)}] (y) {$y$};
	\draw[big edge] (a0)  to[out=90,in=270]  (y);
\end{tikzpicture}
\raisebox{1.5cm}{\large$\rrul$}
\begin{tikzpicture}
 	\node[big site, inner sep=10pt,] (s0) {};
	\node[below right, inner sep=0pt, shift={(0.1,-0.1)}, ] at (s0.north west) {0};
 	\node[right=1.2 of s0, big site, inner sep=10pt,] (s1) {};
	\node[below right, inner sep=0pt, shift={(0.1,-0.1)}, ] at (s1.north west) {1};
	\node[fit=(s1), draw, rounded corners=2,inner sep=15pt,] (a0) {};
	\node[below right, inner sep=0pt, shift={(0.1,-0.1)}, ] at (a0.north west) {Agent};
	\node[fit=(s0)(a0), draw, rounded corners=2,inner sep=10pt,] (building) {};
	\node[below right, inner sep=0pt, shift={(0.1,-0.1)}, ] at (building.north west) {Building};
	\node[big region, fit=(building), inner sep=10pt,] (r0) {};
	\node[below right, inner sep=0pt, shift={(0.1,-0.1)}, ] at (r0.north west) {0};

	\node[above=0.24] at (building|-,|-r0.north) (x) {$x$};
	\draw[big edge] (building)  to[out=90,in=270]  (x);
	\node[above=0.2] at (a0|-,|-r0.north) (y) {$y$};
	\draw[big edge] (a0)  to[out=90,in=270]  (y);
\end{tikzpicture}
\captionsetup{justification=centering}
\caption{\rr{enter\_building}: $\text{Building}_x|\text{Agent}_{y}\rrul\text{Building}_x.(id|\text{Agent}_{y})$}
\label{fig:react-enter-building}
\end{subfigure}

\begin{subfigure}{\linewidth}
\centering
\begin{tikzpicture}
 	\node[big site, inner sep=10pt,] (s0) {};
	\node[below right, inner sep=0pt, shift={(0.1,-0.1)}, ] at (s0.north west) {0};
	\node[fit=(s0), draw, rounded corners=2,inner sep=15pt,] (street) {};
	\node[below right, inner sep=0pt, shift={(0.1,-0.1)}, ] at (street.north west) {Street};
 	\node[right=0.9 of street, big site, inner sep=10pt,] (s1) {};
	\node[below right, inner sep=0pt, shift={(0.1,-0.1)}, ] at (s1.north west) {1};
	\node[fit=(s1), draw, rounded corners=2,inner sep=15pt,] (a0) {};
	\node[below right, inner sep=0pt, shift={(0.1,-0.1)}, ] at (a0.north west) {Agent};
	\node[big region, fit=(street)(a0), inner sep=10pt,] (r0) {};
	\node[below right, inner sep=0pt, shift={(0.1,-0.1)}, ] at (r0.north west) {0};

	\node[above=0.7 of street, shift={(0,0)}] (x) {$x$};
	\draw[big edge] (street)  to[out=90,in=270]  (x);
	\node[above=0.6 of a0, shift={(0,0)}] (y) {$y$};
	\draw[big edge] (a0)  to[out=90,in=270]  (y);
\end{tikzpicture}
\raisebox{1.5cm}{\large$\rrul$}
\begin{tikzpicture}
 	\node[big site, inner sep=10pt,] (s0) {};
	\node[below right, inner sep=0pt, shift={(0.1,-0.1)}, ] at (s0.north west) {0};
 	\node[right=0.9 of s0, big site, inner sep=10pt,] (s1) {};
	\node[below right, inner sep=0pt, shift={(0.1,-0.1)}, ] at (s1.north west) {1};
	\node[fit=(s1), draw, rounded corners=2,inner sep=15pt,] (a0) {};
	\node[below right, inner sep=0pt, shift={(0.1,-0.1)}, ] at (a0.north west) {Agent};
	\node[fit=(s0)(a0), draw, rounded corners=2,inner sep=10pt,] (street) {};
	\node[below right, inner sep=0pt, shift={(0.1,-0.1)}, ] at (street.north west) {Street};
	\node[big region, fit=(street), inner sep=10pt,] (r0) {};
	\node[below right, inner sep=0pt, shift={(0.1,-0.1)}, ] at (r0.north west) {0};

	\node[above=0.25] at (street|-,|-r0.north) (x) {$x$};
	\draw[big edge] (street)  to[out=90,in=270]  (x);
	\node[above=0.2] at (a0|-,|-r0.north) (y) {$y$};
	\draw[big edge] (a0)  to[out=90,in=270]  (y);
\end{tikzpicture}
\captionsetup{justification=centering}
\caption{\rr{enter\_street}: Street$_x|$Agent$_{y}\rrul$Street$_x.(id|$Agent$_{y})$}
\label{fig:react-enter-street}
\end{subfigure}

\begin{subfigure}{\linewidth}
\centering
\begin{tikzpicture}
 	\node[big site, inner sep=10pt,] (s0) {};
	\node[below right, inner sep=0pt, shift={(0.1,-0.1)}, ] at (s0.north west) {0};
 	\node[right=0.7 of s0, big site, inner sep=10pt,] (s1) {};
	\node[below right, inner sep=0pt, shift={(0.1,-0.1)}, ] at (s1.north west) {1};
	\node[fit=(s1), draw, rounded corners=2,inner sep=15pt,] (a0) {};
	\node[below right, inner sep=0pt, shift={(0.1,-0.1)}, ] at (a0.north west) {Agent};
	\node[right=0.2 of a0, shift={(0,0)},draw, rounded corners=2] (JunctionLeft) {Junction};
	\node[fit=(s0)(a0)(JunctionLeft), draw, rounded corners=2,inner sep=10pt,] (streetLeft) {};
	\node[below right, inner sep=0pt, shift={(0.1,-0.1)}, ] at (streetLeft.north west) {Street};
	\node[big region, fit=(streetLeft), inner sep=10pt,] (r0) {};
	\node[below right, inner sep=0pt, shift={(0.1,-0.1)}, ] at (r0.north west) {0};

	\node[above=0.2 of r0, shift={(0,0)}] (x1) {$x_1$};
	\draw[big edge] (streetLeft)  to[out=90,in=270]  (x1);
	\node[above=0.9 of a0, shift={(0,0)}] (y) {$y$};
	\draw[big edge] (a0)  to[out=90,in=270]  (y);

	\node[right=1.5 of streetLeft, shift={(0,0)},draw, rounded corners=2] (JunctionRight) {Junction};
	\node[right=0.2 of JunctionRight, big site, inner sep=10pt,] (s2) {};
	\node[below right, inner sep=0pt, shift={(0.1,-0.1)}, ] at (s2.north west) {2};
	\node[fit=(JunctionRight)(s2), draw, rounded corners=2,inner sep=15pt,] (streetRight) {};
	\node[below right, inner sep=0pt, shift={(0.1,-0.1)}, ] at (streetRight.north west) {Street};
	\node[big region, fit=(streetRight), inner sep=10pt,] (r1) {};
	\node[below right, inner sep=0pt, shift={(0.1,-0.1)}, ] at (r1.north west) {1};

	\node[above=0.2] at (streetRight|-,|-r0.north) (x2) {$x_2$};
	\draw[big edge] (streetRight)  to[out=90,in=270]  (x2);

	\node[above=1.8] at ($(streetLeft.east)!0.5!(streetRight.west)$) (w) {$w$};
	\draw[big edge] (JunctionLeft)  to[out=0,in=270]  (w);
	\draw[big edge] (JunctionRight)  to[out=180,in=270]  (w);
\end{tikzpicture}

\rotatebox[origin=c]{270}{\large$\rrul$}

\begin{tikzpicture}
 	\node[big site, inner sep=10pt,] (s0) {};
	\node[below right, inner sep=0pt, shift={(0.1,-0.1)}, ] at (s0.north west) {0};
	\node[right=0.2 of s0, shift={(0,0)},draw, rounded corners=2] (JunctionLeft) {Junction};
	\node[fit=(s0)(JunctionLeft), draw, rounded corners=2,inner sep=15pt,] (streetLeft) {};
	\node[below right, inner sep=0pt, shift={(0.1,-0.1)}, ] at (streetLeft.north west) {Street};
	\node[big region, fit=(streetLeft), inner sep=10pt,] (r0) {};
	\node[below right, inner sep=0pt, shift={(0.1,-0.1)}, ] at (r0.north west) {0};

	\node[right=1.3 of streetLeft, shift={(0,0)},draw, rounded corners=2] (JunctionRight) {Junction};
 	\node[right=0.7 of JunctionRight, big site, inner sep=10pt,] (s1) {};
	\node[below right, inner sep=0pt, shift={(0.1,-0.1)}, ] at (s1.north west) {1};
	\node[fit=(s1), draw, rounded corners=2,inner sep=15pt,] (a0) {};
	\node[below right, inner sep=0pt, shift={(0.1,-0.1)}, ] at (a0.north west) {Agent};
	\node[right=0.2 of a0, big site, inner sep=10pt,] (s2) {};
	\node[below right, inner sep=0pt, shift={(0.1,-0.1)}, ] at (s2.north west) {2};
	\node[fit=(JunctionRight)(a0)(s2), draw, rounded corners=2,inner sep=8pt,] (streetRight) {};
	\node[below right, inner sep=0pt, shift={(0.1,-0.1)}, ] at (streetRight.north west) {Street};
	\node[big region, fit=(streetRight), inner sep=10pt,] (r1) {};
	\node[below right, inner sep=0pt, shift={(0.1,-0.1)}, ] at (r1.north west) {1};

	\node[above=0.2] at (streetLeft|-,|-r1.north) (x1) {$x_1$};
	\draw[big edge] (streetLeft)  to[out=90,in=270]  (x1);
	\node[above=0.2 of r1, shift={(0,0)}] (x2) {$x_2$};
	\draw[big edge] (streetRight)  to[out=90,in=270]  (x2);
	\node[above=0.2] at (a0|-,|-r1.north) (y) {$y$};
	\draw[big edge] (a0)  to[out=90,in=270]  (y);

	\node[above=1.85] at ($(streetLeft.east)!0.5!(streetRight.west)$) (w) {$w$};
	\draw[big edge] (JunctionLeft)  to[out=0,in=270]  (w);
	\draw[big edge] (JunctionRight)  to[out=180,in=270]  (w);
\end{tikzpicture}
\captionsetup{justification=centering,width=\linewidth}
\caption{\rr{move\_across\_linked\_streets}:\\Street$_{x_1}.(id|$Agent$_{y}|$Junction$_w.1)||$Street$_{x_2}.($Junction$_w.1|id)\rrul$\\Street$_{x_1}.(id|$Junction$_w.1)||$Street$_{x_2}.($Junction$_w.1|$Agent$_{y}|id)$}
\end{subfigure}
\caption{Reaction rules for motion.}
\label{fig:reaction-rules-motion}
\end{figure}

\subsection{Reaction rules for forming communication channels between agents}

Agent nodes can establish communication links with other Agent nodes according to reaction rule \rr{connect\_to\_nearby\_agent} (Figure \ref{fig:react-connect-agent}), by nesting a Contact node in each Agent node and linking them. It is conditional on the absence of an existing pair of linked Contact nodes nested within the Agent nodes, to ensure that no duplicate communication links are established. The nesting of a Contact node, instead of directly linking the Agent nodes, is necessary to model variable number of links despite the fixed aritiy \cite{archibald2024practicalmodellingbigraphs} of the Agent control. In the reaction rules for motion (Figure \ref{fig:reaction-rules-motion}), the sites allow Agent nodes to contain any number of Contact nodes and preserve their links in the reactum, so the established channels between agents persist after the agents move apart, which models communication independent of spatial locality.

\begin{figure}[h!]
\begin{subfigure}{\linewidth}
\centering
\resizebox{\linewidth}{!}{
\begin{tikzpicture}
 	\node[big site, inner sep=10pt,] (s1) {};
	\node[below right, inner sep=0pt, shift={(0.1,-0.1)}, ] at (s1.north west) {1};
	\node[fit=(s1), draw, rounded corners=2,inner sep=15pt,] (a0) {};
	\node[below right, inner sep=0pt, shift={(0.1,-0.1)}, ] at (a0.north west) {Agent};
 	\node[right=0.7 of a0, big site, inner sep=10pt,] (s2) {};
	\node[below right, inner sep=0pt, shift={(0.1,-0.1)}, ] at (s2.north west) {2};
	\node[fit=(s2), draw, rounded corners=2,inner sep=15pt,] (a1) {};
	\node[below right, inner sep=0pt, shift={(0.1,-0.1)}, ] at (a1.north west) {Agent};
 	\node[left=0.2 of a0, big site, inner sep=10pt,] (s0) {};
	\node[below right, inner sep=0pt, shift={(0.1,-0.1)}, ] at (s0.north west) {0};
	\node[fit=(s0)(a0)(a1), draw, rounded corners=2,inner sep=15pt,] (b0) {};
	\node[below right, inner sep=0pt, shift={(0.1,-0.1)}, ] at (b0.north west) {Building};
	\node[big region, fit=(b0), inner sep=10pt,] (r0) {};
	\node[below right, inner sep=0pt, shift={(0.1,-0.1)}, ] at (r0.north west) {0};
	
	\node[above=0.25] at (b0|-,|-r0.north) (id0) {$x$};
	\draw[big edge] (b0)  to[out=90,in=270]  (id0);
	\node[above=0.2] at (a0|-,|-r0.north) (id1) {$y$};
	\draw[big edge] (a0)  to[out=90,in=270]  (id1);
	\node[above=0.2] at (a1|-,|-r0.north) (id2) {$z$};
	\draw[big edge] (a1)  to[out=90,in=270]  (id2);
\end{tikzpicture}
\raisebox{1.6cm}{\large$\rrul$}
\begin{tikzpicture}
 	\node[big site, inner sep=10pt,] (s1) {};
	\node[below right, inner sep=0pt, shift={(0.1,-0.1)}, ] at (s1.north west) {1};
	\node[right=0.3 of s1, draw, rounded corners=2] (Contact0) {Contact};
	\node[fit=(s1)(Contact0), draw, rounded corners=2,inner sep=15pt,] (a0) {};
	\node[below right, inner sep=0pt, shift={(0.1,-0.1)}, ] at (a0.north west) {Agent};
 	\node[right=0.7 of a0, big site, inner sep=10pt,] (s2) {};
	\node[below right, inner sep=0pt, shift={(0.1,-0.1)}, ] at (s2.north west) {2};
	\node[right=0.3 of s2, draw, rounded corners=2] (Contact1) {Contact};
	\node[fit=(s2)(Contact1), draw, rounded corners=2,inner sep=15pt,] (a1) {};
	\node[below right, inner sep=0pt, shift={(0.1,-0.1)}, ] at (a1.north west) {Agent};
 	\node[left=0.2 of a0, big site, inner sep=10pt,] (s0) {};
	\node[below right, inner sep=0pt, shift={(0.1,-0.1)}, ] at (s0.north west) {0};
	\node[fit=(s0)(a0)(a1), draw, rounded corners=2,inner sep=15pt,] (b0) {};
	\node[below right, inner sep=0pt, shift={(0.1,-0.1)}, ] at (b0.north west) {Building};
	\node[big region, fit=(b0), inner sep=10pt,] (r0) {};
	\node[below right, inner sep=0pt, shift={(0.1,-0.1)}, ] at (r0.north west) {0};
	
	\node[above=0.25] at (b0|-,|-r0.north) (id0) {$x$};
	\draw[big edge] (b0)  to[out=90,in=270]  (id0);
	\node[above=0.2] at (a0|-,|-r0.north) (id1) {$y$};
	\draw[big edge] (a0)  to[out=90,in=270]  (id1);
	\node[above=0.25] at (a1|-,|-r0.north) (id2) {$z$};
	\draw[big edge] (a1)  to[out=90,in=270]  (id2);
	\draw[big edge] (Contact0)  to[out=0,in=180]  (Contact1);
\end{tikzpicture}
}
\end{subfigure}
\par\medskip
\begin{subfigure}{\linewidth}
\centering
{if }
\raisebox{-0.6\height}{
\resizebox{0.3\linewidth}{!}{
\begin{tikzpicture}
	\node[draw, rounded corners=2] (Contact0) {Contact};
	\node[big region, fit=(Contact0), inner sep=10pt,] (r0) {};
	\node[below right, inner sep=0pt, shift={(0.1,-0.1)}, ] at (r0.north west) {0};
	\node[right=0.7 of r0, draw, rounded corners=2] (Contact1) {Contact};
	\node[big region, fit=(Contact1), inner sep=10pt,] (r1) {};
	\node[below right, inner sep=0pt, shift={(0.1,-0.1)}, ] at (r1.north west) {1};
	\draw[big edge] (Contact0)  to[out=0,in=180]  (Contact1);
\end{tikzpicture}
}
}
{ does not occur in the parameter}
\end{subfigure}
\centering
\captionsetup{justification=centering,width=\linewidth}
\caption{Reaction rule \rr{connect\_to\_nearby\_agent}: \\$\text{Building}_x.(id|\text{Agent}_y|\text{Agent}_z)\rrul \text{Building}_x./w(id|\text{Agent}_y.(id|\text{Contact}_w.1)|\text{Agent}_z.(id|\text{Contact}_w.1))$\\if $/w(\text{Contact}_w.1||\text{Contact}_w.1)$ does not occur in parameter.}
\label{fig:react-connect-agent}
\end{figure}

\subsection{Reaction rules for communication via spatial names}
\subsubsection{Unicast}
\begin{sloppypar}
Besides communication through established channels between agents, global communication is enabled by the use of spatial names (\S\ref{section:spatial-name}). To send a message to a destination agent, the Agent node is identified by its full spatial name. For instance, consider delivering a message to an agent named "Agent A.William Gates Building.J J Thomson Ave.Cambridge.Cambridgeshire.Cambridgeshire and Peterborough.England.UK". The process begins by creating a Message node and nesting it in the Boundary node corresponding to World. The message is then propagated down the levels of hierarchy, using a sequence of reaction rules to move the Message node into the Boundary nodes of "UK", then "England", "Cambridgeshire and Peterborough", "Cambridgeshire", and "Cambridge", then into the Street node of "JJ Thomson Ave", the Building node of "William Gates Building", and finally into the Agent node of "Agent A". This propagation mechanism is analogous to the way queries traverse DNS domains. For unicast, the last reaction rule must be one that propagates the message into an Agent node. If at any step of the sequence a reaction rule is not applicable, then the spatial name must have not matched any node in the bigraph so the process is aborted and the Message node is deleted using another reaction rule.
\end{sloppypar}

\subsubsection{Multicast}

\begin{figure}[h!]
\resizebox{\linewidth}{!}{
\begin{tikzpicture}[remember picture]
 	\node[big site, inner sep=10pt,] (s0l) {};
	\node[below right, inner sep=0pt, shift={(0.1,-0.1)}, ] at (s0l.north west) {0};
	\node[fit=(s0l), draw, rounded corners=2,inner sep=15pt,text width=1cm] (building) {};
	\node[below right, inner sep=0pt, shift={(0.1,-0.1)}, ] at (building.north west) {Boundary};
 	\node[right=0.8 of building, big site, inner sep=10pt,] (s1l) {};
	\node[below right, inner sep=0pt, shift={(0.1,-0.1)}, ] at (s1l.north west) {1};
	\node[fit=(s1l), draw, rounded corners=2,inner sep=15pt] (a0) {};
	\node[below right, inner sep=0pt, shift={(0.1,-0.1)}, ] at (a0.north west) {Message};
	\node[big region, fit=(building)(a0), inner sep=10pt,] (r1) {};
	\node[below right, inner sep=0pt, shift={(0.1,-0.1)}, ] at (r1.north west) {1};

	\node[below left=0 of building.north west, shift={(-0.9,0)},draw,rounded corners=2] (IDboundary) {ID(<level>)};
	\node[below=0.1 of IDboundary,draw,rounded corners=2,text width=2.3cm] (IDmessage) {ID(<messag-e\_id>)};
	\node[big region, fit=(IDboundary)(IDmessage), inner sep=10pt,] (r0) {};
	\node[below right, inner sep=0pt, shift={(0.1,-0.1)}, ] at (r0.north west) {0};

	\draw[big edge] (building.west|-,|-IDboundary)  to[out=180,in=0]  (IDboundary);
	\draw[big edge] (a0)  to[out=180,in=0]  (IDmessage);
\end{tikzpicture}
\raisebox{1.6cm}{\large$\rrul$}
\begin{tikzpicture}[remember picture]
 	\node[big site, inner sep=10pt,] (s0r) {};
	\node[below right, inner sep=0pt, shift={(0.1,-0.1)}, ] at (s0r.north west) {0};
 	\node[right=1.3 of s0r, big site, inner sep=10pt,] (s1r) {};
	\node[below right, inner sep=0pt, shift={(0.1,-0.1)}, ] at (s1r.north west) {1};
	\node[fit=(s1r), draw, rounded corners=2,inner sep=15pt] (a0) {};
	\node[below right, inner sep=0pt, shift={(0.1,-0.1)}, ] at (a0.north west) {Message};
	\node[fit=(s0r)(a0), draw, rounded corners=2,inner sep=15pt,] (building) {};
	\node[below right, inner sep=0pt, shift={(0.1,-0.1)}, ] at (building.north west) {Boundary};
 	\node[right=0.8 of building, big site, inner sep=10pt,] (s2r) {};
	\node[below right, inner sep=0pt, shift={(0.1,-0.1)}, ] at (s2r.north west) {2};
	\node[fit=(s2r), draw, rounded corners=2,inner sep=15pt] (a1) {};
	\node[below right, inner sep=0pt, shift={(0.1,-0.1)}, ] at (a1.north west) {Message};
	\node[big region, fit=(building)(a1), inner sep=10pt,] (r1) {};
	\node[below right, inner sep=0pt, shift={(0.1,-0.1)}, ] at (r1.north west) {1};

	\node[below left=0 of building.north west, shift={(-0.9,0)},draw,rounded corners=2] (IDboundary) {ID(<level>)};
	\node[below=0.1 of IDboundary,draw,rounded corners=2,text width=2.3cm] (IDmessage) {ID(<messag-e\_id>)};
	\node[below=0.1 of IDmessage,draw,rounded corners=2,text width=2.3cm] (IDmessage1) {ID(<messag-e\_id>)};
	\node[big region, fit=(IDboundary)(IDmessage1), inner sep=10pt,] (r0) {};
	\node[below right, inner sep=0pt, shift={(0.1,-0.1)}, ] at (r0.north west) {0};

	\draw[big edge] (building.west|-,|-IDboundary)  to[out=180,in=0]  (IDboundary);
	\draw[big edge] (a0)  to[out=180,in=0]  (IDmessage);
	\draw[big edge] (a1)  to[out=180,in=0]  (IDmessage1);
\end{tikzpicture}
\begin{tikzpicture}[overlay, remember picture]
	 \draw[big inst map] (s0r) to[out=90, in=90, looseness=0.5] (s0l);
	 \draw[big inst map] (s1r) to[out=-90, in=-90, looseness=0.5] (s1l);
	 \draw[big inst map] (s2r) to[out=-90, in=-90, looseness=0.5] (s1l);
\end{tikzpicture}
}
\centering
\captionsetup{justification=centering}
\captionsetup{width=\textwidth}
\caption{Reaction rule \rr{copy\_message\_into\_boundary}. The blue dashed arrows illustrate the instantiation map \cite{archibald2024practicalmodellingbigraphs} used to duplicate the contents of site 1 in the redex.}
\label{fig:react-copy-message}
\end{figure}
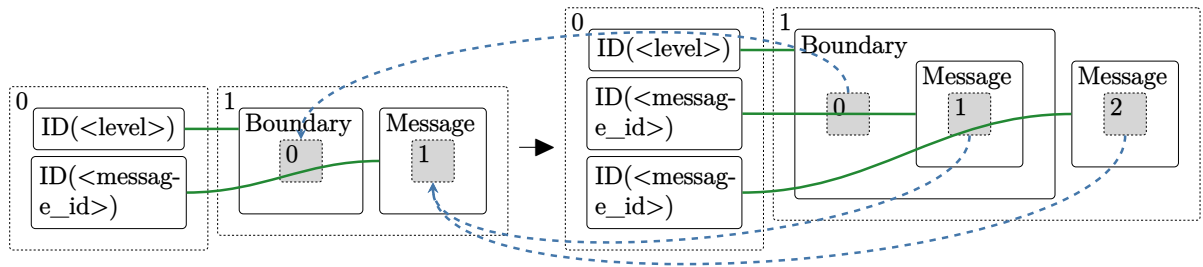

Messages can be sent to all Agent nodes in an area by specifying the area's spatial name. The Message node is created and propagated to the named area in the same way as before. Then, the Message node is copied \cite{archibald2024practicalmodellingbigraphs} into all the descendants of the area using reaction rules such as the one illustrated in Figure \ref{fig:react-copy-message}, reaching all Agent nodes in the area. Finally, another set of reaction rules is applied recursively to delete all Message nodes except those nested in Agent nodes.

\section{Repository overview}
Two repositories were maintained for this dissertation: \texttt{bigraph-tools} and \texttt{bigraph-of-the-world}. The \texttt{bigraph-tools} repository is a fork of the repository for BigraphER \cite{DBLP:conf/cav/SevegnaniC16}. The number of modified lines of code is highlighted in Table \ref{tab:repo-structure}; the modifications implement the algorithmic improvements described in \S \ref{section:bigraphER}. These fixes were submitted as pull requests to the original repository and have been accepted. Meanwhile, all source code in the \texttt{bigraph-of-the-world} repository (Table \ref{tab:botw-repo-structure}) was written from scratch.

\setlength{\tabcolsep}{0em}
\begin{table}[h!]
\centering
\begin{tabularx}{\linewidth}{@{}p{0.35\linewidth}X>{\raggedleft\arraybackslash}p{0.07\linewidth}@{}}
\toprule
\textbf{Repository} &\textbf{Description} & \textbf{Lines} \\
\midrule
\hspace{-0.45cm}\begin{minipage}[t]{1.07\linewidth}\DTsetlength{0.2em}{1em}{0.2em}{0.4pt}{0pt}\setlength{\DTbaselineskip}{14.3pt}\dirtree{%
.1 {\color{blue}bigraph-tools}.
.2 {\color{blue}bigraph}.
.3 {\color{blue}src}.
.4 big.ml\dotfill.
.4 intSet.ml\dotfill.
.4 sparse.ml\dotfill.
}\end{minipage}    
& \:
   \newline \:
   \newline \:
\newline Module that provides operations on bigraphs.
\newline Module that implements finite sets of integers.
\newline Module that implements sparse Boolean matrices.
&\begin{minipage}[t]{\linewidth}
\raggedleft
\:
   \\ \:
   \\ \:
\\ 25
\\ 2
\\ 22
\end{minipage}\\
\bottomrule
\end{tabularx}
\caption{Directory structure for the \texttt{bigraph-tools} repository the modified files, with a count of the number of lines of source code written by me.}
\label{tab:repo-structure}
\end{table}

\begin{table}[h!]
\centering
\begin{tabularx}{\linewidth}{@{}p{0.39\linewidth}X>{\raggedleft\arraybackslash}p{0.07\linewidth}@{}}
\toprule
\textbf{Repository} &\textbf{Description} & \textbf{Lines} \\
\midrule
\hspace{-0.45cm}\begin{minipage}[t]{1.07\linewidth}\DTsetlength{0.2em}{1em}{0.2em}{0.4pt}{0pt}\setlength{\DTbaselineskip}{14.3pt}\dirtree{%
.1 \texttt{\color{blue}bigraph-of-the-world}.
.2 \texttt{\color{blue}bin}.
.3 \texttt{botw.ml}\dotfill \\.
.2 \texttt{\color{blue}data}\dotfill.
.2 \texttt{\color{blue}experiments}.
.3 \texttt{plot\_results.ipynb}\dotfill.
.3 \texttt{run\_experiment.py}\dotfill.
.2 \texttt{\color{blue}lib}.
.3 \texttt{builder.ml}\dotfill \\ \\ \\.
.3 \texttt{hierarchy.ml}\dotfill \\.
.3 \texttt{overpass.ml}\dotfill \\ \\ .
.3 \texttt{query.overpassql}\dotfill \\ \\.
.2 \texttt{\color{blue}output}\dotfill.
.3 \texttt{\color{blue}renders}\dotfill \\.
.2 \texttt{bigraph-of-the-world.opam}\dotfill \\ \\ \\.
.2 \texttt{dune-project}\dotfill.
.2 \texttt{README.md}\dotfill.
}\end{minipage}    
& \:
   \newline \:
   \newline Wrapper on top of the code in \texttt{lib} which launches the executable.
   \newline Folder containing downloaded \texttt{.osm} files.
\newline \:
\newline Notebook to plot the results of experiments.
\newline Script to run experiments for evaluations.
\newline \:
\newline Module that constructs bigraphs of the real world (\S\ref{section:bigraph-implementation}). Depends on \texttt{hierarchy.ml} and \texttt{overpass.ml}, as well as the \texttt{bigraph-tools/bigraph} library.
\newline Module that parses \texttt{.osm} files to derive a space-partitioning hierarchical tree.
\newline Module that queries the Overpass API and downloads an \texttt{.osm} file. Depends on \texttt{query.overpasssql}.
\newline Overpass Query Language source code for extracting geographical data contained within a queried area (\S\ref{section:OSM-implementation}).
\newline Folder containing \texttt{.json} exports of bigraphs.
\newline Folder containing \texttt{.dot} and \texttt{.svg} renders of bigraphs.
\newline Opam package definitions, used for publishing the OCaml library and automating installation of dependencies. Generated based on \texttt{dune-project}.
\newline Project metadata, including dependencies.
\newline Documentation for the repository.
&
\begin{minipage}[t]{\linewidth}
\raggedleft
 \:
\\ \:
\\ 205\\\:\\ \:
\\ \:
\\ 421\:
\\ 286\:
\\ \:
\\ 382
\\ \:
\\ \:
\\ \:
\\ 332
\\ \:
\\ 111
\\ \:
\\ \:
\\102
\end{minipage}\\
\bottomrule
\end{tabularx}
\caption{Directory structure for the \texttt{bigraph-of-the-world} repository highlighting key folders and files, with a count of the number of lines of source code.}
\label{tab:botw-repo-structure}
\end{table}

\section{Extracting geographical data from OpenStreetMap} \label{section:OSM-implementation}

The \texttt{bigraph-of-the-world} repository implements a command-line tool that allows the user to input an administrative area as a parameter, then queries the Overpass API for all buildings, streets and administrative areas contained inside that boundary, before constructing a bigraph of the area. This section explains the query that was built using the Overpass Query Language to extract the required data from OSM. 

\subsection{Querying for buildings inside a boundary}

\begin{figure}[h]
\includegraphics[width=0.75\textwidth]{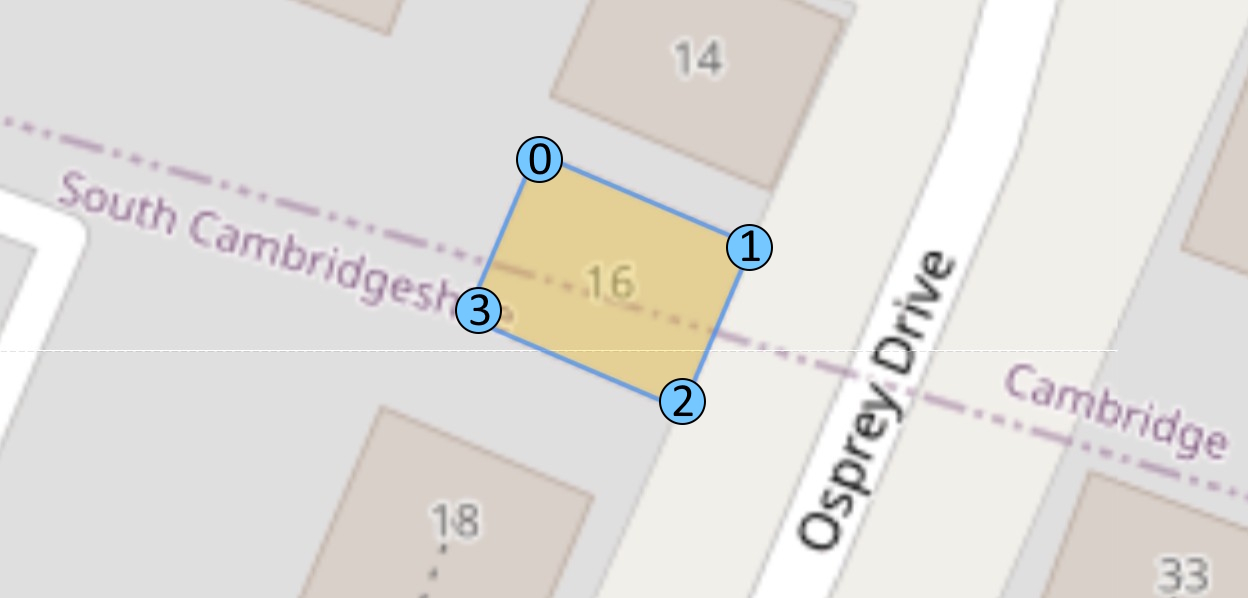}
\centering
\caption{A way element representing a building that crosses the boundary between Cambridge and South Cambridgeshire. The node members of the way element are labelled as blue circles.}
\label{fig:building-on-boundary}
\end{figure}

Buildings on OSM can be found using a filter for node, way and relation elements with a \texttt{building} tag (\S \ref{section:building-tree}). The Overpass Query Language also provides an area filter: it selects all elements that are inside a query area. Nodes are found if they are located inside or on the border of the area. Way elements are found if at least one of its nodes is in the area. Relation elements are found if at least one of its members are in the area. The area filter makes it very convenient to select all buildings inside the input boundary; however, way and relation elements that cross the boundary are included in the results of the area filter for the regions on both sides of the boundary. The problem is illustrated in Figure \ref{fig:building-on-boundary}. Fortunately, node elements that represent buildings are never located on a border because convention on OSM dictates that a node used to describe buildings cannot be part of an administrative boundary.

The hierarchical space-partitioning tree (\S \ref{section:hierarchical-tree}) requires that each building has a single unique position in the tree: a building cannot be in both Cambridge and South Cambridgeshire. In order to unambiguously decide which area a building belongs to, the query is modified to select buildings whose first member node is in the area. For example, the building "16 Osprey Drive" (Figure \ref{fig:building-on-boundary}) is included in the result of the query for Cambridge but not for South Cambridgeshire, because its first member node (numbered 0) is in Cambridge.

\subsection{Querying for streets inside a boundary} \label{section:query-streets}

\begin{figure}[h]
\includegraphics[width=0.7\textwidth]{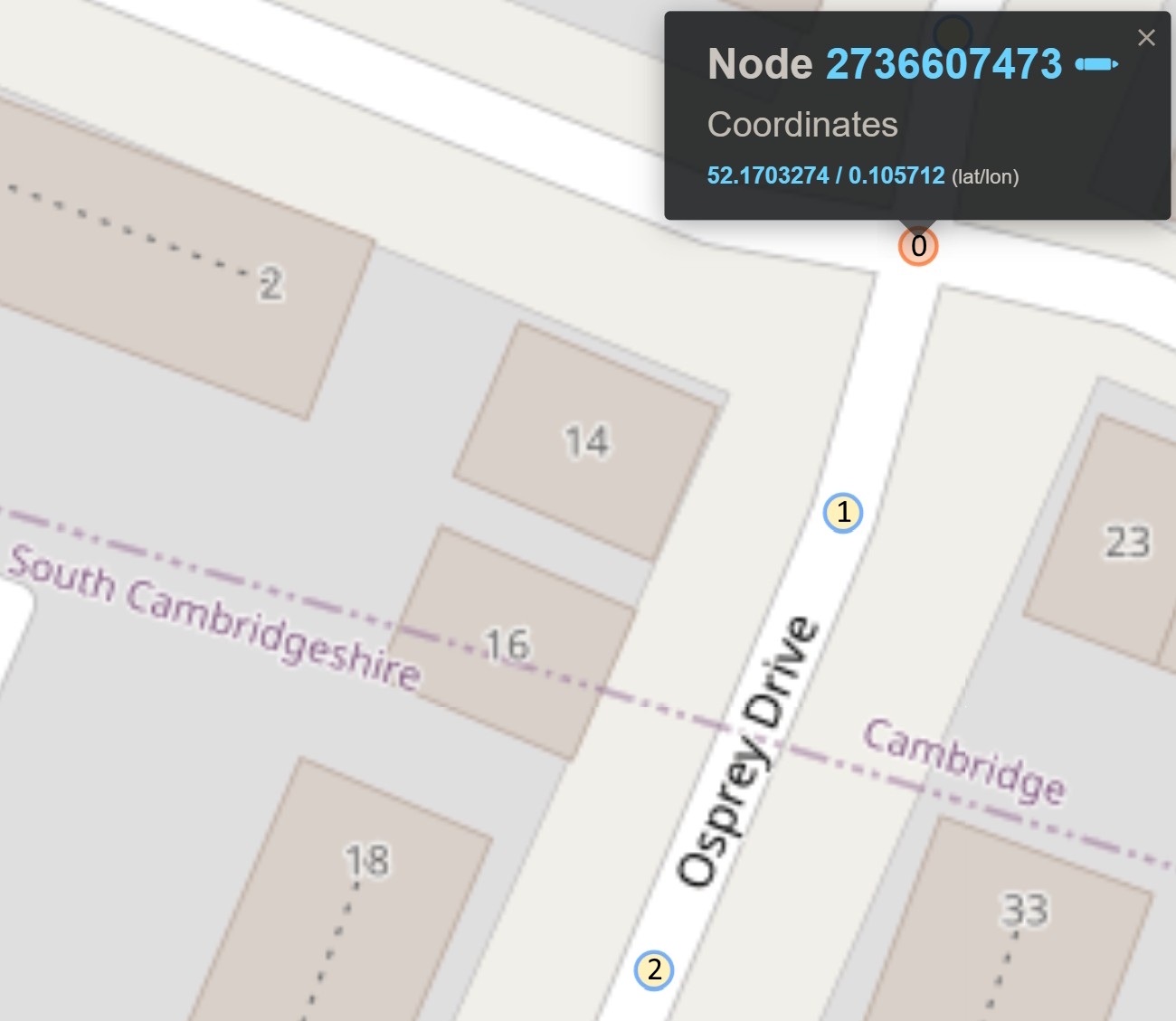}
\centering
\caption{A way element representing a street that crosses the boundary between Cambridge and South Cambridgeshire. The node members of the way element are labelled as circles, and information about the first is displayed.}
\label{fig:street-on-boundary}
\end{figure}

Streets on OSM can be found using a filter for way elements with a \texttt{highway} tag (\S \ref{section:streets-tree}). Again, the area filter can be used to select all streets inside the queried boundary. Streets that cross a boundary shall appear in the regions on both sides of the boundary, following the specification in \S \ref{section:boundaries-tree}.  To allow streets of neighbouring regions to be connected when their bigraphs are combined (\S \ref{composition-world}), the query also needs to extract the nodes that represent intersections between streets inside and outside the boundary. Firstly, the (around:0) filter provided in the Overpass Query Language is used to select the streets that intersect with the boundary. Then, the first node member of these way elements are recorded in the result of the query—these nodes will later be encoded as the outer names of the bigraph of the region. Figure \ref{fig:street-on-boundary} illustrates an example street that crosses an administrative boundary. Both the bigraphs of Cambridge and South Cambridgeshire will contain a Street node that represents "Osprey Drive", which will each nest a Junction node that has an open link with the outer name "node 2736607473". Thus, when the bigraphs of the two regions are combined, their street networks are connected via the link between the Street nodes of "Osprey Drive". 

\subsection{Querying for descendant boundaries of a region}

Administrative boundaries can be found using a filter for relation elements with a \texttt{boundary=adminstrative} tag. It is difficult to obtain just the immediate children boundaries (Figure \ref{fig:child-boundaries-cambridgeshire}) using an Overpass query. Instead, a query is built to obtain the descendant administrative boundaries (Figure \ref{fig:descendant-boundaries-cambridgeshire}); the hierarchy of immediate children is subsequently derived by \texttt{hierarchy.ml}. The area filter can again be used. However, because it selects relations of which at least one member is in the area or on the boundary, the administrative boundaries of neighbouring regions which share parts of a boundary with the queried region are also selected. To obtain only the descendant administrative boundaries contained within the queried boundary, the query is modified to select administrative boundaries which are composed of nodes that are all contained in the queried area.

\section{Constructing bigraphs of the real world} \label{section:bigraph-implementation}

\tikzstyle{startstop} = [rectangle, rounded corners, 
minimum width=3cm, 
minimum height=1cm,
text centered, 
draw=black, 
fill=red!30]

\tikzstyle{io} = [trapezium, 
trapezium stretches=true, 
trapezium left angle=70, 
trapezium right angle=110, 
minimum width=3cm, 
minimum height=1cm, 
align=center,
draw=black, fill=T-Q-PH1]

\tikzstyle{process} = [rectangle, 
minimum width=3cm, 
minimum height=1cm, 
align=center,
draw=black, 
fill=T-Q-PH2]

\tikzstyle{decision} = [diamond, 
minimum width=2.5cm, 
minimum height=1cm, 
text width=2.5cm, 
text centered, 
draw=black, 
fill=T-Q-PH3]
\tikzstyle{arrow} = [very thick,->,>=stealth]

Given a query region, functionality in \texttt{overpass.ml} queries the public Overpass API instance and saves the results in an \texttt{.osm} file. Queries are also made for all descendant boundaries. Then, functionality in \texttt{hierarchy.ml} references the descendant boundaries and \texttt{admin\_level} values in the \texttt{.osm} files to derive a \texttt{Map} from regions to their immediate children boundaries.

\begin{figure}[h!]
\resizebox{\linewidth}{!}{
\begin{tikzpicture}
\node[] (functionName) {\texttt{Builder.build}};
\node[right=0.2 of functionName,draw,fill=T-Q-PH5] (regionParam) {\texttt{region}};
\node (parseOSM) [process, below=1 of functionName.south east] {Parse \texttt{data/region.osm}};
\draw [arrow] (functionName.south east) -- (parseOSM);
\node (outerNames) [draw,right=2.3 of parseOSM,shift={(0,0.75)},align=center, fill=T-Q-PH5] {\texttt{Set} of links that connect\\ to Junctions outside \texttt{region}};
\node (streetBuilding) [draw,right=0.2 of outerNames,align=center, fill=T-Q-PH5] {\texttt{Map} of streets to\\children buildings};
\node (buildingSet) [draw,below=0.2 of outerNames,align=center, fill=T-Q-PH5] {\texttt{Set} of buildings \\w/o \texttt{addr:street}};
\node (streetJunction) [draw,right=0.2 of buildingSet,align=center, fill=T-Q-PH5] {\texttt{Map} of streets to \\children junctions};
\node (data) [draw, rounded corners=2, fit=(outerNames)(streetBuilding)(buildingSet)(streetJunction), inner sep=10pt]{};
\draw [arrow] (parseOSM) -- (data);

\node (dec1) [decision, below=1 of data, inner sep=-1] {Is \texttt{region} subdivided into children boundaries?};
\draw [arrow] (data) -- (dec1);

\node (foreachFull) [process, left=4 of dec1] {For each child in\\\texttt{child\_boundaries[region]}};
\draw [arrow] (dec1) --node[anchor=south east] {yes} (foreachFull);
\coordinate  [below=1 of foreachFull] (belowforeachFull);
\draw [thick] (foreachFull) -- (belowforeachFull);
\node (child2) [process, below=2 of foreachFull,shift={(1.2,0)}] {\texttt{Builder.build}\\\texttt{$\sim$region:child2}};
\draw [arrow] (belowforeachFull) -| (child2);
 	\node[below=2.1 of child2, big site, inner sep=10pt,] (schild2) {};
	\node[fit=(schild2), draw, rounded corners=2,inner sep=15pt,text width=1cm,] (boundarychild2) {};
	\node[below right, inner sep=0pt, shift={(0.1,-0.1)}, ] at (boundarychild2.north west) {Boundary};
	\node[above left=0.25 of boundarychild2.north, shift={(0,0)}] (nodex) {$x$};
	\draw[big edge] (schild2)  to[out=100,in=270]  (nodex);
	\node[above right=0.2 of boundarychild2.north, shift={(0,0)}] (nodey) {$y$};
	\draw[big edge] (schild2)  to[out=80,in=270]  (nodey);
\draw [arrow] (child2) -- (boundarychild2|-,|-nodey.north);

\node (child1) [process, left=0.3 of child2] {\texttt{Builder.build}\\\texttt{$\sim$region:child1}};
\draw [arrow] (belowforeachFull) -| (child1);
 	\node[below=2.1 of child1, big site, inner sep=10pt,] (schild1) {};
	\node[fit=(schild1), draw, rounded corners=2,inner sep=15pt,text width=1cm] (boundarychild1) {};
	\node[below right, inner sep=0pt, shift={(0.1,-0.1)}, ] at (boundarychild1.north west) {Boundary};
	\node[above=0.25 of boundarychild1.north, shift={(0,0)}] (nodex2) {$x$};
	\draw[big edge] (schild1)  to[out=90,in=270]  (nodex2);
\draw [arrow] (child1) -- (boundarychild1|-,|-nodey.north);

\node (child3) [right=0.5 of child2] {\dots};
\draw [arrow] (belowforeachFull) -| (child3);
\node (child3big) [right=1.25 of boundarychild2,big site, inner sep=10pt] {};
\draw [arrow] (child3) -- (child3big);

\node (mergeChildren) [process, below=2] at (foreachFull|-,|-boundarychild2.south) {$merge$ and $nest$ (\S\ref{bigraph-operations})\\into Boundary node};
\coordinate [above=1 of mergeChildren] (abovemergeChildren);
\draw[arrow] (abovemergeChildren) -- (mergeChildren);
\draw [thick] (boundarychild1) |- (abovemergeChildren);
\draw [thick] (boundarychild2) |- (abovemergeChildren);
\draw [thick] (child3big) |- (abovemergeChildren);
 	\node[below=3 of mergeChildren, shift={(0.7,0)},big site, inner sep=10pt,] (schild22) {};
	\node[fit=(schild22), draw, rounded corners=2,inner sep=15pt,text width=1cm,] (boundarychild22) {};
	\node[below right, inner sep=0pt, shift={(0.1,-0.1)}, ] at (boundarychild22.north west) {Boundary};
 	\node[left=1 of boundarychild22, big site, inner sep=10pt,] (schild12) {};
	\node[fit=(schild12), draw, rounded corners=2,inner sep=15pt,text width=1cm] (boundarychild12) {};
	\node[below right, inner sep=0pt, shift={(0.1,-0.1)}, ] at (boundarychild12.north west) {Boundary};
	\node[above=1.65, shift={(0,0)}] at ($(boundarychild12.east)!0.5!(boundarychild22.west)$) (nodex3) {$x$};
	\draw[big edge] (schild12)  to[out=0,in=270]  (nodex3);
	\draw[big edge] (schild22)  to[out=180,in=270]  (nodex3);
	\node (child32big) [right=0.3 of boundarychild22,big site, inner sep=10pt] {};
	\node[fit=(boundarychild12)(boundarychild22)(child32big), draw, rounded corners=2,inner sep=15pt] (boundaryregion) {};
	\node[below right, inner sep=0pt, shift={(0.1,-0.1)}, ] at (boundaryregion.north west) {Boundary};
	\node[above=0.2] at (schild22|-,|-boundaryregion.north) (nodey2) {$y$};
	\draw[big edge] (schild22)  to[out=90,in=270]  (nodey2);
\draw [arrow] (mergeChildren) -- (boundaryregion|-,|-nodey2.north);

\node (build) [process,below=1 of dec1] {Build component bigraphs\\using elementary bigraphs (\S\ref{elementary-bigraphs})\\and bigraph operations (\S\ref{bigraph-operations})};
\draw [arrow] (dec1) --node[anchor=east] {no} (build);
\coordinate  [below=1 of build] (belowbuild);
\draw[thick] (build) -- (belowbuild);

	\node[below=3.5 of build, shift={(-0.7,0)},draw, rounded corners=2] (Junction0HighCross) {Junction};
 	\node[below=0.3 of Junction0HighCross, big site, inner sep=10pt] (SiteHighCross) {};
	\node[fit=(Junction0HighCross)(SiteHighCross), draw, rounded corners=2,inner sep=15pt,] (StreetHighCross) {};
	\node[below right, inner sep=0pt, shift={(0.1,-0.1)}, ] at (StreetHighCross.north west) {Street};
    \node[above=0.25 of StreetHighCross.north, shift={(0,0)}] (nodex5) {$x$};
	\draw[big edge] (Junction0HighCross)  to[out=90,in=270]  (nodex5);
\draw [arrow] (belowbuild) -| (StreetHighCross|-,|-nodex5.north);

 	\node[left=0.8 of StreetHighCross, big site, inner sep=10pt,] (s0) {};
	\node[fit=(s0), draw, rounded corners=2,inner sep=15pt,] (building) {};
	\node[below right, inner sep=0pt, shift={(0.1,-0.1)}, ] at (building.north west) {Building};
\draw [arrow] (belowbuild) -| (building);

	\node[right=1.4 of Junction0HighCross, shift={(0,0.4)}, draw, rounded corners=2] (Junction2) {Junction};
	\node[below=0.3 of Junction2, draw, rounded corners=2] (Junction3) {Junction};
 	\node[below=0.3 of Junction3,big site, inner sep=10pt,] (Site2) {};
	\node[fit=(Junction2)(Site2)(Junction3), draw, rounded corners=2,inner sep=15pt,] (Street2) {};
	\node[below right, inner sep=0pt, shift={(0.1,-0.1)}, ] at (Street2.north west) {Street};
	\node[above left=0.25 of Street2.north, shift={(0,0)}] (nodex4) {$x$};
	\draw[big edge] (Junction2)  to[out=100,in=270]  (nodex4);
	\node[above right=0.2 of Street2.north, shift={(0,0)}] (nodey3) {$y$};
	\draw[big edge] (Junction3)  to[out=80,in=270]  (nodey3);
\draw [arrow] (belowbuild) -| (Street2|-,|-nodey3.north);
	
	\node (others) [right=0.2 of Street2,big site, inner sep=10pt] {};
\draw [arrow] (belowbuild) -| (others);

\node (mergeChildren2) [process, below=2] at (build|-,|-Street2.south) {$merge$ and $nest$ (\S\ref{bigraph-operations})\\into Boundary node};
\coordinate [above=1 of mergeChildren2] (abovemergeChildren2);
\draw[arrow] (abovemergeChildren2) -- (mergeChildren2);
\draw [thick] (building) |- (abovemergeChildren2);
\draw [thick] (StreetHighCross) |- (abovemergeChildren2);
\draw [thick] (Street2) |- (abovemergeChildren2);
\draw [thick] (others) |- (abovemergeChildren2);

	\node[below=3 of mergeChildren2, draw, rounded corners=2,shift={(-1,0)}] (Junction0HighCross2) {Junction};
 	\node[below=0.3 of Junction0HighCross2,  big site, inner sep=10pt] (SiteHighCross2) {};
	\node[fit=(Junction0HighCross2)(SiteHighCross2), draw, rounded corners=2,inner sep=15pt,] (StreetHighCross2) {};
	\node[below right, inner sep=0pt, shift={(0.1,-0.1)}, ] at (StreetHighCross2.north west) {Street};

 	\node[left=0.8 of StreetHighCross2, big site, inner sep=10pt,] (s02) {};
	\node[fit=(s02), draw, rounded corners=2,inner sep=15pt,] (building2) {};
	\node[below right, inner sep=0pt, shift={(0.1,-0.1)}, ] at (building2.north west) {Building};

	\node[right=1.4 of Junction0HighCross2, shift={(0,0.4)}, draw, rounded corners=2] (Junction22) {Junction};
	\node[below=0.3 of Junction22, draw, rounded corners=2] (Junction32) {Junction};
 	\node[below=0.3 of Junction32,big site, inner sep=10pt,] (Site22) {};
	\node[fit=(Junction22)(Site22)(Junction32), draw, rounded corners=2,inner sep=15pt,] (Street22) {};
	\node[below right, inner sep=0pt, shift={(0.1,-0.1)}, ] at (Street22.north west) {Street};
	
	\node (others2) [right=0.2 of Street22,big site, inner sep=10pt] {};
	\node[above=2.35, shift={(0,0)}] at ($(StreetHighCross2.east)!0.5!(Street22.west)$) (nodex6) {$x$};
	\draw[big edge] (Junction0HighCross2)  to[out=0,in=270]  (nodex6);
	\draw[big edge] (Junction22)  to[out=180,in=270]  (nodex6);
	\node[fit=(Street22)(building2)(others2), draw, rounded corners=2,inner sep=10pt,] (boundary) {};
	\node[below right, inner sep=0pt, shift={(0.1,-0.1)}, ] at (boundary.north west) {Boundary};
	\node[above=0.2] at (Junction32|-,|-boundary.north) (nodey4) {$y$};
	\draw[big edge] (Junction32)  to[out=90,in=270]  (nodey4);
\draw[arrow] (mergeChildren2) -- (boundary|-,|-nodey4.north);

\node (closeLinks) [process] at (mergeChildren|-,|-boundary) {Close all links except\\those that connect to\\Junctions outside \texttt{region}};
\draw[arrow] (boundary) -- (closeLinks);
\draw[arrow] (boundaryregion) -- (closeLinks);

\node (birgaphregion) [draw,below=1 of closeLinks, fill=T-Q-PH5] {Bigraph of \texttt{region}};
\draw[arrow] (closeLinks) -- (birgaphregion);

\node[draw, fit=(functionName)(child1)(birgaphregion)(data)(boundary), inner sep=10pt,] (function) {};
 
\coordinate [below=1 of birgaphregion] (output) ;
\draw [arrow]  (birgaphregion) -- (output);
\end{tikzpicture}
}
\caption{Flowchart describing execution of the \texttt{Builder.build} function.}
\label{fig:flowchart-build}
\end{figure}
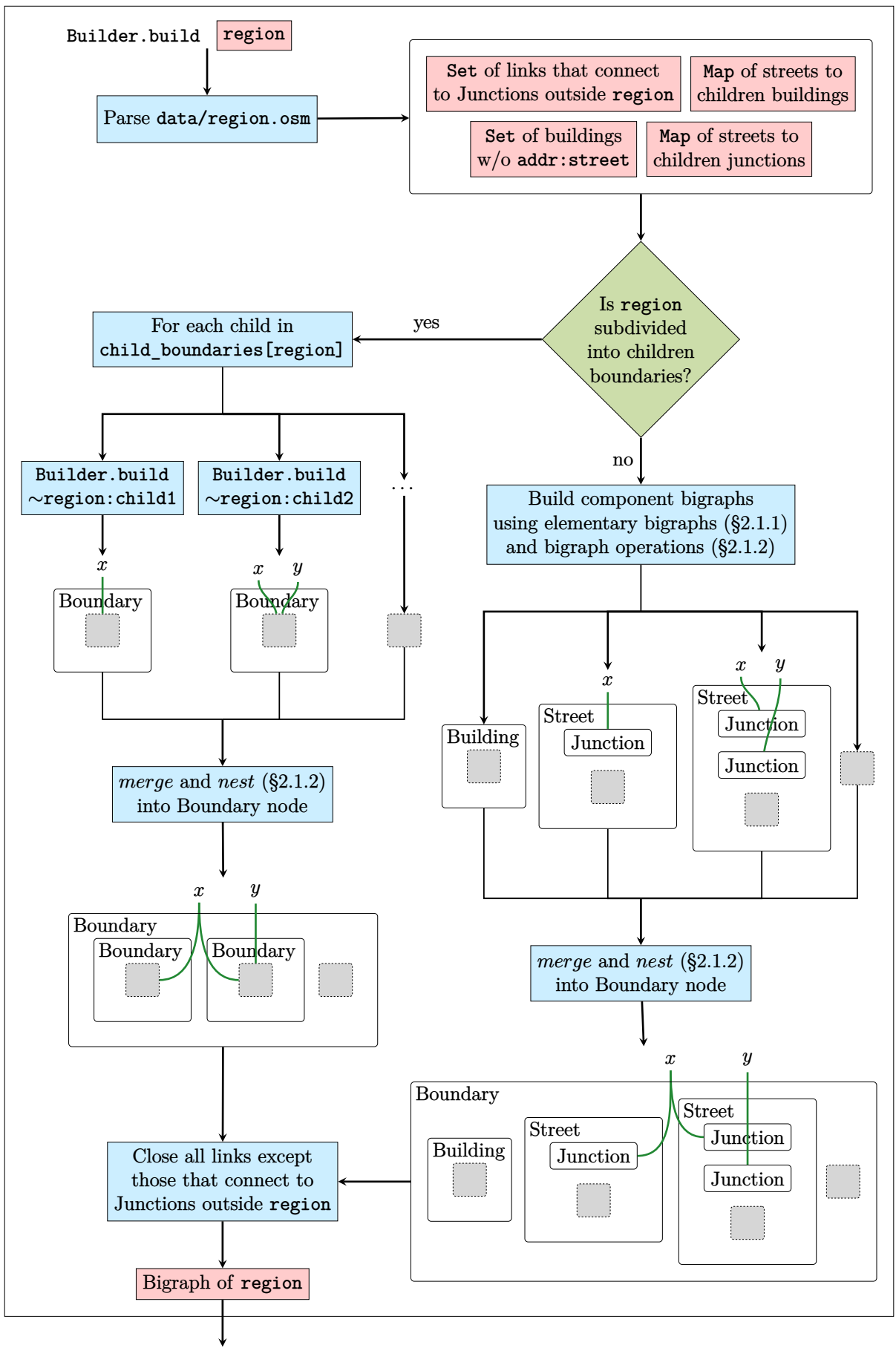

Figure \ref{fig:flowchart-build} illustrates with a flowchart the execution of the \texttt{Builder.build} function. It takes a string parameter, the name of a region. If the region is subdivided, bigraphs are first built for the immediate children boundaries in a divide-and-conquer approach; they are then merged. Otherwise, the \texttt{data/region.osm} file is parsed to derive which streets share a node member with other streets—these are encoded as Junction nodes with named links that can be combined (e.g. name $x$ in Figure \ref{fig:flowchart-build}) in the $merge$ and $nest$ operations (\S\ref{bigraph-operations}). The query results (\S\ref{section:query-streets}) also contain the set of intersections between streets inside and outside the region—only the corresponding outer names must be kept open, all other links are completed and can be closed. Subsequent experiments in \S \ref{section:eval-build} reveal that the closure of completed links after the bigraph of a region is constructed yields up to a 10× speedup in bigraph-building time.

After the bigraph of the query region is built, it is saved to a JSON file, which can be rapidly reloaded for subsequent use.

\section{Improvements to BigraphER tool} \label{section:bigraphER}

The bigraph of Cambridgeshire has 210,768 nodes and 103,274 edges — this is the largest bigraph ever built with BigraphER. The construction of bigraphs of the real world exposed critical performance limitations of BigraphER for constructing and rewriting large bigraphs, necessitating the algorithmic improvements detailed in this section.

\subsection{Representation of sparse matrices}\label{section:sparse-representation}

\begin{figure}[h!]
\includegraphics[width=\textwidth]{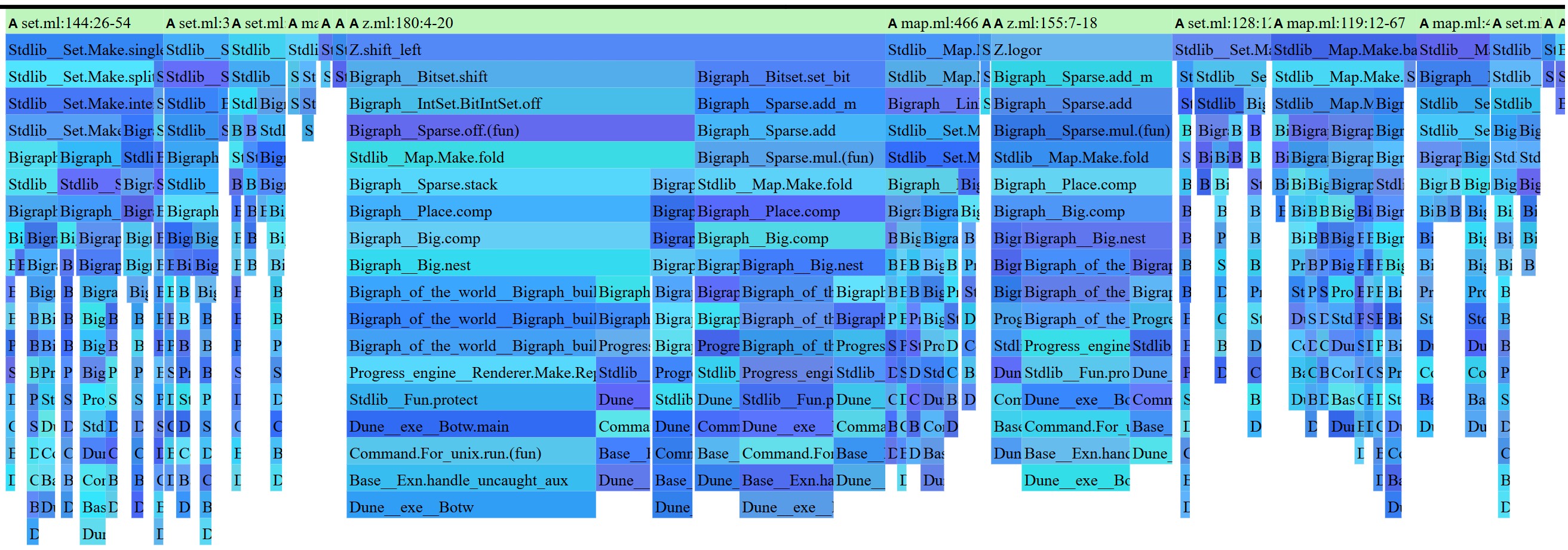}
\centering
\caption{Memory allocations visualised as an icicle chart.}
\label{fig:BitIntSet-iciclegraph}
\end{figure}

While building the bigraph of Cambridgeshire, peak live memory usage reached 11.4GB. Memory usage when constructing the bigraph was profiled using the Memtrace tool \cite{memtrace}, and the memory that remained live near the end of the program is visualised as an icicle chart in Figure \ref{fig:BitIntSet-iciclegraph}. Each function that allocates memory appears at the top, and underneath each function are its callers, scaled according to which callers caused the most allocations.

45\% of memory live near the end of the program was allocated by \texttt{Zarith} library's \texttt{z.ml}; the rest is attributed to the standard OCaml \texttt{Map} and \texttt{Set} libraries. \texttt{Bigraph.Sparse} implements adjacency matrices using standard OCaml \texttt{Map}s from \texttt{int} row indices to \texttt{Zarith} arbitrary-precision integers, used to represent a bitstring that encodes each column as 0 or 1. Place graphs are implemented using \texttt{Bigraph.Sparse}: if the $i^{th}$ node is the parent of the $j^{th}$ node, then row $i$ column$j$ has value 1, else 0. The entire $n \times n$ adjacency matrix, where $n$ is the number of nodes in the place graph, is stored in memory and hence its memory usage is $\mathcal{O}(n^2)$. 

The matrix representing a place graph is sparse: the number of 1 elements is exactly $n$ because each node has exactly one parent in a forest. Therefore, representing the place graph as an adjacency list is more space-efficient, by storing only the 1 elements. An adjacency list can be implemented with a standard OCaml \texttt{Map} from \texttt{int} row indices to standard OCaml \texttt{Set}s of \texttt{int} values. The space complexity of the adjacency list implementation is $\mathcal{O}(n)$.

After implementing sparse matrices using adjacency lists and applying the fix to the \texttt{Bigraph} library, building the bigraph of Cambrigeshire now peaks at 743MB of live memory, a \textbf{16× reduction} from before. More comprehensive results are presented in \S\ref{section:eval-build-memory}.

A comparison can also be made for the time complexity of checking if an edge exists. The original implementation of a sparse matrix in BigraphER used a standard OCaml \texttt{Map} from \texttt{int} row indices to bitstrings. \texttt{Map} access is $\mathcal{O}(\log n)$ and bitstring access is $\mathcal{O}(1)$ so checking for an edge is $\mathcal{O}(\log n)$. The proposed fix implements an adjacency list using a standard OCaml \texttt{Map} from \texttt{int} row indices to standard OCaml \texttt{Set}s of \texttt{int} values, therefore checking for an edge takes $\mathcal{O}(\log^2n)$ time, which is still very quick.

\subsection{Transitive closure of sparse matrices} \label{section:trans-implementation}

\begin{figure}[h]
\includegraphics[width=\linewidth]{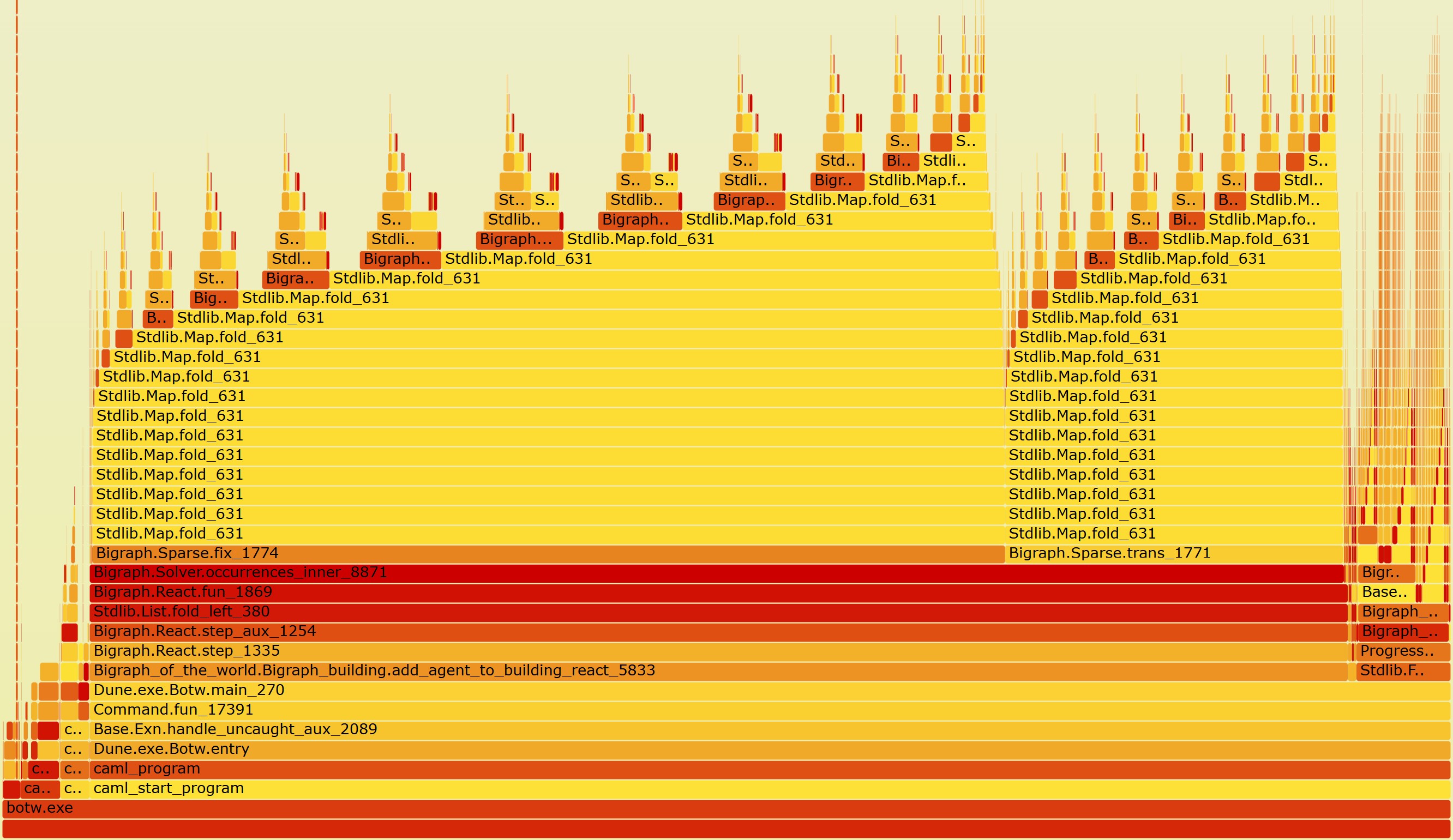}
\centering
\caption{Stack traces visualised as a flame graph.}
\label{fig:old-trans-flamegraph}
\end{figure}

Applying the reaction rule \texttt{leave\_building} (\S\ref{brs-implementation}) once on the bigraph of Cambridgeshire took 5.5 minutes. Stack traces were sampled using the perf tool and visualised as a flame graph \cite{flamegraph} in Figure \ref{fig:old-trans-flamegraph}. The profiling results indicate that the hottest code paths occur at calls to \texttt{Bigraph.Sparse.trans} and \texttt{Bigraph.Sparse.fix}, identifying them as the primary contributors to the prolonged runtime.

\texttt{Bigraph.Sparse.fix} is a helper function of \texttt{Bigraph.Sparse.trans} (Listing \ref{fig:trans-code-old}), narrowing the problem down to the \texttt{Bigraph.Sparse.trans} function. It computes the transitive closure of an adjacency matrix of a directed graph: the result indicates whether a node is reachable from another via zero or more steps in the graph. The transitive closure is used in the matching algorithm for finding occurrences of the redex of a reaction rule in a bigraph: there is an occurrence if the bigraph can be decomposed into the redex and a context, and no node in the context has an ancestor in the redex.

\begin{lstfloat}
\begin{minted}[
framesep=2mm,
baselinestretch=1,
bgcolor=lightgray!25,
fontsize=\small,
xleftmargin=1.5em,
linenos
]{ocaml}
let trans m0 =
  let rec fix m acc =
    let m' = mul m0 m in
    if equal m m' then acc else fix m' (sum m' acc)
  in
  fix m0 m0
\end{minted}
\centering
\captionof{listing}{\texttt{Bigraph.Sparse.trans} function, using the naive method.}
\label{fig:trans-code-old}
\end{lstfloat}

BigraphER's implementation (Listing \ref{fig:trans-code-old}) computes the transitive closure of a matrix using the naive method of repeatedly multiplying the matrix by itself until it reaches a fixed point. Assuming matrix access is constant time, matrix multiplication of $n \times n$ matrices is typically $\mathcal{O}(n^3)$ and it could take $n$ iterations to propagate reachability through $n$ nodes, so the time complexity of the naive method is $\mathcal{O}(n^4)$.

The problem of computing transitive closure has been widely studied. Warshall's \cite{10.1145/321105.321107} famous algorithm computes transitive closure on the Boolean semiring; the algorithm was previously introduced by Roy \cite{roy1959transitivite}. Penn \cite{PENN200672} showed that the all-pairs-shortest-path problem for sparse matrices can be efficiently solved by running Dijkstra from each of the $n$ sources. Since the place graph is a forest and is represented by a sparse boolean matrix, a simple $n \times$Depth First Search (DFS) algorithm (Listing \ref{fig:trans-code-new}) can be used to compute the transitive closure. The core idea is to perform DFS from a source node, iterating over all $n$ nodes. Assuming matrix access is constant time, this algorithm runs in $\mathcal{O}(n(n+m))$ time, where $m$ is the number of edges, or $\mathcal{O}(n^3)$ for general DAGs. Ioannidis and Ramakrishnan \cite{transitive-closure} provided a proof of correctness for the $n \times$DFS algorithm and an analysis of complexity that reaches the same conclusion.

\begin{lstfloat}
\begin{minted}[
xleftmargin=1.5em,
framesep=2mm,
baselinestretch=1,
bgcolor=lightgray!25,
fontsize=\small,
linenos
]{ocaml}
let trans m0 =
  let compute_reachability_from source _ closure = 
    let rec dfs stack_to_explore closure = 
      match stack_to_explore with
      | [] -> closure
      | current::stack_to_explore -> 
        let children = chl closure current in
        let stack_to_explore,closure = 
          let update_reachability child (stack_to_explore,closure) = 
            if mem closure source child then (stack_to_explore,closure) 
            else (child::stack_to_explore, add source child closure) 
          in
          IntSet.fold update_reachability children (stack_to_explore,closure) 
        in
        dfs stack_to_explore closure 
    in
    dfs (IntSet.to_list (chl closure source)) closure 
  in
  M_int.fold compute_reachability_from m0.r_major m0
\end{minted}
\centering
\captionof{listing}{Improved \texttt{Bigraph.Sparse.trans} function, using the $n \times$DFS algorithm.}
\label{fig:trans-code-new}
\end{lstfloat}

The asymptotic analysis can be made more precise for BigraphER's application on place graphs. Each node in the place graph has exactly one parent so the number of edges $m$ is exactly $n$.  Therefore, the naive method takes $\mathcal{O}(n^4)$ time while the $n \times$DFS algorithm takes $\mathcal{O}(n(n+m))=\mathcal{O}(n^2)$ time. Specifically for bigraphs of the real world, the place graph has a maximum depth of 13 (\S\ref{eval-hierarchy}). Therefore, reachability is propagated through within 13 iterations, so time complexity of the naive method is $\mathcal{O}(n^3)$.  The $n \times$DFS algorithm still has time complexity $\mathcal{O}(n^2)$.

After implementing the $n \times$DFS transitive closure algorithm in \texttt{Bigraph.Sparse.trans}, applying the reaction rule \texttt{leave\_building} (\S\ref{brs-implementation}) once on the bigraph of Cambridgeshire now takes 3.4s, a \textbf{97× speedup}.

\S\ref{section:fixes-sharing} verifies that the improvements proposed here also apply for bigraphs with sharing.

\chapter{Evaluation} \label{section:eval}

\textit{In this chapter, the bigraphs of the real world are qualitatively evaluated. The tool developed to generate bigraphs of the real world is evaluated empirically, and the algorithmic improvements made to bigraph-building tools are quantified.}

\section{Qualities of bigraphs of the real world}\label{section:qualities}

\subsection*{Completeness and uniformity of coverage}

This dissertation set out with the requirement (\S \ref{section:requirements}) that all named buildings must be included in the bigraph of the real world. This criteria was met: all buildings in OSM that have either a \texttt{name} or both \texttt{addr:housenumber} and \texttt{addr:street} tags are represented in the bigraph (\S\ref{section:building-tree}). The bigraph of Cambridgeshire contains 70,323 buildings—30.6\% of the 229,595 annotated buildings in the area. The bigraph of the whole world, if constructured, would contain 85,428,813 buildings—about 13\% of the 642,991,959 buildings annotated in OSM. Building coverage in OSM itself has been found to exceed 80\% in developed areas (where 16\% of the urban population live), but falls below 20\% elsewhere (48\% of the urban population) \cite{building-coverage}. This coverage improves year-on-year due to the OSM mapping community's effort (Figure \ref{fig:building-chronology}), and will directly translate to greater completeness of bigraphs of the real world.

The requirement for the complete road network to be included was also fulfilled (\S\ref{section:streets-tree}). Street coverage in OSM is very complete \cite{10.1371/journal.pone.0180698}: of 185 countries studied, 77 are more than 95\% complete; but countries such as Kiribati, Afghanistan, Egypt and China are less than one third complete. This is steadily improving (Figure \ref{fig:highway-chronology}).

\subsection*{Hierarchical depth and partition granularity} \label{eval-hierarchy}

Bigraphs of the real world encode a hierarchy of containment that has a maximum depth of 13: the root is the "world", followed by administrative boundaries with \rr{admin\_level} values ranging from 2 to 11, then the street and lastly the building. The hierachical tree (\S\ref{section:hierarchical-tree}) partitions the world with fine granularity, using the 718,429 administrative boundaries in OSM. The coverage of administrative boundaries is attributed to the voting and discussion by the mapping community to determine boundaries that are widely recognized and best reflect on-the-ground realities—combining local knowledge with standards such as ISO 3166 and the Nomenclature of Territorial Units for Statistics. Furthermore, buildings are grouped according to their \texttt{addr:street} tag, of which there are 4,007,075 unique values. The specification for the space-partitioning tree of the world strikes a balance between the criteria for partition granularity and the requirements for recognisability and generalisability by using administrative areas and streets, which are standard components in addressing systems.

\subsection*{Ability to unify models}

\begin{figure}[h]
\includegraphics[width=0.65\textwidth]{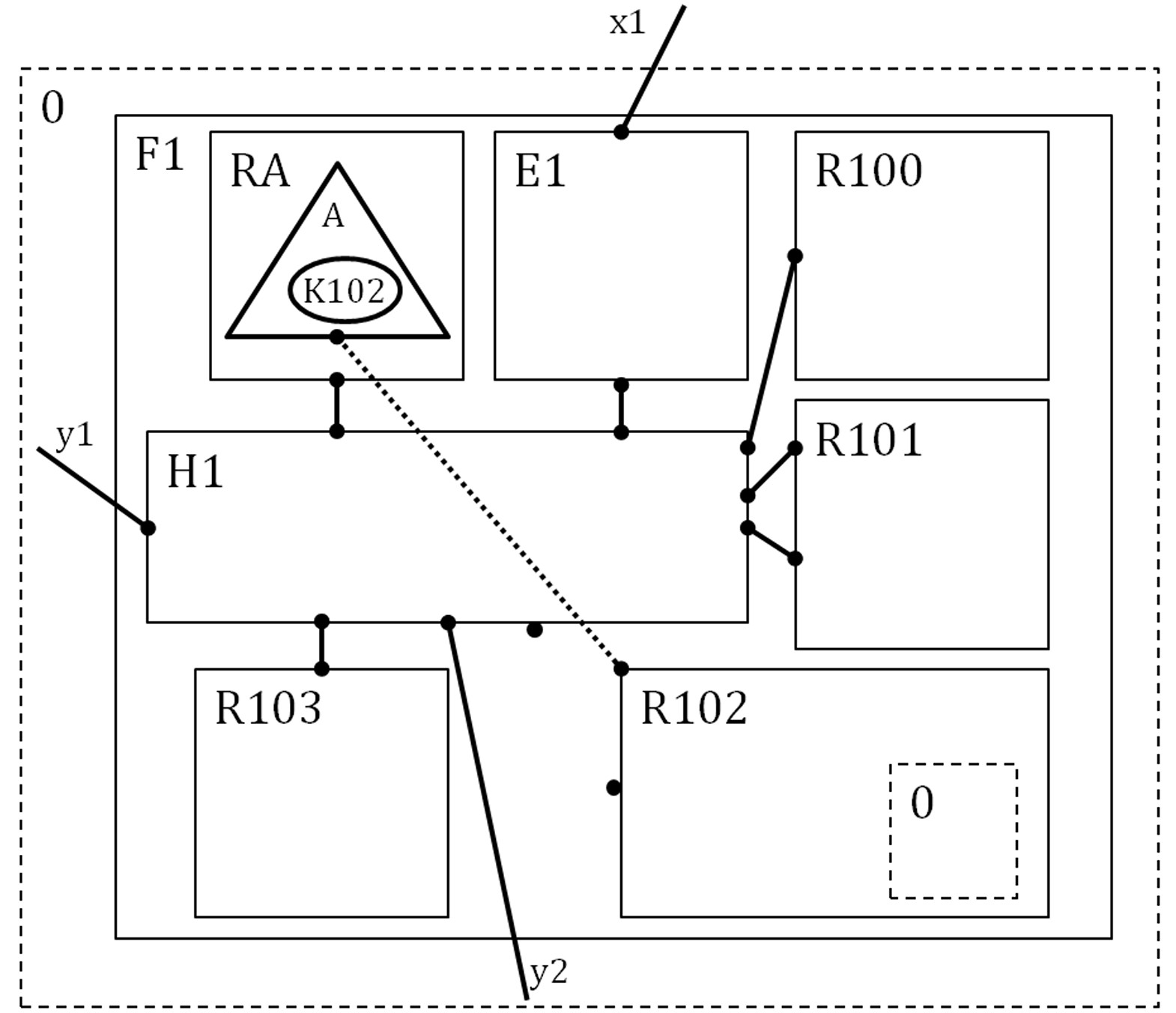}
\centering
\caption{An indoor bigraph. The {F1} node represents the first floor, {H1} the hallway, {E1} the elevator, {RA} the reception area, and the remaining {R} nodes represent individual rooms.~\cite{10.1007/978-3-642-33024-7_17}}
\label{fig:indoor-bigraph}
\end{figure}

Bigraphs of the real world model the complete outdoor space of the world, and thereby acts as a unifying interface for indoor bigraphs (e.g. Figure \ref{fig:indoor-bigraph}) constructed by Walton and Worboys \cite{10.1007/978-3-642-33024-7_17}—they can immediately be nested into the Building nodes. An indoor bigraph encodes accessibility in its link graph; its reaction rules allows an agent to move between two places that are linked, which differs from the reaction rules for motion in this dissertation (\S \ref{section:react-motion}) and Milner's introduction \cite{10.5555/1540607}. Regardless, both sets of reactions rules can be combined in a BRS to allow movement around indoor and outdoor space.

\begin{figure}[h]
\includegraphics[width=0.83\textwidth]{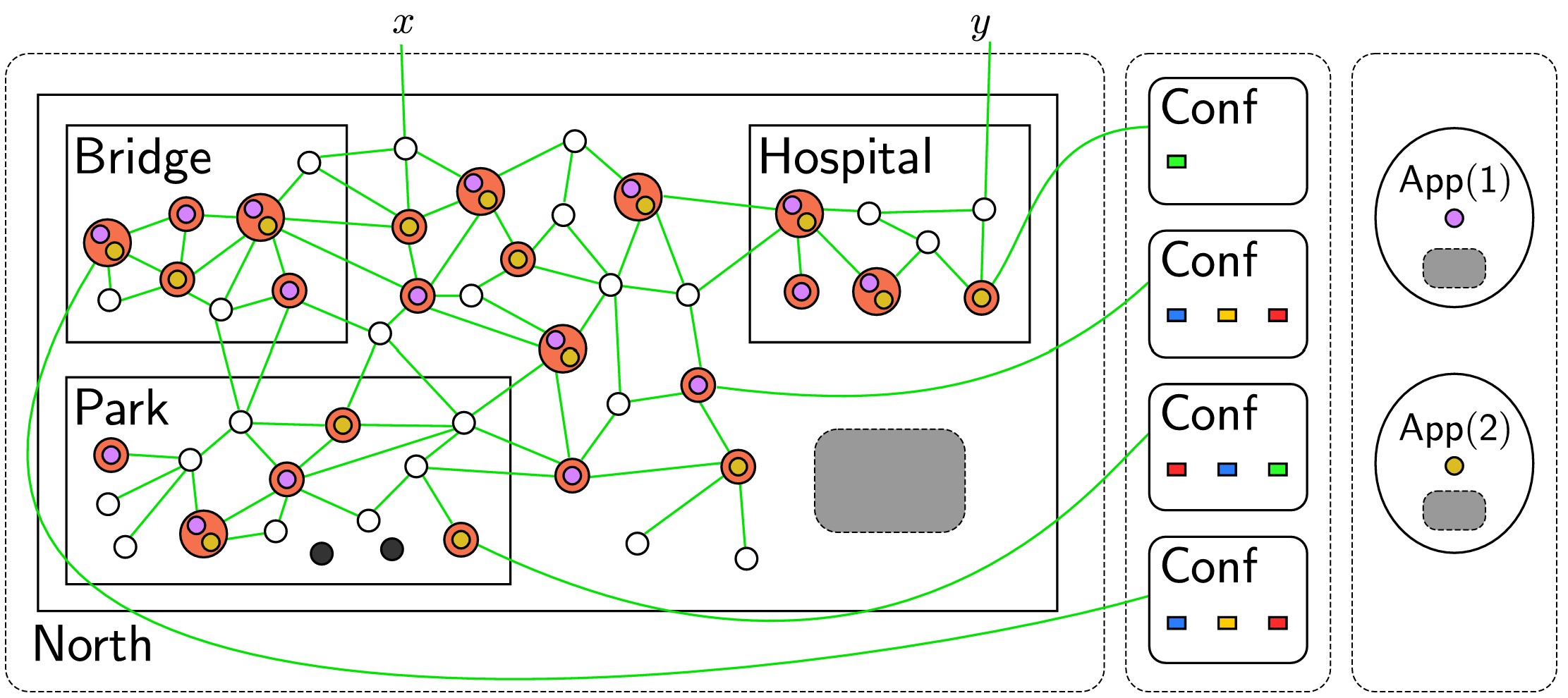}
\centering
\caption{Bigraph model of a city-wide sensor network infrastructure. The three regions in the bigraph correspond, from left to right, to the Physical, Data, and Service perspectives, respectively.~\cite{8595061}}
\label{fig:large-sensor-network}
\end{figure}

Bigraphs of the real world can also ground the physical entities of all bigraphs from past and future work to their physical location in the world, to provide spatial context to the rest of the world. For example, in a case-study for the modelling and verification of large-scale sensor networks \cite{8595061}, a bigraph was built consisting three parallel perspectives: Physical, Data, and Service (Figure \ref{fig:large-sensor-network}). The Physical perspective contained a hierarchy that abstracted the location of the geographical places and physical sensors around Queen Elizabeth Olympic Park in London—this can be immediately substituted by a bigraph of London constructed by the deliverables of this dissertation. The Data and Service perspectives can be added as parallel regions to the bigraph of the real world.

Extensive research has demonstrated that bigraph’s universal formalism can be used to model a wide range of applications. These bigraph models can then be unified by a bigraph of the real world, which maps all physical entities to their geographical locations and gives insight to their spatial relationships with entities of other models.

This section evaluated the properties of bigraphs of the real world and presented evidence that they fulfil all the requirements set out in \S \ref{section:requirements}. The following sections present a quantitative evaluation of the tool developed to build and transform bigraphs of the real world, demonstrating that the theory is supported by a practical and efficient implementation.

\section{Time and memory required to build bigraphs} \label{section:eval-build}
Building bigraphs of the real world involves parsing OSM data for a queried region and encoding buildings, streets and boundaries into the bigraph (\S \ref{section:bigraph-implementation}). This section first presents the peak memory usage when building a bigraph of the real world, highlighting the effect of the improvements made to BigraphER (\S \ref{section:sparse-representation}), followed by an evaluation of the time it takes to build it.

\subsection*{Memory usage} \label{section:eval-build-memory}
\begin{figure}[h!]
\includegraphics[width=\textwidth]{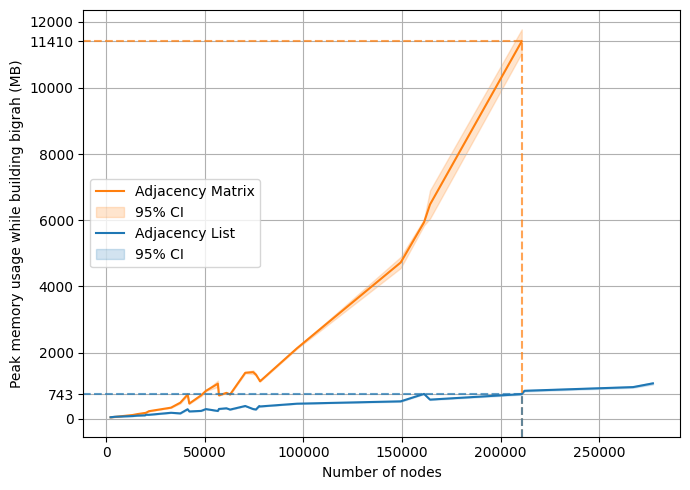}
\centering
\caption{Plot of the mean peak live memory usage while building a bigraph as a function of the number of nodes, comparing the original adjacency matrix and the proposed adjacency list sparse matrix representations. The 95\% confidence interval is represented in the shaded area about the mean.}
\label{fig:memory-graph}
\end{figure}

The first experiment measures the peak memory usage of constructing bigraphs of the real world, comparing the original adjacency matrix sparse matrix representation and the proposed adjacency list implementation (\S \ref{section:sparse-representation}). Bigraphs were constructed for 30 selected regions from around the world (See Table \ref{table:bigraph-stats} for details). The experiment was repeated for 30 runs to obtain accurate averages, and memory usage was measured using the maximum resident set size reported by the \texttt{/usr/bin/time} tool. Attempts to construct bigraphs bigger than that of Cambridgeshire, which has 210,768 nodes, failed when using the adjacency matrix implementation as the application received a kill signal for using excessive memory. Figure \ref{fig:memory-graph} illustrates the results, which show that the peak memory usage of the original adjacency matrix implementation grows quickly—asymptotic analysis conducted in \S \ref{section:sparse-representation} suggests the memory complexity is  $\mathcal{O}(n^2)$. The results show, for the proposed adjacency list implementation, a linear relationship between the memory usage and the number of nodes, which agrees with the asymptotic analysis conducted in \S \ref{section:sparse-representation}. The bigraph of Cambridgeshire is built with a mean peak memory usage of 730MB with the improved adjacency list representation of sparse matrices, a \textbf{16× reduction} from 11.41GB.

All subsequent experiments are conducted using the improved sparse matrix representation.

\subsection*{Execution time} 
\begin{figure}[h!]
\includegraphics[width=\textwidth]{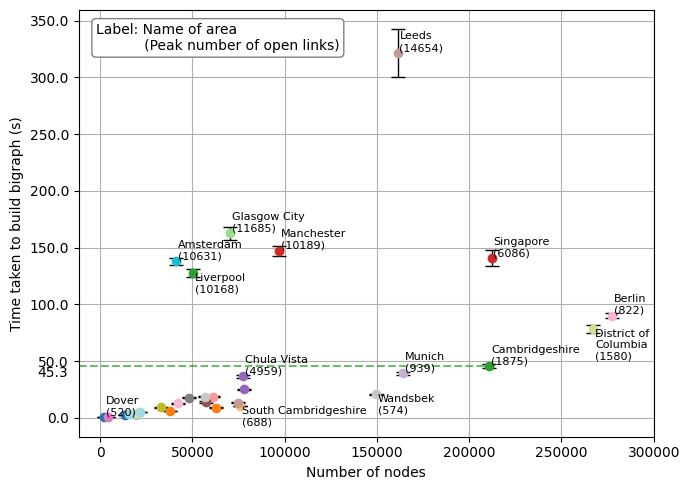}
\centering
\caption{Scatter plot of the mean bigraph construction time versus the number of nodes. Each data point is labelled with the corresponding area name and peak number of open links during construction, with error bars indicating the standard error.}
\label{fig:scatter-plot-building-time}
\end{figure}

The next experiment measures the time taken to build a bigraph of the real world. The same experiment setup is used and execution time is measured. Figure \ref{fig:scatter-plot-building-time} illustrates the results in a scatter plot—it shows no consistent relationship between the time taken to build a bigraph and the number of nodes. The time required to build a bigraph of Singapore is three times that of Cambridgeshire, despite both having a similar number of nodes and edges (Table \ref{table:sg-cambridgeshire}). 

\begin{table}[h!]
\setlength{\tabcolsep}{2pt}
\resizebox{\linewidth}{!}{
\begin{tabular} { llllllll}
\toprule
Area & Nodes & Edges & Peak open links & Subdivisions & Streets & Buildings & Junctions \\
\midrule
Cambridgeshire & 210,768 & 103,274  & 1,875 & 241 & 16,190 & 69,754 & 38,396 \\
Singapore & 212,140 & 104,497 & 6,086 & 12 & 11,917 & 70,981 & 46,318 \\
\bottomrule
\end{tabular}
}
\caption{Key statistics of the bigraphs of Cambridgeshire and Singapore.}
\label{table:sg-cambridgeshire}
\end{table}

My hypothesis is that the time taken to build a bigraph depends on the number of open links maintained while constructing the bigraph (\S\ref{section:bigraph-implementation}). In the step where the $merge$ and $nest$ operations (\S \ref{bigraph-operations}) are applied to Street and Boundary nodes, open links with matching outer names are combined, linking Junction nodes together. When the bigraph of a region is completed, links that correspond to intersections between streets inside and out the region must be kept open; the rest can be closed since they will not connect to any more Junction nodes. Since Cambridgeshire is subdivided into 241 areas, a divide-and-conquer approach is taken to first build the bigraphs for its smaller subdivisions, close the unnecessary links for each, before merging them. The number of open links maintained peaks at 1,875. Meanwhile, Singapore is a densely-populated but small country, subdivided into only 12 areas; as a result, the live set of open links peaks at a higher value of 6,086. It is not merely the number of subdivisions that matters: Leeds, despite having 38 subdivisions, contains one exceptionally large area (Figure \ref{fig:leeds}). As a result, the number of open links peaks at 14,654 during bigraph construction, leading to the longest build time, as shown in Figure \ref{fig:scatter-plot-building-time}. The experiment results reveals that the requirement for partition granularity of bigraphs of the real world (\S \ref{eval-hierarchy}) also plays a key role in reducing the time taken to build them. 

The bigraph of Cambridgeshire is constructed in \textbf{45.3s} on average. The following section shows how bigraphs of the real world can be loaded rapidly for real-time applications.

\section{Time and memory required to load bigraphs} \label{sect:eval-load}

\begin{figure}[h!]
\includegraphics[width=0.95\textwidth]{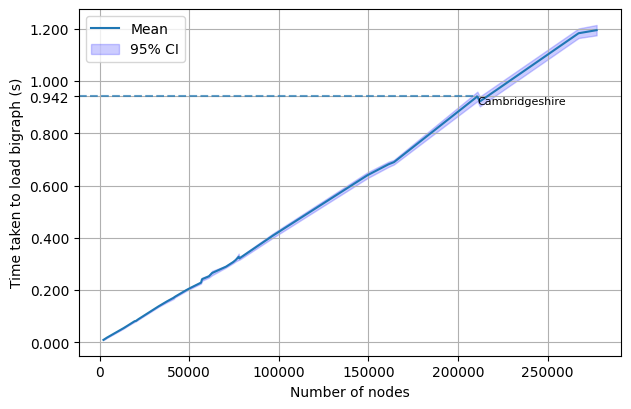}
\centering
\caption{Plot of the time taken to load a bigraph from a JSON file. The 95\% confidence interval is represented in the shaded area about the mean.}
\label{fig:load-json-time}
\end{figure}

\begin{figure}[h!]
\includegraphics[width=0.95\textwidth]{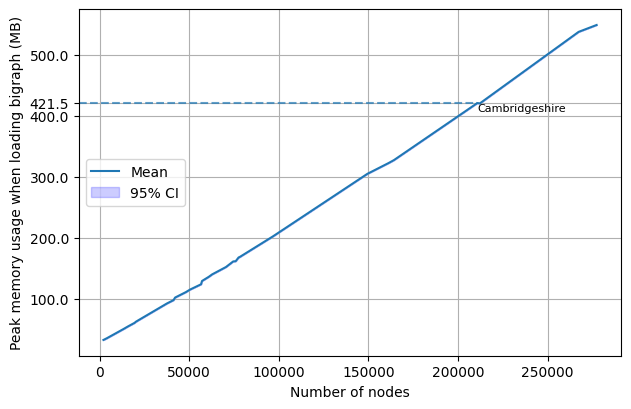}
\centering
\caption{Plot of the peak memory usage when loading a bigraph from a JSON file. The 95\% confidence interval is represented in the shaded area about the mean.}
\label{fig:load-json-memory}
\end{figure}

After bigraphs are constructed from parsing OSM data, they are exported to JSON files which can be reloaded subsequently. Here, the experiment measures the time taken and memory required to load a bigaph from a JSON file instead of building it from scratch. Bigraphs were loaded for 30 regions (Table \ref{table:bigraph-stats}) and repeated for 100 runs. Figures \ref{fig:load-json-time} and \ref{fig:load-json-memory} show that the time taken and memory usage scale linearly with the number of nodes. It takes \textbf{only 0.942s} of execution time and 421.5MB of memory to load the bigraph of Cambridgeshire from a JSON file. Additionally, the JSON export of the bigraph of Cambridgeshire is only \textbf{27MB}; the file size also scales linearly to the number of nodes in the bigraph.

\section{Time and memory required to apply reaction rules} \label{sect:eval-transform}

\subsection*{Time taken to apply a single reaction rule}

\begin{figure}[h!]
\includegraphics[width=\textwidth]{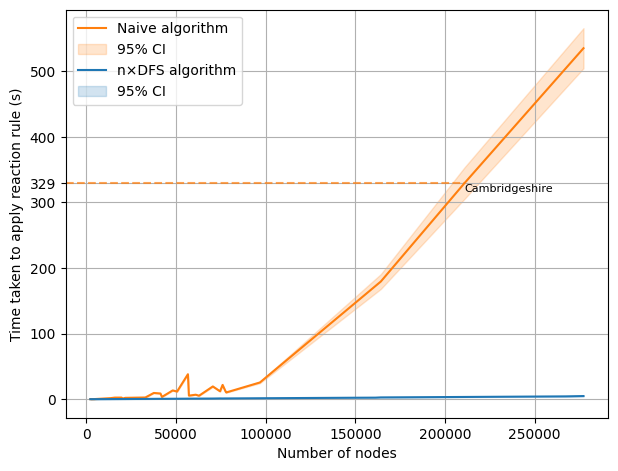}
\centering
\caption{Plot of time taken to apply the reaction rule \texttt{leave\_building} as a function of the number of nodes, comparing the naive method of computing transitive closure and the $n\times DFS$ algorithm. The 95\% confidence interval is represented in the shaded area about the mean.}
\label{fig:trans-compare}
\end{figure}

\begin{figure}[h!]
\includegraphics[width=0.85\textwidth]{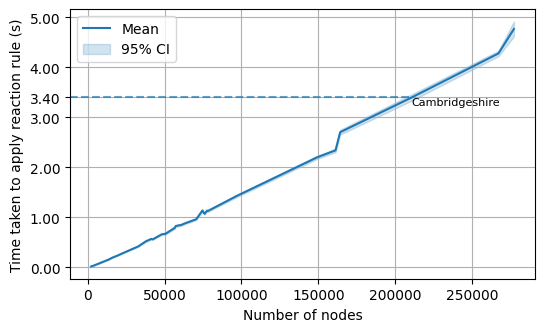}
\centering
\caption{Plot of time taken to apply the reaction rule \texttt{leave\_building} as a function of the number of nodes for the $n\times DFS$ algorithm. The 95\% confidence interval is represented in the shaded area about the mean.}
\label{fig:time-react}
\end{figure}

The next experiment measures the time taken to apply a single reaction rule on bigraphs of the real world, comparing the original naive method of computing transitive closure of sparse matrices against the proposed $n\times DFS$ algorithm (\S \ref{section:trans-implementation}). Firstly, bigraphs for 30 select regions (Table \ref{table:bigraph-stats}) were loaded; an Agent node was then added inside a random Building node in each bigraph. The reaction rule \texttt{leave\_building} (\S \ref{section:react-motion}) was then applied once and execution time is measured. The experiment is repeated for 30 runs with the Agent node starting in a random Building node each time to get accurate averages. The two algorithms are compared in Figure \ref{fig:trans-compare}, and the results for the $n\times DFS$ algorithm are plotted again in Figure \ref{fig:time-react} for clarity.

Figure \ref{fig:trans-compare} shows that when using the original naive transitive closure method, the execution time grows quickly—asymptotic analysis in \S \ref{section:trans-implementation} suggests that the time complexity is $\mathcal{O}(n^3)$. After switching to the proposed $n\times DFS$ algorithm, computing the transitive closure of sparse matrices no longer is a major contributor to execution time (See Figure \ref{fig:new-trans-flamegraph} for a flame graph of the stack trace); the time taken to apply a reaction rule now scales linearly with the number of nodes in the bigraph, as seen in Figure \ref{fig:time-react}. With the $n\times DFS$ algorithm, it only takes \textbf{3.40s} to apply the reaction rule \texttt{leave\_building} on the bigraph of Cambridgeshire, a \textbf{97× reduction} from 329s.

All subsequent experiments are conducted using the $n\times DFS$ algorithm to compute the transitive closure of sparse matrices.

\subsection*{Memory usage when applying a single reaction rule}

\begin{figure}[h!]
\includegraphics[width=0.85\textwidth]{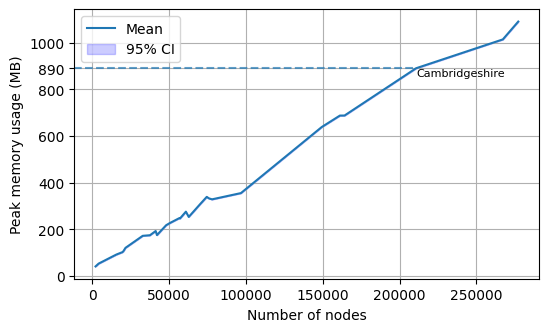}
\centering
\caption{Plot of peak memory usage when applying \texttt{leave\_building} against the number of nodes. The 95\% confidence interval is represented in the shaded area about the mean.}
\label{fig:memory-react-once}
\end{figure}

In the previous experiment where the reaction rule \texttt{leave\_building} was applied once, peak memory usage was also recorded. Figure \ref{fig:memory-react-once} illustrates the results and shows a linear relationship between the peak memory usage and the number of nodes in the bigraph.

\subsection*{Difference in execution times of reaction rules} \label{section:eval-reacts}

\begin{figure}[h!]
\begin{subfigure}{\linewidth}
\includegraphics[width=\textwidth]{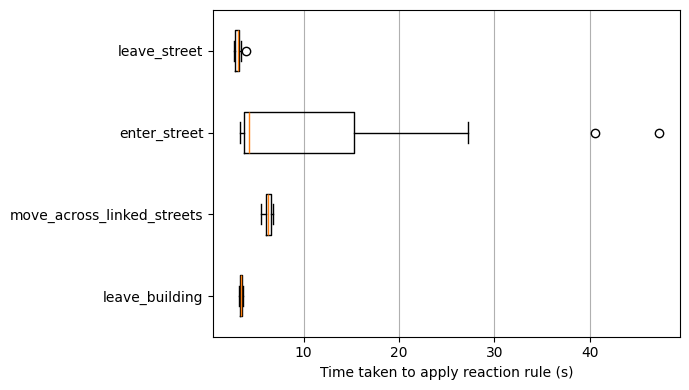}
\centering
\end{subfigure}

\begin{subfigure}{\linewidth}
\includegraphics[width=\textwidth]{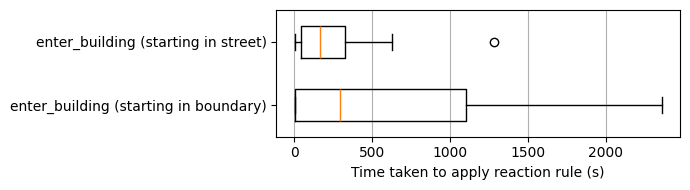}
\centering
\end{subfigure}
\caption{Box and whisker plots showing the distribution of execution times for applying the reaction rules for motion. The bottom panel details \texttt{enter\_building}, contrasting scenarios where the agent starts within a Street node versus a Boundary node. Note the different x-axis scale used in the bottom panel to accommodate the wider range of execution times for entering buildings.}
\label{fig:reactions-for-motion}
\end{figure}

The last experiment investigates whether there are significant differences in the execution times of the reaction rules of motion (\S \ref{section:react-motion}). The bigraph of Cambridgeshire was first loaded; an Agent node is then added to a random Building node in the bigraph. The reaction rules are applied one at a time and execution time for each is measured. The experiment is repeated for 20 runs, with the Agent node starting in a random Building node each time for accurate average results. Figure \ref{fig:reactions-for-motion} illustrates the results.

\begin{figure}[h!]
\centering
\begin{subfigure}{\linewidth}
\centering
\begin{tikzpicture}
 	\node[big site, inner sep=10pt,] (s0) {};
	\node[below right, inner sep=0pt, shift={(0.1,-0.1)}, ] at (s0.north west) {0};
 	\node[right=0.8 of s0, big site, inner sep=10pt,] (s1) {};
	\node[below right, inner sep=0pt, shift={(0.1,-0.1)}, ] at (s1.north west) {1};
	\node[fit=(s1), draw, rounded corners=2,inner sep=15pt,] (building) {};
	\node[below right, inner sep=0pt, shift={(0.1,-0.1)}, ] at (building.north west) {Building};
 	\node[right=0.8 of building, big site, inner sep=10pt,] (s2) {};
	\node[below right, inner sep=0pt, shift={(0.1,-0.1)}, ] at (s2.north west) {2};
	\node[fit=(s2), draw, rounded corners=2,inner sep=15pt,] (a0) {};
	\node[below right, inner sep=0pt, shift={(0.1,-0.1)}, ] at (a0.north west) {Agent};
	\node[fit=(s0)(building)(a0), draw, rounded corners=2, inner sep=8pt,] (street) {};
	\node[below right, inner sep=0pt, shift={(0.1,-0.1)}, ] at (street.north west) {Street};
	\node[big region, fit=(street), inner sep=10pt,] (r0) {};
	\node[below right, inner sep=0pt, shift={(0.1,-0.1)}, ] at (r0.north west) {0};

	\node[above=0.6 of street, shift={(0,0)}] (x) {$x$};
	\draw[big edge] (street)  to[out=90,in=270]  (x);
	\node[above=0.9 of building, shift={(0,0)}] (y) {$y$};
	\draw[big edge] (building)  to[out=90,in=270]  (y);
	\node[above=0.9 of a0, shift={(0,0)}] (z) {$z$};
	\draw[big edge] (a0)  to[out=90,in=270]  (z);
\end{tikzpicture}
\raisebox{1.6cm}{\large$\rrul$}
\begin{tikzpicture}
 	\node[big site, inner sep=10pt,] (s0) {};
	\node[below right, inner sep=0pt, shift={(0.1,-0.1)}, ] at (s0.north west) {0};
 	\node[right=0.8 of s0, big site, inner sep=10pt,] (s1) {};
	\node[below right, inner sep=0pt, shift={(0.1,-0.1)}, ] at (s1.north west) {1};
 	\node[right=1.2 of s1, big site, inner sep=10pt,] (s2) {};
	\node[below right, inner sep=0pt, shift={(0.1,-0.1)}, ] at (s2.north west) {2};
	\node[fit=(s2), draw, rounded corners=2,inner sep=15pt,] (a0) {};
	\node[below right, inner sep=0pt, shift={(0.1,-0.1)}, ] at (a0.north west) {Agent};
	\node[fit=(s1)(a0), draw, rounded corners=2,inner sep=10pt,] (building) {};
	\node[below right, inner sep=0pt, shift={(0.1,-0.1)}, ] at (building.north west) {Building};
	\node[fit=(s0)(building), draw, rounded corners=2, inner sep=8pt,] (street) {};
	\node[below right, inner sep=0pt, shift={(0.1,-0.1)}, ] at (street.north west) {Street};
	\node[big region, fit=(street), inner sep=10pt,] (r0) {};
	\node[below right, inner sep=0pt, shift={(0.1,-0.1)}, ] at (r0.north west) {0};

	\node[above=0.25] at (street|-,|-r0.north) (x) {$x$};
	\draw[big edge] (street)  to[out=90,in=270]  (x);
	\node[above=0.2] at (building|-,|-r0.north) (y) {$y$};
	\draw[big edge] (building)  to[out=90,in=270]  (y);
	\node[above=0.25] at (a0|-,|-r0.north) (z) {$z$};
	\draw[big edge] (a0)  to[out=90,in=270]  (z);
\end{tikzpicture}
\captionsetup{justification=centering}
\caption{\texttt{enter\_building\_from\_street}:\\$\text{Street}_x.(id|\text{Building}_y|\text{Agent}_{z})\rrul\text{Street}_x.(id|\text{Building}_y.(id|\text{Agent}_{z})$}
\label{fig:street-to-building}
\end{subfigure}

\begin{subfigure}{\linewidth}
\centering
\begin{tikzpicture}
 	\node[big site, inner sep=10pt,] (s0) {};
	\node[below right, inner sep=0pt, shift={(0.1,-0.1)}, ] at (s0.north west) {0};
 	\node[right=0.8 of s0, big site, inner sep=10pt,] (s1) {};
	\node[below right, inner sep=0pt, shift={(0.1,-0.1)}, ] at (s1.north west) {1};
	\node[fit=(s1), draw, rounded corners=2,inner sep=15pt,] (building) {};
	\node[below right, inner sep=0pt, shift={(0.1,-0.1)}, ] at (building.north west) {Building};
 	\node[right=0.8 of building, big site, inner sep=10pt,] (s2) {};
	\node[below right, inner sep=0pt, shift={(0.1,-0.1)}, ] at (s2.north west) {2};
	\node[fit=(s2), draw, rounded corners=2,inner sep=15pt,] (a0) {};
	\node[below right, inner sep=0pt, shift={(0.1,-0.1)}, ] at (a0.north west) {Agent};
	\node[fit=(s0)(building)(a0), draw, rounded corners=2, inner sep=15pt,] (street) {};
	\node[below right, inner sep=0pt, shift={(0.1,-0.1)}, ] at (street.north west) {Boundary};
	\node[big region, fit=(street), inner sep=10pt,] (r0) {};
	\node[below right, inner sep=0pt, shift={(0.1,-0.1)}, ] at (r0.north west) {0};

	\node[above=0.6 of street, shift={(0,0)}] (x) {$x$};
	\draw[big edge] (street)  to[out=90,in=270]  (x);
	\node[above=0.9 of building, shift={(0,0)}] (y) {$y$};
	\draw[big edge] (building)  to[out=90,in=270]  (y);
	\node[above=0.9 of a0, shift={(0,0)}] (z) {$z$};
	\draw[big edge] (a0)  to[out=90,in=270]  (z);
\end{tikzpicture}
\raisebox{1.6cm}{\large$\rrul$}
\begin{tikzpicture}
 	\node[big site, inner sep=10pt,] (s0) {};
	\node[below right, inner sep=0pt, shift={(0.1,-0.1)}, ] at (s0.north west) {0};
 	\node[right=0.8 of s0, big site, inner sep=10pt,] (s1) {};
	\node[below right, inner sep=0pt, shift={(0.1,-0.1)}, ] at (s1.north west) {1};
 	\node[right=1.2 of s1, big site, inner sep=10pt,] (s2) {};
	\node[below right, inner sep=0pt, shift={(0.1,-0.1)}, ] at (s2.north west) {2};
	\node[fit=(s2), draw, rounded corners=2,inner sep=15pt,] (a0) {};
	\node[below right, inner sep=0pt, shift={(0.1,-0.1)}, ] at (a0.north west) {Agent};
	\node[fit=(s1)(a0), draw, rounded corners=2,inner sep=10pt,] (building) {};
	\node[below right, inner sep=0pt, shift={(0.1,-0.1)}, ] at (building.north west) {Building};
	\node[fit=(s0)(building), draw, rounded corners=2, inner sep=15pt,] (street) {};
	\node[below right, inner sep=0pt, shift={(0.1,-0.1)}, ] at (street.north west) {Boundary};
	\node[big region, fit=(street), inner sep=10pt,] (r0) {};
	\node[below right, inner sep=0pt, shift={(0.1,-0.1)}, ] at (r0.north west) {0};

	\node[above=0.25] at (street|-,|-r0.north) (x) {$x$};
	\draw[big edge] (street)  to[out=90,in=270]  (x);
	\node[above=0.2] at (building|-,|-r0.north) (y) {$y$};
	\draw[big edge] (building)  to[out=90,in=270]  (y);
	\node[above=0.25] at (a0|-,|-r0.north) (z) {$z$};
	\draw[big edge] (a0)  to[out=90,in=270]  (z);
\end{tikzpicture}
\captionsetup{justification=centering}
\caption{\texttt{enter\_building\_from\_boundary}:\\$\text{Boundary}_x.(id|\text{Building}_y|\text{Agent}_{z})\rrul\text{Boundary}_x.(id|\text{Building}_y.(id|\text{Agent}_{z})$}
\label{fig:boundary-to-building}
\end{subfigure}
\caption{Alternative reaction rules that allow an agent to enter a building.}
\label{fig:alternative-rules}
\end{figure}
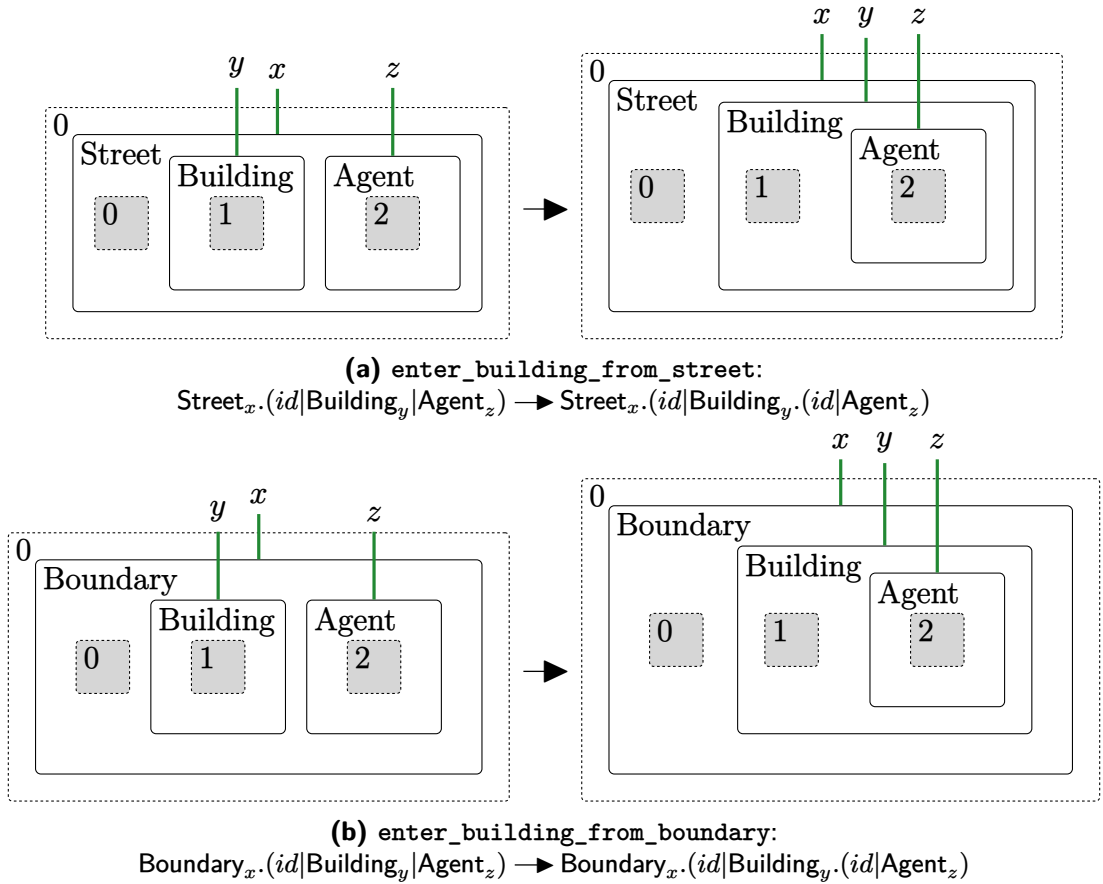

\begin{figure}[h!]
\begin{subfigure}{\linewidth}
\includegraphics[width=\textwidth]{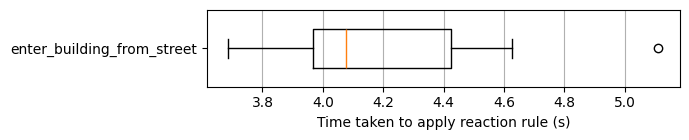}
\centering
\end{subfigure}

\begin{subfigure}{\linewidth}
\includegraphics[width=\textwidth]{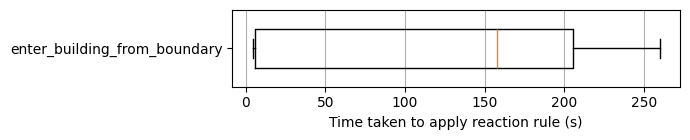}
\centering
\end{subfigure}
\caption{Box and whisker plots showing the distribution of execution times for applying the reaction rules \texttt{enter\_building\_from\_street} and \texttt{enter\_building\_from\_boundary}. Note the different x-axis scales.}
\label{fig:alternative-rules-boxplot}
\end{figure}

The experimental results show that \texttt{enter\_building} takes extraordinarily long to execute—almost a thousand times slower than the other reaction rules. Figure \ref{fig:alternative-rules} shows two alternative reaction rules that have the same effect as \texttt{enter\_building}. They are tested with the same experiment setup as before; the results, illustrated in Figure \ref{fig:alternative-rules-boxplot}, show that they execute much more quickly. This calls for the reaction rule \texttt{enter\_building} to be replaced by \texttt{enter\_building\_from\_street} and \texttt{enter\_building\_from\_boundary}. Even so, the BRS for motion does not yet support real-time transformation of bigraphs: while most rules execute in 4-10s, the median time taken to apply \texttt{enter\_building\_from\_boundary} is 150s.  The stack trace (Figure \ref{fig:sat-solver}) reveals that the SAT solving for matching an instance of the redex in the bigraph dominates the execution time. A deeper investigation into how BigraphER encodes the subgraph matching process as a constraint satisfaction problem is required to diagnose the underlying issue. Additionally, the Glasgow Subgraph Solver \cite{DBLP:conf/gg/McCreeshP020} shows promise in speeding up the matching process—a binding to BigraphER is in the works.

\chapter{Conclusion}

\textit{This chapter concludes the dissertation with a summary of its achievements, and a discussion of lessons learnt and future work.}

\section{Summary of achievements}
This project was a success: not only were all the requirements met, a range of extensions were completed.

It proposed a specification for organising all named buildings, streets and administrative boundaries in the world into a \textbf{hierarchical space-partitioning tree} and provides a methodology for building it using the comprehensive data available on OSM. 

The hierarchical tree was represented in a \textbf{bigraph of the real world} which models all physical locations and includes complete street connectivity. It captures not only containment and adjacency of physical locations and objects, but also non-spatial relationships through linking. Reaction rules were defined to transform bigraphs of the real world, enabling the modelling of \textbf{motion and communication of agents}. This dissertation also demonstrated that bigraphs of the real world can \textbf{unify all bigraph models} from past and future work, giving insight to the spatial relationships between entities of different models.

The methodology for building the hierarchy space-partitioning tree of the world also yields a specification and implementation for the \textbf{assignment of unique spatial names to all physical locations and objects}, effectively creating a geographic addressing system with global coverage. Unicast and multicast communications via spatial names were modelled in the bigraph of the real world.

This dissertation delivered a tool that can generate bigraphs representing regions from \textbf{any part of the world} (See Table \ref{table:bigraph-stats} for details of bigraphs built for 30 cities from around the world). Experiment results give evidence that the tool is both efficient and practical for building and transforming bigraphs of the real world, which are of unprecedented scale. This is made possible by the \textbf{algorithmic improvements} contributed to state-of-the-art bigraph-building tools, which resulted in up to a \textbf{97× speedup} and a \textbf{16× reduction in memory usage}.

\section{Lessons learnt}
This project took me on a joyride: I delved deep into the universal formalism of bigraphs to model dynamic systems, navigated the complexities of OSM's vast geographical database, coded an implementation of my theoretical work in OCaml, and applied my understanding of data structures and algorithms to contribute performance improvements to open-source tools. My ideas and contributions connected me with computer scientists  researching bigraphs outside Cambridge. I also learnt that rapid prototyping can offer a more sustainable approach—particularly in long-term projects that involve data exploration and development of new theoretical frameworks.

\section{Future work}
This dissertation opens the doors to experiments in unifying bigraph models of different behaviours using the foundation of a bigraph of the real world. Ongoing work by Achibald et al. \cite{Archibald2024} to unify digital twins of the various transport systems at the Port of Dover will demonstrate a real-world application of the unifying capability of bigraphs of the real world. 

Milner \cite{bigraph-sem-notes} opined that BRS modularity is a critical research direction; it could make real-time transformations to bigraphs of the real world tractable as the BRS would not need to include the whole world. Plato-graphical bigraph models \cite{plato} point to a potential solution: they model context-aware systems by splitting the BRS into a context, proxy and agent. To query or alter the context, agents must use the proxy as a sensor and actuator. Future work could experiment with using a local component of the bigraph as a proxy to the context of the whole world.

\label{lastcontentpage} %

\newpage
\addcontentsline{toc}{chapter}{Bibliography}
\bibliographystyle{plainurl} %
\bibliography{refs} %

\begin{thebibliography}{10}

\bibitem{DBLP:journals/fac/AlbalweAS24}
Maram Albalwe, Blair Archibald, and Michele Sevegnani.
\newblock Modelling and analysing routing protocols diagrammatically with
  bigraphs.
\newblock {\em Formal Aspects Comput.}, 36(3):17:1--17:25, 2024.
\newblock \href {https://doi.org/10.1145/3685934} {\path{doi:10.1145/3685934}}.

\bibitem{archibald2024practicalmodellingbigraphs}
Blair Archibald, Muffy Calder, and Michele Sevegnani.
\newblock Practical modelling with bigraphs, 2024.
\newblock URL: \url{https://arxiv.org/abs/2405.20745}, \href
  {https://arxiv.org/abs/2405.20745} {\path{arXiv:2405.20745}}.

\bibitem{Archibald2024}
Blair Archibald, Paul Harvey, and Michele Sevegnani.
\newblock A digital twinning approach to decarbonisation: Research challenges.
\newblock In {\em 1st International Workshop on Low Carbon Computing}, December
  2024.

\bibitem{barabasi2016network}
Albert-László Barabási and Márton Pósfai.
\newblock {\em Network science}.
\newblock Cambridge University Press, Cambridge, 2016.
\newblock URL: \url{http://barabasi.com/networksciencebook/}.

\bibitem{10.1371/journal.pone.0180698}
Christopher Barrington-Leigh and Adam Millard-Ball.
\newblock The world’s user-generated road map is more than 80\% complete.
\newblock {\em PLOS ONE}, 12(8):1--20, 08 2017.
\newblock \href {https://doi.org/10.1371/journal.pone.0180698}
  {\path{doi:10.1371/journal.pone.0180698}}.

\bibitem{plato}
Lars Birkedal, Søren Debois, Ebbe Elsborg, Thomas Hildebrandt, and Henning
  Niss.
\newblock Bigraphical models of context-aware systems.
\newblock pages 187--201, 03 2006.
\newblock \href {https://doi.org/10.1007/11690634_13}
  {\path{doi:10.1007/11690634_13}}.

\bibitem{wireless-home-networks}
Muffy Calder, Alexandros Koliousis, Michele Sevegnani, and Joseph Sventek.
\newblock Real-time verification of wireless home networks using bigraphs with
  sharing.
\newblock {\em Science of Computer Programming}, 80:288–310, 02 2014.
\newblock \href {https://doi.org/10.1016/j.scico.2013.08.004}
  {\path{doi:10.1016/j.scico.2013.08.004}}.

\bibitem{networking}
Muffy Calder and Michele Sevegnani.
\newblock Modelling {IEEE} 802.11 {CSMA}/{CA} {RTS}/{CTS} with stochastic
  bigraphs with sharing.
\newblock {\em Formal Aspects of Computing}, 26(3):537--561, 2014.
\newblock URL: \url{http://dx.doi.org/10.1007/s00165-012-0270-3}, \href
  {https://doi.org/10.1007/s00165-012-0270-3}
  {\path{doi:10.1007/s00165-012-0270-3}}.

\bibitem{osm-xml}
Patrick Ferris.
\newblock osm\_xml, 2024.
\newblock [Online; accessed 15-April-2025].
\newblock URL: \url{https://github.com/geocaml/osm_xml}.

\bibitem{10.1145/3626111.3628210}
Ryan Gibb, Anil Madhavapeddy, and Jon Crowcroft.
\newblock Where on earth is the spatial name system?
\newblock In {\em Proceedings of the 22nd ACM Workshop on Hot Topics in
  Networks}, HotNets '23, page 79–86, New York, NY, USA, 2023. Association
  for Computing Machinery.
\newblock \href {https://doi.org/10.1145/3626111.3628210}
  {\path{doi:10.1145/3626111.3628210}}.

\bibitem{flamegraph}
Brendan Gregg.
\newblock Flame graph, 2025.
\newblock [Online; accessed 15-April-2025].
\newblock URL: \url{https://github.com/brendangregg/FlameGraph}.

\bibitem{building-coverage}
Benjamin Herfort, Sven Lautenbach, Joao De~Albuquerque, Jennings Anderson, and
  Alexander Zipf.
\newblock A spatio-temporal analysis investigating completeness and
  inequalities of global urban building data in openstreetmap.
\newblock {\em Nature Communications}, 14, 07 2023.
\newblock \href {https://doi.org/10.1038/s41467-023-39698-6}
  {\path{doi:10.1038/s41467-023-39698-6}}.

\bibitem{transitive-closure}
Yannis Ioannidis and Raghu Ramakrishnan.
\newblock Efficient transitive closure algorithms.
\newblock pages 382--394, 01 1988.

\bibitem{KRIVINE200873}
Jean Krivine, Robin Milner, and Angelo Troina.
\newblock Stochastic bigraphs.
\newblock {\em Electronic Notes in Theoretical Computer Science}, 218:73--96,
  2008.
\newblock Proceedings of the 24th Conference on the Mathematical Foundations of
  Programming Semantics (MFPS XXIV).
\newblock URL:
  \url{https://www.sciencedirect.com/science/article/pii/S1571066108004003},
  \href {https://doi.org/10.1016/j.entcs.2008.10.006}
  {\path{doi:10.1016/j.entcs.2008.10.006}}.

\bibitem{memtrace}
Jane Street~Group LLC.
\newblock Memtrace, 2024.
\newblock [Online; accessed 15-April-2025].
\newblock URL: \url{https://github.com/janestreet/memtrace}.

\bibitem{DBLP:conf/gg/McCreeshP020}
Ciaran McCreesh, Patrick Prosser, and James Trimble.
\newblock The glasgow subgraph solver: Using constraint programming to tackle
  hard subgraph isomorphism problem variants.
\newblock In Fabio Gadducci and Timo Kehrer, editors, {\em Graph Transformation
  - 13th International Conference, {ICGT} 2020, Held as Part of {STAF} 2020,
  Bergen, Norway, June 25-26, 2020, Proceedings}, volume 12150 of {\em Lecture
  Notes in Computer Science}, pages 316--324. Springer, 2020.
\newblock \href {https://doi.org/10.1007/978-3-030-51372-6\_19}
  {\path{doi:10.1007/978-3-030-51372-6\_19}}.

\bibitem{bigraph-notes}
Robin Milner.
\newblock Lecture notes on bigraphs: a model for mobile agents, 2008.
\newblock [Online; accessed 15-April-2025].
\newblock URL: \url{https://www.cl.cam.ac.uk/archive/rm135/Bigraphs-Notes.pdf}.

\bibitem{bigraph-sem-notes}
Robin Milner.
\newblock Seminar notes on developments in bigraphs, 2009.
\newblock [Online; accessed 15-April-2025].
\newblock URL:
  \url{https://www.cl.cam.ac.uk/archive/rm135/Bigraphs-Seminars.pdf}.

\bibitem{10.5555/1540607}
Robin Milner.
\newblock {\em The Space and Motion of Communicating Agents}.
\newblock Cambridge University Press, USA, 1st edition, 2009.

\bibitem{PENN200672}
Gerald Penn.
\newblock Efficient transitive closure of sparse matrices over closed
  semirings.
\newblock {\em Theoretical Computer Science}, 354(1):72--81, 2006.
\newblock Algebraic Methods in Language Processing.
\newblock URL:
  \url{https://www.sciencedirect.com/science/article/pii/S0304397505008546},
  \href {https://doi.org/10.1016/j.tcs.2005.11.008}
  {\path{doi:10.1016/j.tcs.2005.11.008}}.

\bibitem{roy1959transitivite}
Bernard Roy et~al.
\newblock Transitivit{\'e} et connexit{\'e}.
\newblock {\em CR Acad. Sci. Paris}, 249(216-218):182, 1959.

\bibitem{SEVEGNANI201543}
Michele Sevegnani and Muffy Calder.
\newblock Bigraphs with sharing.
\newblock {\em Theoretical Computer Science}, 577:43--73, 2015.
\newblock URL:
  \url{https://www.sciencedirect.com/science/article/pii/S0304397515001085},
  \href {https://doi.org/10.1016/j.tcs.2015.02.011}
  {\path{doi:10.1016/j.tcs.2015.02.011}}.

\bibitem{DBLP:conf/cav/SevegnaniC16}
Michele Sevegnani and Muffy Calder.
\newblock Bigraph{ER}: Rewriting and analysis engine for bigraphs.
\newblock In {\em Computer Aided Verification - 28th International Conference,
  {CAV} 2016, Toronto, ON, Canada, July 17-23, 2016, Proceedings, Part {II}},
  pages 494--501, 2016.
\newblock \href {https://doi.org/10.1007/978-3-319-41540-6\_27}
  {\path{doi:10.1007/978-3-319-41540-6\_27}}.

\bibitem{8595061}
Michele Sevegnani, Milan Kabac, Muffy Calder, and Julie McCann.
\newblock Modelling and verification of large-scale sensor network
  infrastructures.
\newblock In {\em 2018 23rd International Conference on Engineering of Complex
  Computer Systems (ICECCS)}, pages 71--81, 2018.
\newblock \href {https://doi.org/10.1109/ICECCS2018.2018.00016}
  {\path{doi:10.1109/ICECCS2018.2018.00016}}.

\bibitem{ocamlformat}
Josh Berdine Hugo Heuzard Guillaume~Petiot Tarides, Facebook and Jules
  Aguillon.
\newblock ocamlformat, 2025.
\newblock [Online; accessed 15-April-2025].
\newblock URL:
  \url{https://github.com/ocaml-ppx/ocamlformat?tab=readme-ov-file}.

\bibitem{10.1007/978-3-642-03832-7_22}
Lisa Walton and Michael Worboys.
\newblock An algebraic approach to image schemas for geographic space.
\newblock In Kathleen~Stewart Hornsby, Christophe Claramunt, Michel Denis, and
  G{\'e}rard Ligozat, editors, {\em Spatial Information Theory}, pages
  357--370, Berlin, Heidelberg, 2009. Springer Berlin Heidelberg.

\bibitem{10.1007/978-3-642-33024-7_17}
Lisa~A. Walton and Michael Worboys.
\newblock A qualitative bigraph model for indoor space.
\newblock In Ningchuan Xiao, Mei-Po Kwan, Michael~F. Goodchild, and Shashi
  Shekhar, editors, {\em Geographic Information Science}, pages 226--240,
  Berlin, Heidelberg, 2012. Springer Berlin Heidelberg.

\bibitem{10.1145/321105.321107}
Stephen Warshall.
\newblock A theorem on boolean matrices.
\newblock {\em J. ACM}, 9(1):11–12, January 1962.
\newblock \href {https://doi.org/10.1145/321105.321107}
  {\path{doi:10.1145/321105.321107}}.

\end{thebibliography}

\appendix

\chapter{Additional bigraph formalism}

\section{Reactive system of agents manipulating their environment} \label{section:brs-lock}
\S\ref{section:brs} defined a BRS modelling the motion of agents. Reaction rules can also be defined to model agents manipulating their environment, such as locking a room. Walton and Worboys \cite{10.1007/978-3-642-03832-7_22} proposed 3 possibilities of modelling a locked room: 
\begin{enumerate}
\item the Room node is contained in a Lock node that acts as a barrier
\item the Room node is linked to a Lock node
\item the Room node contains a Lock node
\end{enumerate}
The first method is incorrect: it does not prevent the application of the reaction rule for leaving a room: the agent will exit the room and be contained in the Lock node. The second method is plausible, but requires changing the arity of the Room node and to add a link. The third method is demonstrated here. Let Lock, OpenLock and Key be control of arity 1. A locked room is represented by a Room node containing a Lock node; an unlocked room is represented by a Room node with a OpenLock node. The reaction rule \rr{lock\_room} (Figure \ref{fig:react-lock-room}) allows an agent holding a key to a room to lock it from outside. It requires the agent and the room to be in the same place (presumably a building). The outer name $x$ allows the Room node to be linked to Key nodes carried by other Agent nodes, allowing multiple agents to have the authority to lock/unlock a room.

\begin{figure}[h]
\centering
\begin{tikzpicture}
	\node[big site, inner sep=10pt,] (s0) {};
	\node[below right, inner sep=0pt, shift={(0.1,-0.1)}, ] at (s0.north west) {0};
	\node[above=0.2 of s0,draw, rounded corners=2] (l0) {OpenLock};
	\node[fit=(s0)(l0), draw, rounded corners=2,inner sep=15pt] (ro0) {};
	\node[below right, inner sep=0pt, shift={(0.1,-0.1)}] at (ro0.north west) {Room};
	\node[right=1.4 of l0,draw, rounded corners=2] (k0) {Key};
	\node[below=0.2 of k0, big site, inner sep=10pt,] (s1) {};
	\node[below right, inner sep=0pt, shift={(0.1,-0.1)}, ] at (s1.north west) {1};
	\node[fit=(k0)(s1), draw, rounded corners=2,inner sep=15pt,] (a0) {};
	\node[below right, inner sep=0pt, shift={(0.1,-0.1)}, ] at (a0.north west) {Agent};
	\node[big region, fit=(ro0)(a0), inner sep=15pt,] (r0) {};
	\node[below right, inner sep=0pt, shift={(0.1,-0.1)}, ] at (r0.north west) {0};

	\node[above=0.25] at (ro0|-,|-r0.north) (x) {$x$};
	\draw[big edge] (ro0)  to[out=90,in=270]  (x);
	\node[above=2.05] at ($(ro0.east)!0.5!(a0.west)$) (y) {$y$};
	\draw[big edge] (l0)  to[out=0,in=270]  (y);
	\draw[big edge] (k0)  to[out=180,in=270]  (y);
	\node[above=0.25] at (a0|-,|-r0.north) (z) {$z$};
	\draw[big edge] (a0)  to[out=90,in=270]  (z);
\end{tikzpicture}
\raisebox{1.5cm}{\large$\rrul$}
\begin{tikzpicture}
	\node[big site, inner sep=10pt,] (s0) {};
	\node[below right, inner sep=0pt, shift={(0.1,-0.1)}, ] at (s0.north west) {0};
	\node[above=0.2 of s0,draw, rounded corners=2] (l0) {Lock};
	\node[fit=(s0)(l0), draw, rounded corners=2,inner sep=15pt] (ro0) {};
	\node[below right, inner sep=0pt, shift={(0.1,-0.1)}] at (ro0.north west) {Room};
	\node[right=1.4 of l0,draw, rounded corners=2] (k0) {Key};
	\node[below=0.2 of k0, big site, inner sep=10pt,] (s1) {};
	\node[below right, inner sep=0pt, shift={(0.1,-0.1)}, ] at (s1.north west) {1};
	\node[fit=(k0)(s1), draw, rounded corners=2,inner sep=15pt,] (a0) {};
	\node[below right, inner sep=0pt, shift={(0.1,-0.1)}, ] at (a0.north west) {Agent};
	\node[big region, fit=(ro0)(a0), inner sep=15pt,] (r0) {};
	\node[below right, inner sep=0pt, shift={(0.1,-0.1)}, ] at (r0.north west) {0};

	\node[above=0.25] at (ro0|-,|-r0.north) (x) {$x$};
	\draw[big edge] (ro0)  to[out=90,in=270]  (x);
	\node[above=2.05] at ($(ro0.east)!0.5!(a0.west)$) (y) {$y$};
	\draw[big edge] (l0)  to[out=0,in=270]  (y);
	\draw[big edge] (k0)  to[out=180,in=270]  (y);
	\node[above=0.25] at (a0|-,|-r0.north) (z) {$z$};
	\draw[big edge] (a0)  to[out=90,in=270]  (z);
\end{tikzpicture}
\captionsetup{justification=centering}
\caption{Reaction rule \rr{lock\_room}: $\text{Room}_x.(\text{OpenLock}_y.1|id)|\text{Agent}_{z}.(\text{Key}_y.1|id)\rrul\text{Room}_x.(\text{Lock}_y.1|id)|\text{Agent}_{z}.(\text{Key}_y.1|id$}
\label{fig:react-lock-room}
\end{figure}

\begin{figure}[h]
\centering
\begin{subfigure}{\linewidth}
\centering
\begin{tikzpicture}
 	\node[big site, inner sep=10pt,] (s0) {};
	\node[below right, inner sep=0pt, shift={(0.1,-0.1)}, ] at (s0.north west) {0};
 	\node[right=0.9 of s0, big site, inner sep=10pt,] (s1) {};
	\node[below right, inner sep=0pt, shift={(0.1,-0.1)}, ] at (s1.north west) {1};
	\node[fit=(s1), draw, rounded corners=2,inner sep=15pt,] (a0) {};
	\node[below right, inner sep=0pt, shift={(0.1,-0.1)}, ] at (a0.north west) {Agent};
	\node[fit=(s0)(a0), draw, rounded corners=2,inner sep=15pt,] (room) {};
	\node[below right, inner sep=0pt, shift={(0.1,-0.1)}, ] at (room.north west) {Room};
	\node[big region, fit=(room), inner sep=15pt,] (r0) {};
	\node[below right, inner sep=0pt, shift={(0.1,-0.1)}, ] at (r0.north west) {0};

	\node[above=0.25 of r0, shift={(0,0)}] (x) {$x$};
	\draw[big edge] (room)  to[out=90,in=270]  (x);
	\node[above=0.2] at (a0|-,|-r0.north) (z) {$z$};
	\draw[big edge] (a0)  to[out=90,in=270]  (z);
\end{tikzpicture}
\raisebox{2cm}{\large$\rrul$}
\begin{tikzpicture}
 	\node[big site, inner sep=10pt,] (s0) {};
	\node[below right, inner sep=0pt, shift={(0.1,-0.1)}, ] at (s0.north west) {0};
	\node[fit=(s0), draw, rounded corners=2,inner sep=15pt,] (room) {};
	\node[below right, inner sep=0pt, shift={(0.1,-0.1)}, ] at (room.north west) {Room};
 	\node[right=0.9 of room, big site, inner sep=10pt,] (s1) {};
	\node[below right, inner sep=0pt, shift={(0.1,-0.1)}, ] at (s1.north west) {1};
	\node[fit=(s1), draw, rounded corners=2,inner sep=15pt,] (a0) {};
	\node[below right, inner sep=0pt, shift={(0.1,-0.1)}, ] at (a0.north west) {Agent};
	\node[big region, fit=(room)(a0), inner sep=15pt,] (r0) {};
	\node[below right, inner sep=0pt, shift={(0.1,-0.1)}, ] at (r0.north west) {0};

	\node[above=0.25] at (room|-,|-r0.north) (x) {$x$};
	\draw[big edge] (room)  to[out=90,in=270]  (x);
	\node[above=0.2] at (a0|-,|-r0.north) (z) {$z$};
	\draw[big edge] (a0)  to[out=90,in=270]  (z);
\end{tikzpicture}
\end{subfigure}

\begin{subfigure}{\linewidth}
\centering
{if }
\raisebox{-.3\height}{
\begin{tikzpicture}
	\node[draw, rounded corners=2] (l0) {Lock};
	\node[big region, fit=(l0), inner sep=15pt,] (r0) {};
	\node[below right, inner sep=0pt, shift={(0.1,-0.1)}, ] at (r0.north west) {0};
	\node[above=0.2] at (l0|-,|-r0.north) (y) {$y$};
	\draw[big edge] (l0)  to[out=90,in=270]  (y);
\end{tikzpicture}
}
{ does not occur in the parameter}
\end{subfigure}
\captionsetup{justification=centering}
\caption{Modified reaction rule \rr{leave\_room}: $\text{Room}_x.(id|\text{Agent}_{z})\rrul\ \text{Room}_x|\text{Agent}_{z}$\\if Lock$_y.1$ does not occur in the parameter}
\label{fig:react-leave-room-modified}
\end{figure}
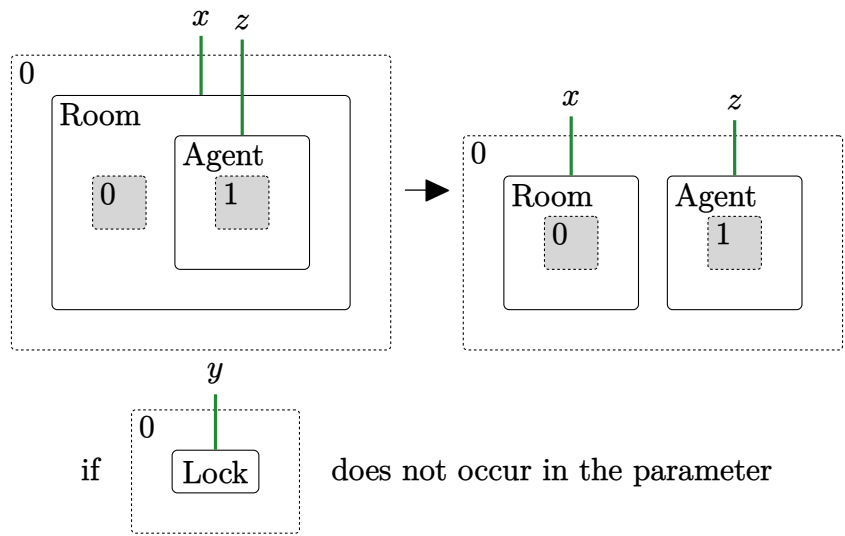

The reaction rule \texttt{leave\_room} (Figure \ref{fig:react-leave-room}) must be modified as illustrated in Figure \ref{fig:react-leave-room-modified} so that it does not apply for locked rooms. The reaction rules \texttt{leave\_room}, \texttt{lock\_room} and their inverse rules can be used in a BRS that models agents moving in and out of rooms, and locking and unlocking them.

\section{Bigraphs with sharing} \label{section:sharing}

Bigraphs with sharing \cite{SEVEGNANI201543} extend Milner's theory of bigraphs by relaxing the restriction on place graphs from a forest to a directed acyclic graph, allowing a node/site to have multiple parents. This enables the placing of bigraphs to represent locations that can intersect and overlap, which is a natural model for domains such as wireless signalling (Figure \ref{fig:sharing-signals}) and audio communication. The operations on bigraphs introduced in \S \ref{bigraph-operations} still apply as described.

\begin{figure}[h!]
\centering
\begin{subfigure}{\linewidth}
\begin{tikzpicture}
	\node[draw, rounded corners=2] (Machine0) {Machine(0)};
	\node[draw,ellipse,minimum width=275pt,minimum height=200pt] at (Machine0) (Signal0){};
	\node[below=0.2, shift={(-1.4,0)}, rotate=8,inner sep=0pt] at (Signal0.north) {Signal(0)};
	\node[above left=0.4 of Signal0.east, draw, rounded corners=2] (Machine1) {Machine(1)};
	\node[draw,ellipse,minimum width=275pt,minimum height=200pt] at (Machine1) (Signal1){};
	\node[below=0.1, shift={(0,0)},inner sep=0pt] at (Signal1.north) {Signal(1)};
	\node[below left=0.4 of Signal1.east, draw, rounded corners=2] (Machine2) {Machine(2)};
	\node[draw,ellipse,minimum width=275pt,minimum height=200pt] at (Machine2) (Signal2){};
	\node[below=0.2, shift={(1.4,0)}, rotate=-8,inner sep=0pt] at (Signal2.north) {Signal(2)};
	\node[big region,fit=(Signal0)(Signal1)(Signal2), inner sep=8pt] (r0) {};
	\node[below right, inner sep=0pt, shift={(0.1,-0.1)}, ] at (r0.north west) {0};
\end{tikzpicture}
\centering
\caption{Bigraph}
\label{fig:sharing-signals-bigraph}
\end{subfigure}

\begin{subfigure}{\linewidth}
\begin{forest}
for tree={edge = {-latex}}
[0,big region
	[Signal(0), name=s0
		[Machine(0), name=m0]
	]
	[Signal(1), name=s1
		[Machine(1), name=m1]
	]
	[Signal(2), name=s2
		[Machine(2), name=m2]
	]
]
\draw[-latex] (s0) -- (m1);
\draw[-latex] (s2) -- (m1);
\draw[-latex] (s1) -- (m0);
\draw[-latex] (s1) -- (m2);
\end{forest}
\centering
\caption{Place graph}
\label{fig:sharing-signals-place-graph}
\end{subfigure}
\caption{Wireless network.}
\label{fig:sharing-signals}
\end{figure}

\section{Effect of improvements to BigraphER for bigraphs with sharing} \label{section:fixes-sharing}
\S\ref{section:sparse-representation} proposes changing the representation of sparse matrices to an adjacency list. Since the place graph of a bigraph with sharing is a DAG that can have up to $n(n-1)/2$ edges, the adjacency list representation uses $\mathcal{O}(n^3)$ space. However these bigraphs are used to model real networks, most of which are sparse \cite{barabasi2016network}. Therefore, the number of edges is more likely to be $\mathcal{O}(n)$, so the asymptotic analysis conducted in \S\ref{section:sparse-representation} hold true for bigraphs with sharing.

\S\ref{section:trans-implementation} proposes using a $n\times DFS$ algorithm for calculating transitive closure of a sparse matrix. The algorithm is also correct for bigraphs with sharing as their place graphs are DAGs. Re-looking at the asymptotic analysis for bigraphs with sharing, the naive method still takes $\mathcal{O}(n^4)$ time, but the $n \times$DFS algorithm now takes $\mathcal{O}(n(n+m))=\mathcal{O}(n^3)$ time. Again, since these bigraphs are usually used to model sparse networks, the number of edges is more likely to be $\mathcal{O}(n)$, so the asymptotic analysis conducted in \S\ref{section:trans-implementation}  hold true for bigraphs with sharing.

\chapter{OpenStreetMap data visualisations} \label{appendix-osm}

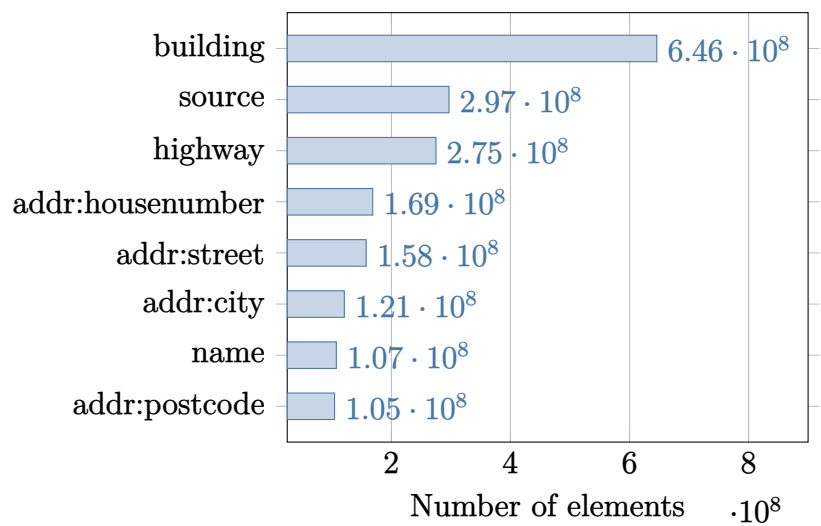
\begin{figure}[h]
\centering
\begin{tikzpicture}
\begin{axis}[
    xbar,
    xlabel={Number of elements},
    symbolic y coords={addr:postcode,name,addr:city,addr:street,addr:housenumber,highway, source, building},
    ytick=data,
    nodes near coords={\pgfmathprintnumber{\pgfplotspointmeta}},
    xmajorgrids=true,
    xmax=900000000,
    tick label style={/pgf/number format/fixed}
]
\addplot coordinates {
    (104529579,addr:postcode)
    (107484726,name)
    (120893860,addr:city)
    (157591473,addr:street)
    (168654989,addr:housenumber)
    (274868533,highway)
    (296729285,source)
    (645811488,building)
};
\end{axis}
\end{tikzpicture}
\captionsetup{belowskip=-15pt}
\caption[Caption for LOF]{Top 8 most common keys used in OSM\footnotemark.}
\label{fig:common-keys}
\end{figure}
\footnotetext{data from \url{https://taginfo.openstreetmap.org/keys}}
\begin{figure}[h!]
\includegraphics[width=0.4\linewidth]{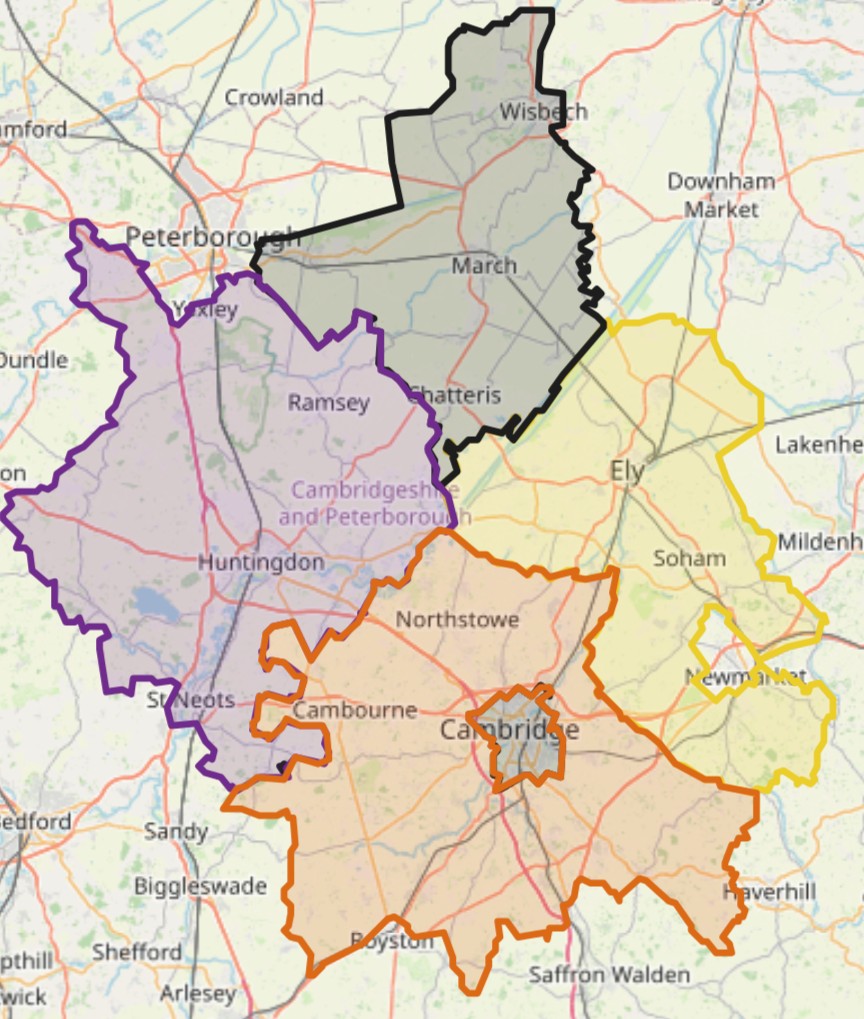}
\centering
\captionsetup{width=\linewidth}
\caption[Caption for LOF]{The 5 immediate children boundaries of Cambridgeshire, namely South Cambrigdeshire (orange), Cambridge (surrounded by South Cambridgeshire), Huntingdonshire (purple), Fenland (black) and East Cambridgeshire (yellow)\footnotemark.}
\label{fig:child-boundaries-cambridgeshire}
\end{figure}
\footnotetext{image taken from \url{https://osm-boundaries.com/map}}
\pagebreak

\begin{figure}[h!]
\begin{minipage}{.49\textwidth}
\includegraphics[width=\linewidth]{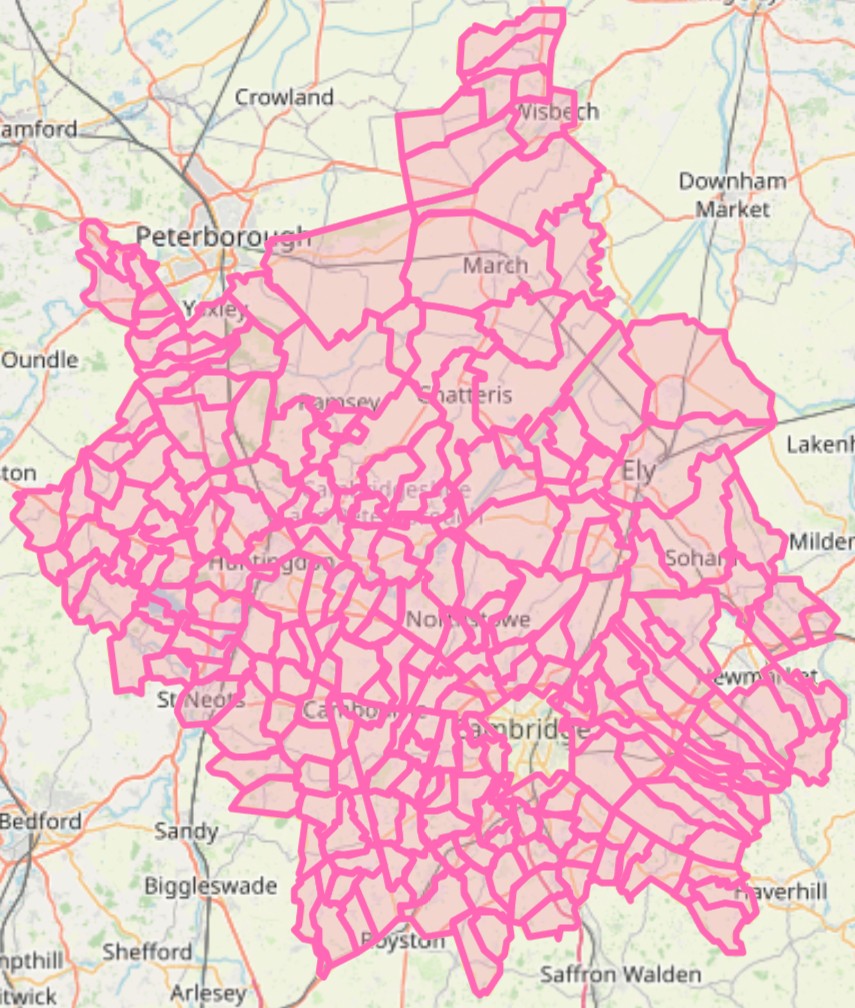}
\centering
\captionsetup{width=\linewidth}
\captionof{figure}[Caption for LOF]{The 241 descendant boundaries of Cambridgeshire\footnotemark[\value{footnote}].}
\label{fig:descendant-boundaries-cambridgeshire}
\end{minipage}
\hfill
\begin{minipage}{.49\textwidth}
\includegraphics[width=\linewidth]{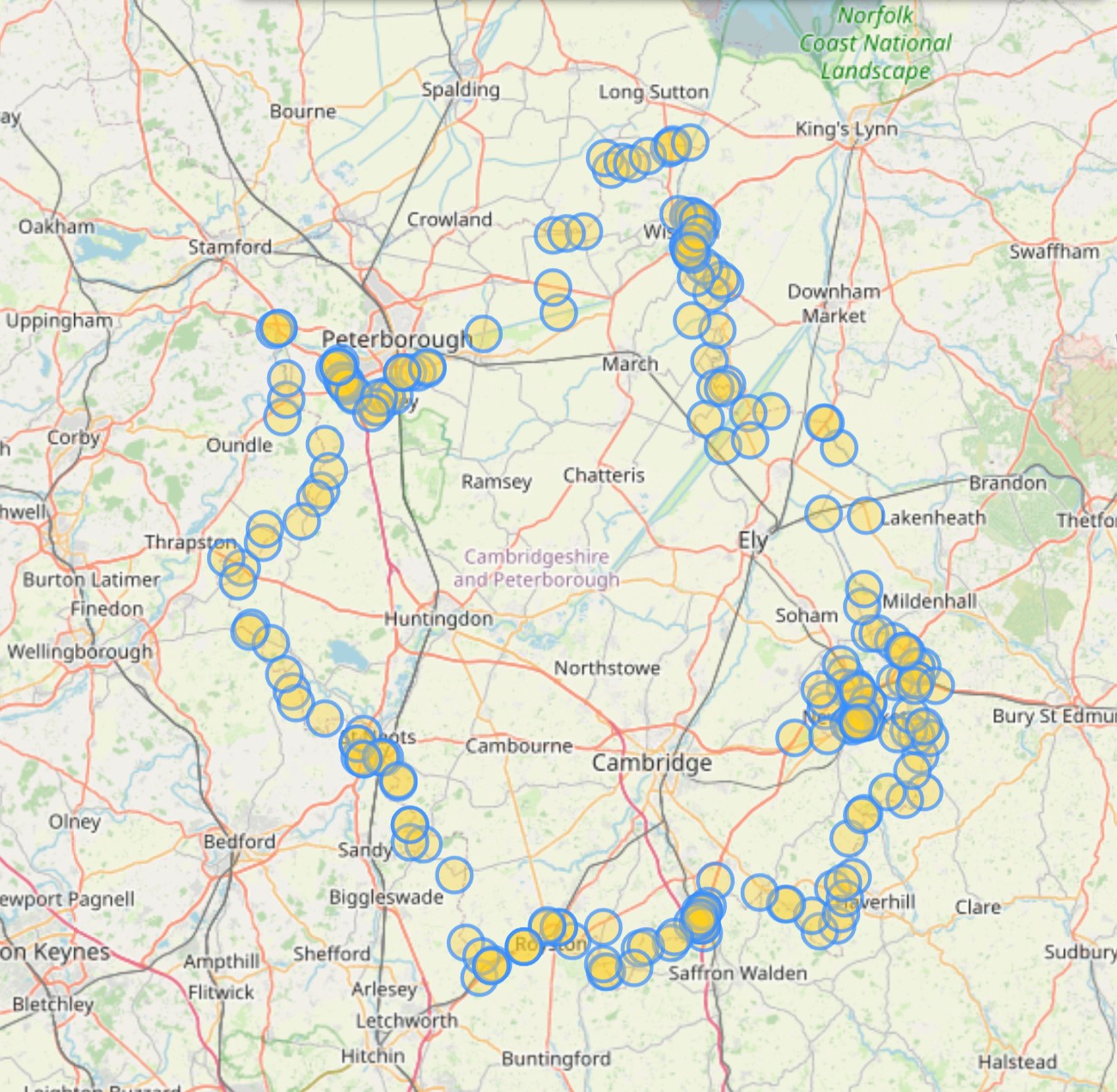}
\centering
\captionsetup{width=\linewidth}
\captionof{figure}[Caption for LOF]{The 225 intersections between streets inside and outside Cambridgeshire, labelled with circles\footnotemark.}
\label{fig:cambridgeshire-outer-names}
\end{minipage}
\end{figure}
\footnotetext{image taken from the results of running the query for outer names (\S\ref{section:query-streets}) on \url{https://overpass-turbo.eu}}

\begin{figure}[h!]
\includegraphics[width=0.8\linewidth]{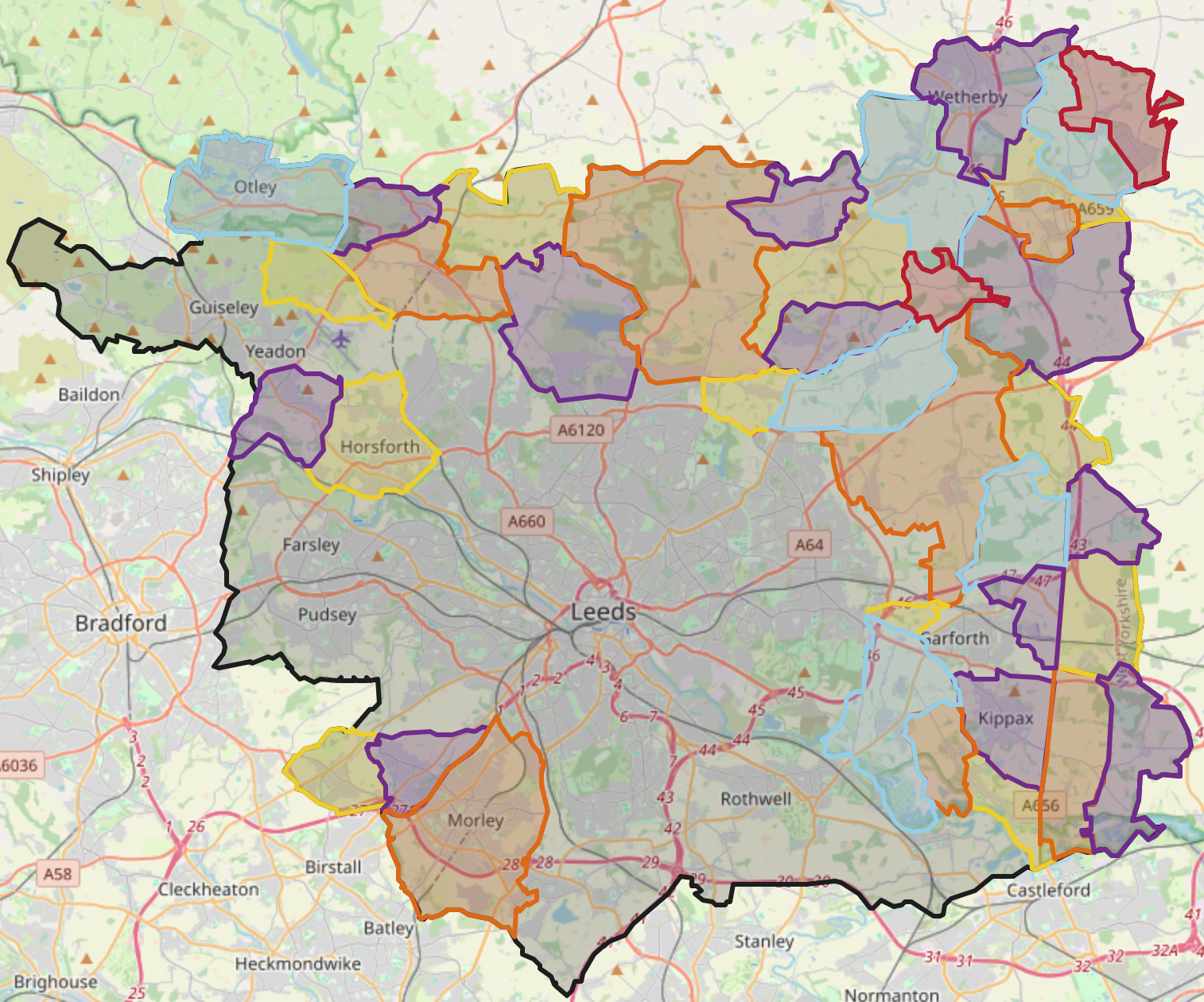}
\centering
\captionsetup{width=\linewidth}
\caption[Caption for LOF]{The descendant administrative boundaries contained in Leeds.}
\label{fig:leeds}
\end{figure}
\newpage

\begin{figure}[h!]
\includegraphics[width=\linewidth]{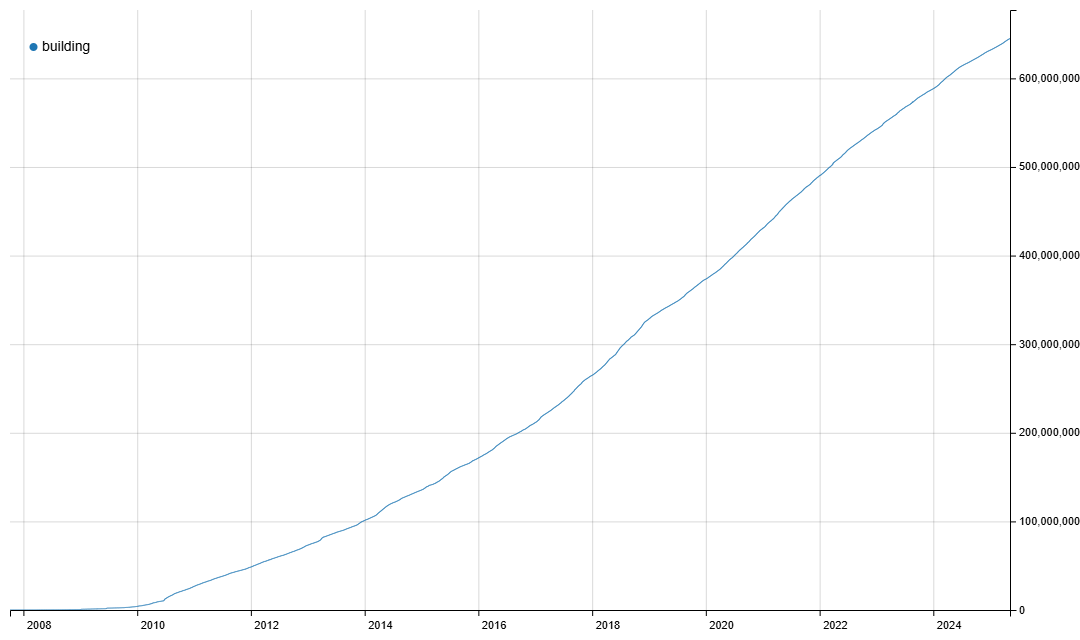}
\centering
\caption[Caption for LOF]{Chronology of the use of the \texttt{building} tag on OSM\footnotemark.}
\label{fig:building-chronology}
\end{figure}
\footnotetext{image taken from \url{https://taghistory.raifer.tech/}}

\begin{figure}[h!]
\includegraphics[width=\linewidth]{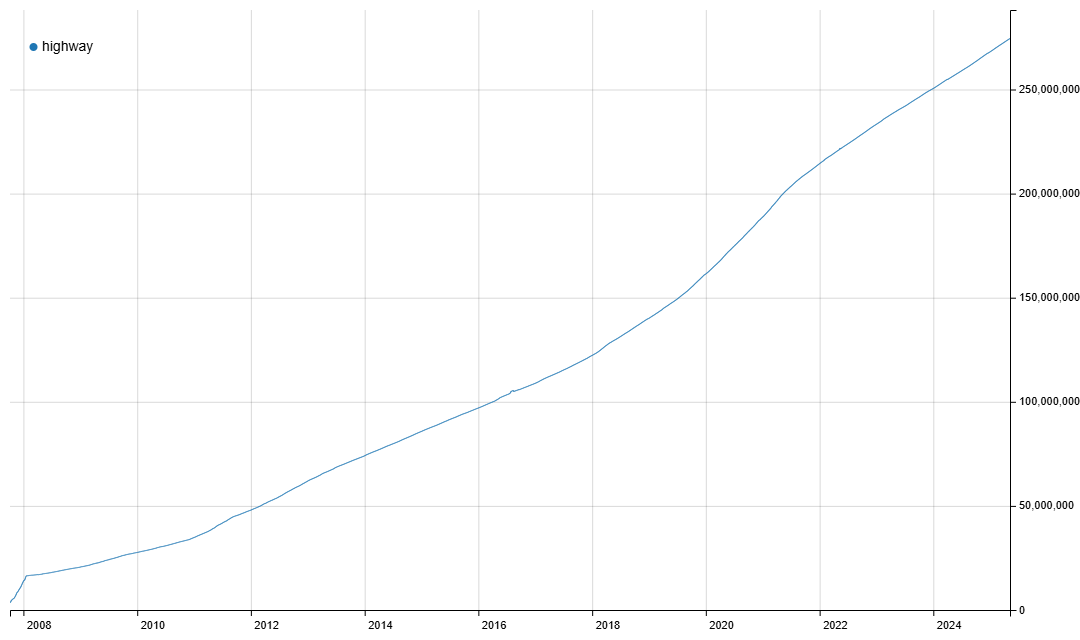}
\centering
\caption[Caption for LOF]{Chronology of the use of the \texttt{highway} tag on OSM\footnotemark[\value{footnote}].}
\label{fig:highway-chronology}
\end{figure}

\chapter{More on bigraphs of the real world}

\section{Visualisation of the combination of bigraphs} \label{section:results-of-combining}
\begin{figure}[h!]
\begin{tikzpicture}
 	\node[big site, inner sep=10pt,] (SiteM11Cam) {};
	\node[below right, inner sep=0pt, shift={(0.1,-0.1)}, ] at (SiteM11Cam.north west) {4};
	\node[right=0.3 of SiteM11Cam, shift={(0,0)},draw, rounded corners=2] (JunctionM11Cam) {Junction};
	\node[fit=(SiteM11Cam)(JunctionM11Cam), draw, rounded corners=2,inner sep=15pt,] (StreetM11Cam) {};
	\node[below right, inner sep=0pt, shift={(0.1,-0.1)}, ] at (StreetM11Cam.north west) {Street};
 	\node[left=0.3 of StreetM11Cam, big site, inner sep=10pt,] (SiteIckleton) {};
	\node[below right, inner sep=0pt, shift={(0.1,-0.1)}, ] at (SiteIckleton.north west) {3};
	\node[fit=(SiteIckleton)(StreetM11Cam), draw, rounded corners=2,inner sep=15pt] (BoundaryIckleton) {};
	\node[below right, inner sep=0pt, shift={(0.1,-0.1)}, ] at (BoundaryIckleton.north west) {Boundary};
 	\node[left=0.3 of BoundaryIckleton, big site, inner sep=10pt,] (SiteSouthCamb) {};
	\node[below right, inner sep=0pt, shift={(0.1,-0.1)}, ] at (SiteSouthCamb.north west) {2};
	\node[fit=(SiteSouthCamb)(BoundaryIckleton), draw, rounded corners=2,inner sep=15pt] (BoundarySouthCamb) {};
	\node[below right, inner sep=0pt, shift={(0.1,-0.1)}, ] at (BoundarySouthCamb.north west) {Boundary};
 	\node[left=0.3 of BoundarySouthCamb, big site, inner sep=10pt,] (SiteCamb) {};
	\node[below right, inner sep=0pt, shift={(0.1,-0.1)}, ] at (SiteCamb.north west) {1};
	\node[fit=(SiteCamb)(BoundarySouthCamb), draw, rounded corners=2,inner sep=15pt] (BoundaryCamb) {};
	\node[below right, inner sep=0pt, shift={(0.1,-0.1)}, ] at (BoundaryCamb.north west) {Boundary};
	\node[big region,fit=(BoundaryCamb),inner sep=15pt,] (r1) {};
	\node[below right, inner sep=0pt, shift={(0.1,-0.1)}, ] at (r1.north west) {1};

	\node[below left, draw, rounded corners=2,text width=2.1cm] at (-6.8,|- BoundaryCamb.north) (IDCambridgeshire) {ID( Cambridgeshire)};
	\draw[big edge] (IDCambridgeshire)  to[out=0,in=180]  (BoundaryCamb.west|-,|-IDCambridgeshire);
	\node[below=0.1 of IDCambridgeshire, draw, rounded corners=2,text width=2.1cm] (IDSouthCambridgeshire) {ID(South Cambridgeshire)};
	\draw[big edge] (IDSouthCambridgeshire)  to[out=0,in=180]  (BoundarySouthCamb.west|-,|-IDSouthCambridgeshire);
	\node[below=0.1 of IDSouthCambridgeshire,draw, rounded corners=2,text width=2.1cm] (IDIckleton) {ID(Ickleton)};
	\draw[big edge] (IDIckleton)  to[out=0,in=180]  (BoundaryIckleton);
	\node[below=0.1 of IDIckleton,draw, rounded corners=2,text width=2.1cm] (IDM11Cam) {ID(M11)};
	\draw[big edge] (IDM11Cam)  to[out=0,in=180]  (StreetM11Cam);
 	\node[below=0.3 of IDM11Cam, big site, inner sep=10pt,] (SiteID0) {};
	\node[below right, inner sep=0pt, shift={(0.1,-0.1)}, ] at (SiteID0.north west) {0};
	\node[big region,fit=(IDCambridgeshire)(SiteID0),inner sep=15pt,] (r0) {};
	\node[below right, inner sep=0pt, shift={(0.1,-0.1)}, ] at (r0.north west) {0};

	\node[below=0.7 of r0, draw, rounded corners=2,text width=2.1cm] (IDEssex) {ID(Essex)};
	\node[below=0.1 of IDEssex, draw, rounded corners=2,text width=2.1cm] (IDUttlesford) {ID( Uttlesford)};
	\node[below=0.1 of IDUttlesford,draw, rounded corners=2,text width=2.1cm] (IDGreat) {ID(Great Chesterford)};
	\node[below=0.1 of IDGreat,draw, rounded corners=2,text width=2.1cm] (IDM11Essex) {ID(M11)};
 	\node[below=0.3 of IDM11Essex, big site, inner sep=10pt,] (SiteID1) {};
	\node[below right, inner sep=0pt, shift={(0.1,-0.1)}, ] at (SiteID1.north west) {5};
	\node[big region,fit=(IDEssex)(SiteID1),inner sep=15pt,] (r2) {};
	\node[below right, inner sep=0pt, shift={(0.1,-0.1)}, ] at (r2.north west) {2};

 	\node[below=6.2 of SiteM11Cam,big site, inner sep=10pt,] (SiteM11Essex) {};
	\node[below right, inner sep=0pt, shift={(0.1,-0.1)}, ] at (SiteM11Essex.north west) {9};
	\node[right=0.3 of SiteM11Essex, shift={(0,0)},draw, rounded corners=2] (JunctionM11Essex) {Junction};
	\node[fit=(SiteM11Essex)(JunctionM11Essex), draw, rounded corners=2,inner sep=15pt,] (StreetM11Essex) {};
	\node[below right, inner sep=0pt, shift={(0.1,-0.1)}, ] at (StreetM11Essex.north west) {Street};
 	\node[left=0.3 of StreetM11Essex, big site, inner sep=10pt,] (SiteGreat) {};
	\node[below right, inner sep=0pt, shift={(0.1,-0.1)}, ] at (SiteGreat.north west) {8};
	\node[fit=(SiteGreat)(StreetM11Essex), draw, rounded corners=2,inner sep=15pt] (BoundaryGreat) {};
	\node[below right, inner sep=0pt, shift={(0.1,-0.1)}, ] at (BoundaryGreat.north west) {Boundary};
 	\node[left=0.3 of BoundaryGreat, big site, inner sep=10pt,] (SiteUttlesford) {};
	\node[below right, inner sep=0pt, shift={(0.1,-0.1)}, ] at (SiteUttlesford.north west) {7};
	\node[fit=(SiteUttlesford)(BoundaryGreat), draw, rounded corners=2,inner sep=15pt] (BoundaryUttlesford) {};
	\node[below right, inner sep=0pt, shift={(0.1,-0.1)}, ] at (BoundaryUttlesford.north west) {Boundary};
 	\node[left=0.3 of BoundaryUttlesford, big site, inner sep=10pt,] (SiteEssex) {};
	\node[below right, inner sep=0pt, shift={(0.1,-0.1)}, ] at (SiteEssex.north west) {6};
	\node[fit=(SiteEssex)(BoundaryUttlesford), draw, rounded corners=2,inner sep=15pt] (BoundaryEssex) {};
	\node[below right, inner sep=0pt, shift={(0.1,-0.1)}, ] at (BoundaryEssex.north west) {Boundary};
	\node[big region,fit=(BoundaryEssex),inner sep=15pt,] (r3) {};
	\node[below right, inner sep=0pt, shift={(0.1,-0.1)}, ] at (r3.north west) {3};

	\draw[big edge] (IDEssex)  to[out=0,in=180]  (BoundaryEssex.west|-,|-IDEssex);
	\draw[big edge] (IDUttlesford)  to[out=0,in=180]  (BoundaryUttlesford.west|-,|-IDUttlesford);
	\draw[big edge] (IDGreat)  to[out=0,in=180]  (BoundaryGreat);
	\draw[big edge] (IDM11Essex)  to[out=0,in=180]  (StreetM11Essex);

	\node[above=0.3 of r1] (node215742) {node 215742};
	\draw[big edge] (JunctionM11Cam)  to[out=90,in=270]  (node215742);
	\draw[big edge] (JunctionM11Essex)  to[out=90,in=270]  (node215742);
\end{tikzpicture}
\centering
\caption{Parallel product of the bigraphs for Cambridgeshire and Essex.}
\label{fig:cambridgeshire_essex}
\end{figure}
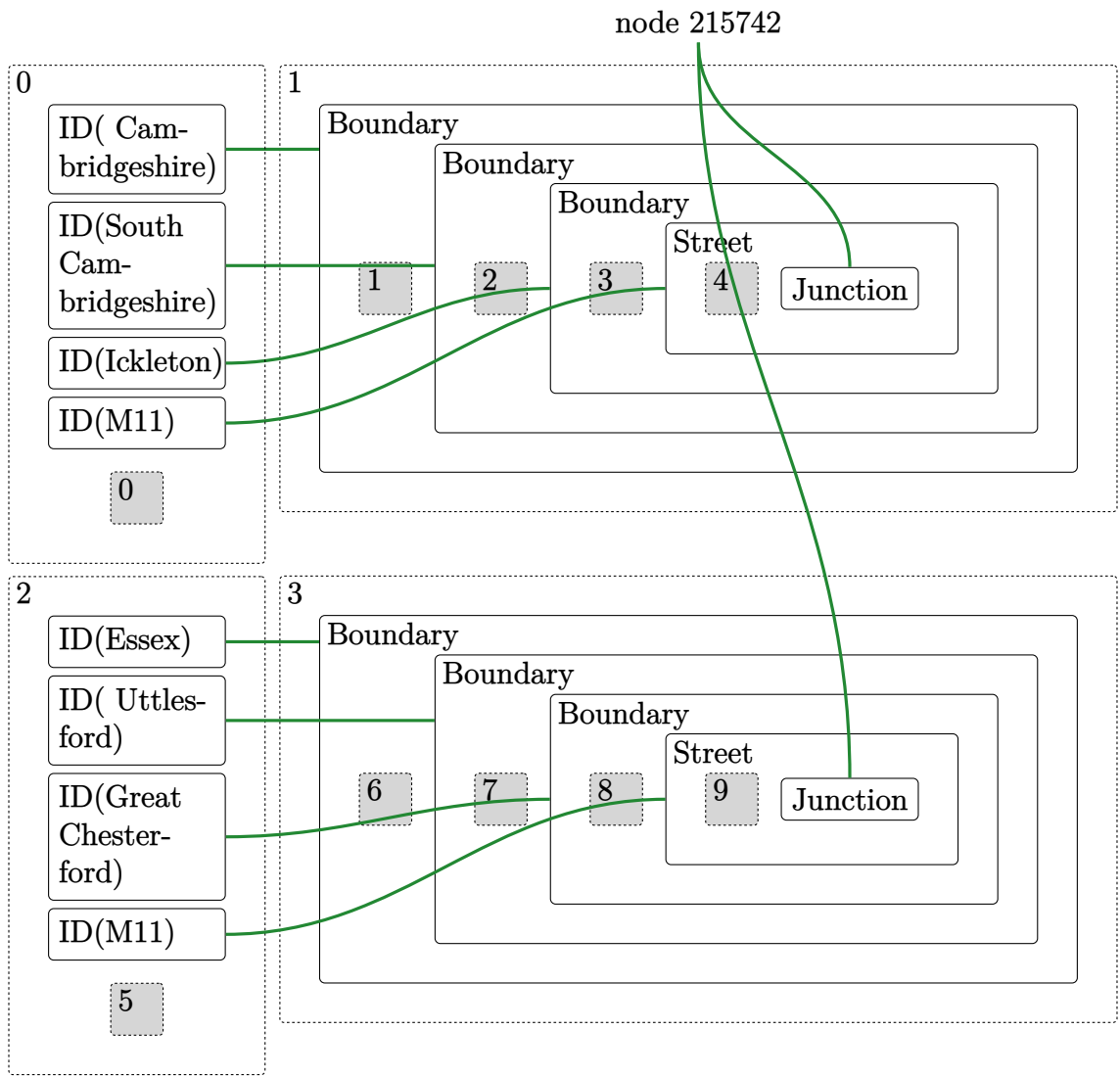

\S\ref{composition-world} explained the methodology for combining bigraphs of different regions. Figure \ref{fig:cambridgeshire_essex} illustrates the parallel product of bigraphs of Cambridgeshire and Essex.
\pagebreak

\begin{figure}[h!]
\begin{subfigure}{\linewidth}
\begin{tikzpicture}[remember picture]
 	\node[big site, inner sep=10pt,] (SiteCambridgeshireAndPeterborough) {};
	\node[below right, inner sep=0pt, shift={(0.1,-0.1)}, ] at (SiteCambridgeshireAndPeterborough.north west) {5};
	\node[fit=(SiteCambridgeshireAndPeterborough), draw, rounded corners=2,inner sep=15pt, text width=1cm] (BoundaryCambridgeshireAndPeterborough) {};
	\node[below right, inner sep=0pt, shift={(0.1,-0.1)}, ] at (BoundaryCambridgeshireAndPeterborough.north west) {Boundary};
	\node[above=1.3 of SiteCambridgeshireAndPeterborough, big site, inner sep=10pt,] (SiteEssex) {};
	\node[below right, inner sep=0pt, shift={(0.1,-0.1)}, ] at (SiteEssex.north west) {4};
	\node[fit=(SiteEssex), draw, rounded corners=2,inner sep=15pt, text width=1cm] (BoundaryEssex) {};
	\node[below right, inner sep=0pt, shift={(0.1,-0.1)}, ] at (BoundaryEssex.north west) {Boundary};
	\node[left=0.3, big site, inner sep=10pt,] at ($(BoundaryEssex.west)!0.5!(BoundaryCambridgeshireAndPeterborough.west)$) (SiteEngland) {};
	\node[below right, inner sep=0pt, shift={(0.1,-0.1)}, ] at (SiteEngland.north west) {3};
	\node[fit=(BoundaryCambridgeshireAndPeterborough)(BoundaryEssex)(SiteEngland), draw, rounded corners=2,inner sep=15pt,] (BoundaryEngland) {};
	\node[below right, inner sep=0pt, shift={(0.1,-0.1)}, ] at (BoundaryEngland.north west) {Boundary};
	\node[left=0.3 of BoundaryEngland, big site, inner sep=10pt,] (SiteUK) {};
	\node[below right, inner sep=0pt, shift={(0.1,-0.1)}, ] at (SiteUK.north west) {2};
	\node[fit=(BoundaryEngland)(SiteUK), draw, rounded corners=2,inner sep=15pt,] (BoundaryUK) {};
	\node[below right, inner sep=0pt, shift={(0.1,-0.1)}, ] at (BoundaryUK.north west) {Boundary};
	\node[left=0.3 of BoundaryUK, big site, inner sep=10pt,] (SiteWorld) {};
	\node[below right, inner sep=0pt, shift={(0.1,-0.1)}, ] at (SiteWorld.north west) {1};
	\node[fit=(BoundaryUK)(SiteWorld), draw, rounded corners=2,inner sep=15pt,] (BoundaryWorld) {};
	\node[below right, inner sep=0pt, shift={(0.1,-0.1)}, ] at (BoundaryWorld.north west) {Boundary};
	\node[below right,draw, rounded corners=2,text width=3.4cm] at (-10.7, |- BoundaryWorld.north) (IDWorld) {ID(World)};
	\draw[big edge] (IDWorld)  to[out=0,in=180]  (BoundaryWorld.west|-,|-IDWorld);
	\node[below=0.1 of IDWorld,draw, rounded corners=2,text width=3.4cm] (IDUK) {ID(UK)};
	\draw[big edge] (IDUK)  to[out=0,in=180]  (BoundaryUK.west|-,|-IDUK);
	\node[below=0.1 of IDUK,draw, rounded corners=2,text width=3.4cm]  (IDEngland) {ID(England)};
	\draw[big edge] (IDEngland)  to[out=0,in=180]  (BoundaryEngland.west|-,|-IDEngland);
	\node[below=0.1 of IDEngland,draw, rounded corners=2,text width=3.4cm] (IDEssex) {ID(Essex)};
	\draw[big edge] (IDEssex)  to[out=0,in=180]  (BoundaryEssex);
	\node[below=0.1 of IDEssex,draw, rounded corners=2,text width=3.4cm] (IDCambridgeshireAndPeterborough) {ID(Cambridgeshire and Peterborough)};
	\draw[big edge] (IDCambridgeshireAndPeterborough)  to[out=0,in=180] (BoundaryCambridgeshireAndPeterborough.west|-,|-IDCambridgeshireAndPeterborough);
	\node[below=0.1 of IDCambridgeshireAndPeterborough, big site, inner sep=10pt,] (SiteID) {};
	\node[below right, inner sep=0pt, shift={(0.1,-0.1)}, ] at (SiteID.north west) {0};
	\node[big region,fit=(IDWorld)(SiteID),inner sep=15pt,] (r0) {};
	\node[below right, inner sep=0pt, shift={(0.1,-0.1)}, ] at (r0.north west) {0};
	\node[big region,fit=(BoundaryWorld),inner sep=15pt,] (r1) {};
	\node[below right, inner sep=0pt, shift={(0.1,-0.1)}, ] at (r1.north west) {1};
	
	\node[above=0.3 of r1] (node215742) {node 215742};
	\draw[big edge] (SiteEssex)  to[out=90,in=270]  (node215742);
\end{tikzpicture}
\centering
\caption{$Context$ bigraph}
\label{fig:contextual-bigraph}
\end{subfigure}

\begin{subfigure}{\linewidth}
\centering
\begin{tikzpicture}[remember picture]
 	\node[big site, inner sep=10pt,] (s0) {};
	\node[below right, inner sep=0pt, shift={(0.1,-0.1)}, ] at (s0.north west) {0};
	\node[right=0.3 of s0,big site, inner sep=10pt,] (s6Connector) {};
	\node[below right, inner sep=0pt, shift={(0.1,-0.1)}, ] at (s6Connector.north west) {6};
	\node[big region, fit=(s0)(s6Connector), inner sep=15pt,] (r0Connector) {};
	\node[below right, inner sep=0pt, shift={(0.1,-0.1)}, ] at (r0Connector.north west) {0};
	\node[right=0.7 of r0Connector,big site, inner sep=10pt,] (s1) {};
	\node[below right, inner sep=0pt, shift={(0.1,-0.1)}, ] at (s1.north west) {1};
	\node[big region, fit=(s1), inner sep=15pt,] (r1) {};
	\node[below right, inner sep=0pt, shift={(0.1,-0.1)}, ] at (r1.north west) {1};
	\node[right=0.7 of r1,big site, inner sep=10pt,] (s2) {};
	\node[below right, inner sep=0pt, shift={(0.1,-0.1)}, ] at (s2.north west) {2};
	\node[big region, fit=(s2), inner sep=15pt,] (r2) {};
	\node[below right, inner sep=0pt, shift={(0.1,-0.1)}, ] at (r2.north west) {2};
	\node[right=0.7 of r2,big site, inner sep=10pt,] (s3) {};
	\node[below right, inner sep=0pt, shift={(0.1,-0.1)}, ] at (s3.north west) {3};
	\node[big region, fit=(s3), inner sep=15pt,] (r3) {};
	\node[below right, inner sep=0pt, shift={(0.1,-0.1)}, ] at (r3.north west) {3};
	\node[right=0.7 of r3,big site, inner sep=10pt,] (s4) {};
	\node[below right, inner sep=0pt, shift={(0.1,-0.1)}, ] at (s4.north west) {4};
	\node[big region, fit=(s4), inner sep=15pt,] (r4) {};
	\node[below right, inner sep=0pt, shift={(0.1,-0.1)}, ] at (r4.north west) {4};
 	\node[right=0.7 of r4, big site, inner sep=10pt,] (s5) {};
	\node[below right, inner sep=0pt, shift={(0.1,-0.1)}, ] at (s5.north west) {5};
	\node[right=0.3 of s5,big site, inner sep=10pt,] (s7Connector) {};
	\node[below right, inner sep=0pt, shift={(0.1,-0.1)}, ] at (s7Connector.north west) {7};
	\node[big region, fit=(s5)(s7Connector), inner sep=15pt,] (r5Connector) {};
	\node[below right, inner sep=0pt, shift={(0.1,-0.1)}, ] at (r5Connector.north west) {5};
\end{tikzpicture}
\caption{Placing $\phi$}
\label{fig:connector}
\end{subfigure}

\begin{subfigure}{\linewidth}
\centering
\begin{tikzpicture}[remember picture]
	\node[big site, inner sep=10pt,] (s0) {};
	\node[below right, inner sep=0pt, shift={(0.1,-0.1)}, ] at (s0.north west) {0};
	\node[big region, fit=(s0), inner sep=10pt,] (r0) {};
	\node[below right, inner sep=0pt, shift={(0.1,-0.1)}, ] at (r0.north west) {0};
	\node[right=0.5 of r0,big site, inner sep=10pt,] (s1) {};
	\node[below right, inner sep=0pt, shift={(0.1,-0.1)}, ] at (s1.north west) {1};
	\node[big region, fit=(s1), inner sep=10pt,] (r1) {};
	\node[below right, inner sep=0pt, shift={(0.1,-0.1)}, ] at (r1.north west) {1};
	\node[right=0.5 of r1,big site, inner sep=10pt,] (s2) {};
	\node[below right, inner sep=0pt, shift={(0.1,-0.1)}, ] at (s2.north west) {2};
	\node[big region, fit=(s2), inner sep=10pt,] (r2) {};
	\node[below right, inner sep=0pt, shift={(0.1,-0.1)}, ] at (r2.north west) {2};
	\node[right=0.5 of r2,big site, inner sep=10pt,] (s3) {};
	\node[below right, inner sep=0pt, shift={(0.1,-0.1)}, ] at (s3.north west) {3};
	\node[big region, fit=(s3), inner sep=10pt,] (r3) {};
	\node[below right, inner sep=0pt, shift={(0.1,-0.1)}, ] at (r3.north west) {3};
	\node[right=0.5 of r3,big site, inner sep=10pt,] (s4) {};
	\node[below right, inner sep=0pt, shift={(0.1,-0.1)}, ] at (s4.north west) {4};
	\node[big region, fit=(s4), inner sep=10pt,] (r4) {};
	\node[below right, inner sep=0pt, shift={(0.1,-0.1)}, ] at (r4.north west) {4};
 	\node[right=0.5 of r4, big site, inner sep=10pt,] (s5) {};
	\node[below right, inner sep=0pt, shift={(0.1,-0.1)}, ] at (s5.north west) {5};
	\node[big region, fit=(s5), inner sep=10pt,] (r5) {};
	\node[below right, inner sep=0pt, shift={(0.1,-0.1)}, ] at (r5.north west) {5};

	\node[right=0.5 of r5,draw, rounded corners=2,text width=2.1cm, shift={(0,0.5)},] (IDCambridgeshire) {ID( Cambridgeshire)};
 	\node[below=0.3 of IDCambridgeshire, big site, inner sep=10pt,] (s6) {};
	\node[below right, inner sep=0pt, shift={(0.1,-0.1)}, ] at (s6.north west) {6};
	\node[big region, fit=(IDCambridgeshire)(s6), inner sep=10pt,] (r6) {};
	\node[below right, inner sep=0pt, shift={(0.1,-0.1)}, ] at (r6.north west) {6};

	\node[right=1.2 of r6,, big site, inner sep=10pt,] (SiteCambridgeshire) {};
	\node[below right, inner sep=0pt, shift={(0.1,-0.1)}, ] at (SiteCambridgeshire.north west) {7};
	\node[fit=(SiteCambridgeshire), draw, rounded corners=2,inner sep=15pt, text width=1cm,] (BoundaryCambridgeshire) {};
	\node[below right, inner sep=0pt, shift={(0.1,-0.1)}, ] at (BoundaryCambridgeshire.north west) {Boundary};
	\node[big region, fit=(BoundaryCambridgeshire), inner sep=10pt,] (r7) {};
	\node[below right, inner sep=0pt, shift={(0.1,-0.1)}, ] at (r7.north west) {7};
	\draw[big edge] (IDCambridgeshire)  to[out=0,in=180] (BoundaryCambridgeshire);

	\node[above=0.3 of r7] (node215742) {node 215742};
	\draw[big edge] (SiteCambridgeshire)  to[out=90,in=270]  (node215742);
\end{tikzpicture}
\caption{$id || id || id || id || id || id || Cambridgeshire$}
\label{fig:bigraph-Cambridgeshire-and-sites}
\end{subfigure}
\centering
\caption{Bigraphs to connect Cambridgeshire to the rest of the world. Arrows indicate the joining of the $i^{th}$ root with the $i^{th}$ site during nesting, showing how the ID and Physical regions of $Cambridgshire$ are nested into the correct sites in $Context$.}
\end{figure}
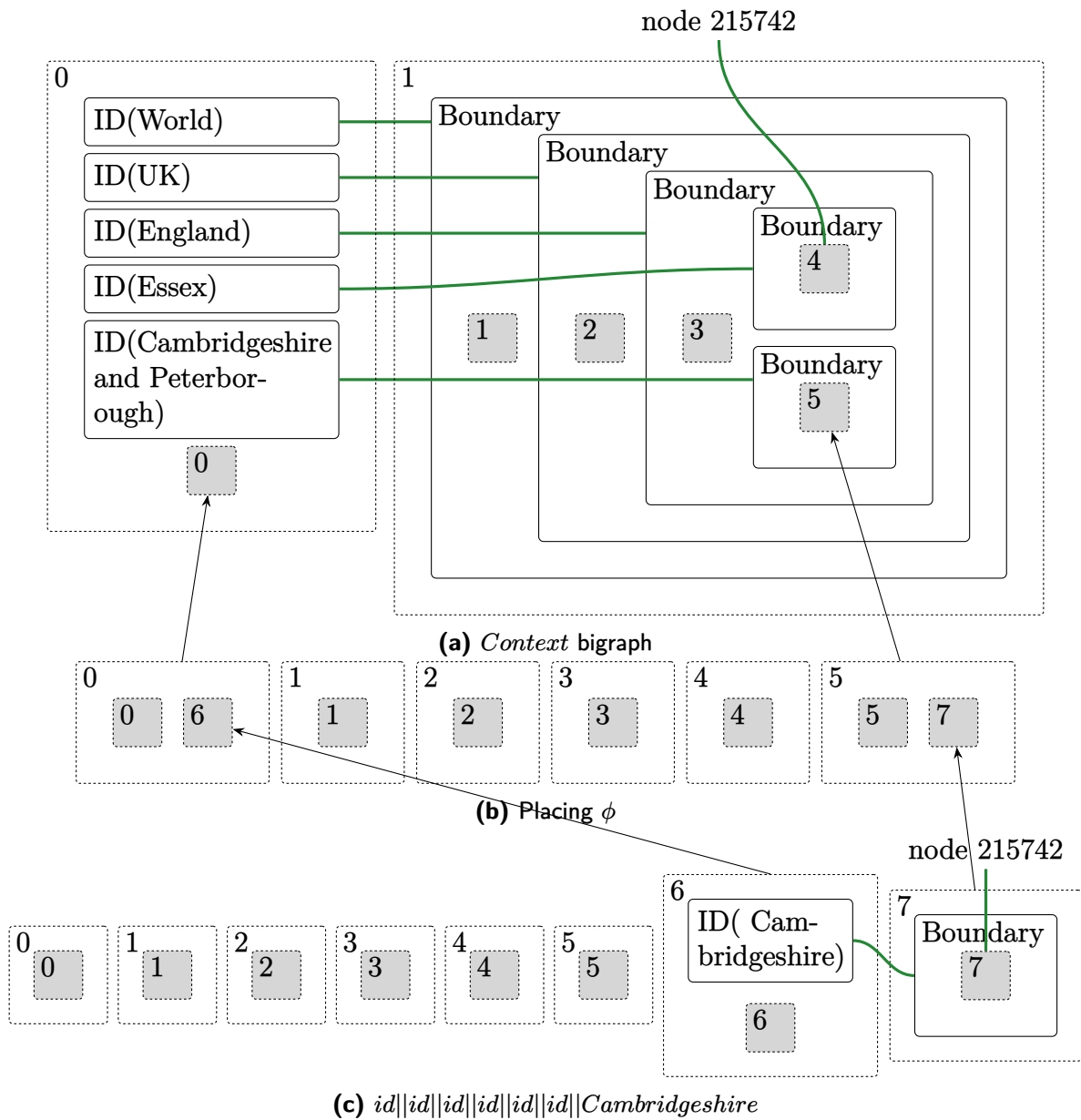
\begin{tikzpicture}[overlay, remember picture]
          \draw[-Stealth] (r7) -- (s7Connector);
          \draw[-Stealth] (r6.north) -- (s6Connector);
          \draw[-Stealth] (r0Connector) -- (SiteID);
          \draw[-Stealth] (r5Connector) -- (SiteCambridgeshireAndPeterborough);	
\end{tikzpicture}

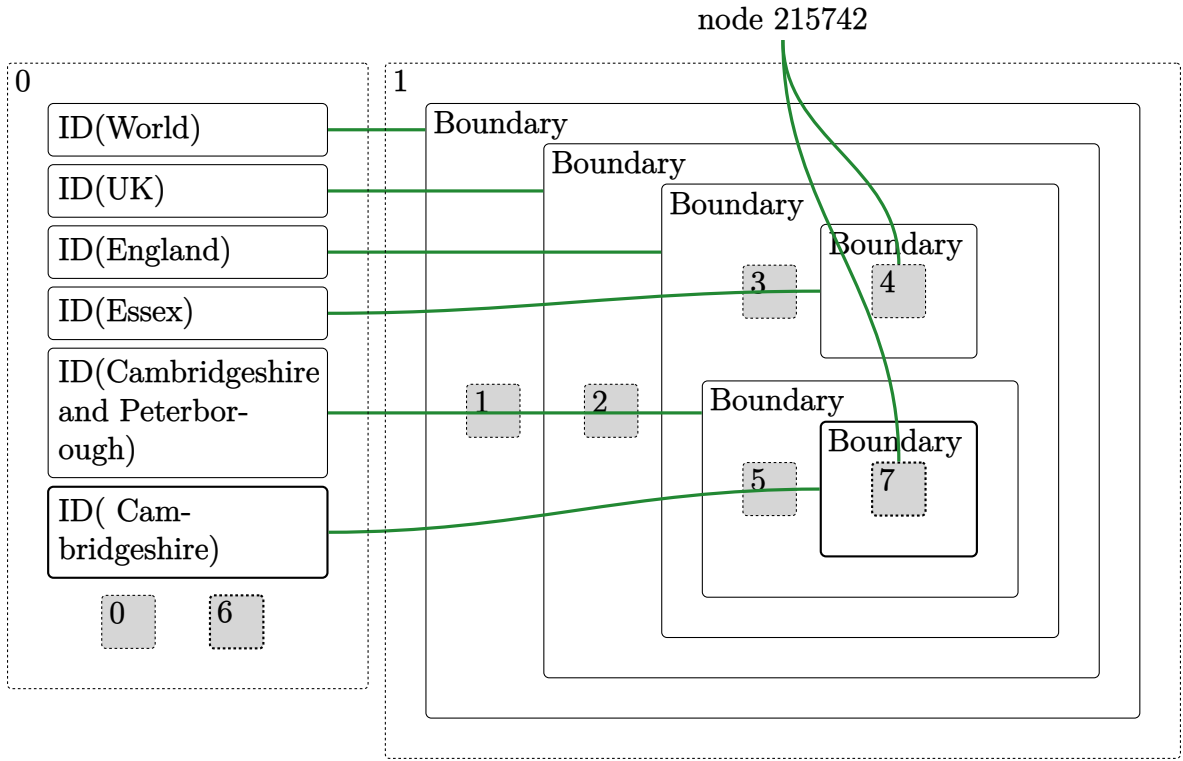
\begin{figure}[h]
\begin{tikzpicture}
	\node[big site, inner sep=10pt,thick] (SiteCambridgeshire) {};
	\node[below right, inner sep=0pt, shift={(0.1,-0.1)}, ] at (SiteCambridgeshire.north west) {7};
	\node[fit=(SiteCambridgeshire), draw, rounded corners=2,inner sep=15pt, text width=1cm,thick] (BoundaryCambridgeshire) {};
	\node[below right, inner sep=0pt, shift={(0.1,-0.1)}, ] at (BoundaryCambridgeshire.north west) {Boundary};
 	\node[left=0.3 of BoundaryCambridgeshire, big site, inner sep=10pt,] (SiteCambridgeshireAndPeterborough) {};
	\node[below right, inner sep=0pt, shift={(0.1,-0.1)}, ] at (SiteCambridgeshireAndPeterborough.north west) {5};
	\node[fit=(SiteCambridgeshireAndPeterborough)(BoundaryCambridgeshire), draw, rounded corners=2,inner sep=15pt] (BoundaryCambridgeshireAndPeterborough) {};
	\node[below right, inner sep=0pt, shift={(0.1,-0.1)}, ] at (BoundaryCambridgeshireAndPeterborough.north west) {Boundary};
	\node[above=1.9 of SiteCambridgeshire, big site, inner sep=10pt,] (SiteEssex) {};
	\node[below right, inner sep=0pt, shift={(0.1,-0.1)}, ] at (SiteEssex.north west) {4};
	\node[fit=(SiteEssex), draw, rounded corners=2,inner sep=15pt, text width=1cm] (BoundaryEssex) {};
	\node[below right, inner sep=0pt, shift={(0.1,-0.1)}, ] at (BoundaryEssex.north west) {Boundary};
	\node[above=1.9 of SiteCambridgeshireAndPeterborough, big site, inner sep=10pt,] (SiteEngland) {};
	\node[below right, inner sep=0pt, shift={(0.1,-0.1)}, ] at (SiteEngland.north west) {3};
	\node[fit=(BoundaryCambridgeshireAndPeterborough)(BoundaryEssex)(SiteEngland), draw, rounded corners=2,inner sep=15pt,] (BoundaryEngland) {};
	\node[below right, inner sep=0pt, shift={(0.1,-0.1)}, ] at (BoundaryEngland.north west) {Boundary};
	\node[left=0.3 of BoundaryEngland, big site, inner sep=10pt,] (SiteUK) {};
	\node[below right, inner sep=0pt, shift={(0.1,-0.1)}, ] at (SiteUK.north west) {2};
	\node[fit=(BoundaryEngland)(SiteUK), draw, rounded corners=2,inner sep=15pt,] (BoundaryUK) {};
	\node[below right, inner sep=0pt, shift={(0.1,-0.1)}, ] at (BoundaryUK.north west) {Boundary};
	\node[left=0.3 of BoundaryUK, big site, inner sep=10pt,] (SiteWorld) {};
	\node[below right, inner sep=0pt, shift={(0.1,-0.1)}, ] at (SiteWorld.north west) {1};
	\node[fit=(BoundaryUK)(SiteWorld), draw, rounded corners=2,inner sep=15pt,] (BoundaryWorld) {};
	\node[below right, inner sep=0pt, shift={(0.1,-0.1)}, ] at (BoundaryWorld.north west) {Boundary};
	\node[below right,draw, rounded corners=2,text width=3.4cm] at (-11.2, |- BoundaryWorld.north) (IDWorld) {ID(World)};
	\draw[big edge] (IDWorld)  to[out=0,in=180]  (BoundaryWorld.west|-,|-IDWorld);
	\node[below=0.1 of IDWorld,draw, rounded corners=2,text width=3.4cm] (IDUK) {ID(UK)};
	\draw[big edge] (IDUK)  to[out=0,in=180]  (BoundaryUK.west|-,|-IDUK);
	\node[below=0.1 of IDUK,draw, rounded corners=2,text width=3.4cm]  (IDEngland) {ID(England)};
	\draw[big edge] (IDEngland)  to[out=0,in=180]  (BoundaryEngland.west|-,|-IDEngland);
	\node[below=0.1 of IDEngland,draw, rounded corners=2,text width=3.4cm] (IDEssex) {ID(Essex)};
	\draw[big edge] (IDEssex)  to[out=0,in=180]  (BoundaryEssex);
	\node[below=0.1 of IDEssex,draw, rounded corners=2,text width=3.4cm] (IDCambridgeshireAndPeterborough) {ID(Cambridgeshire and Peterborough)};
	\draw[big edge] (IDCambridgeshireAndPeterborough)  to[out=0,in=180] (BoundaryCambridgeshireAndPeterborough.west|-,|-IDCambridgeshireAndPeterborough);	
	\node[below=0.1 of IDCambridgeshireAndPeterborough,draw, rounded corners=2,text width=3.4cm, thick] (IDCambridgeshire) {ID( Cambridgeshire)};
	\draw[big edge] (IDCambridgeshire)  to[out=0,in=180] (BoundaryCambridgeshire);
	\node[below right=0.3 of IDCambridgeshire.south west, big site, inner sep=10pt,shift={(0.5,0)}] (SiteID) {};
	\node[below right, inner sep=0pt, shift={(0.1,-0.1)}, ] at (SiteID.north west) {0};
	\node[right=0.7 of SiteID, big site, inner sep=10pt,thick] (SiteIDCambridgeshire) {};
	\node[below right, inner sep=0pt, shift={(0.1,-0.1)}, ] at (SiteIDCambridgeshire.north west) {6};
	\node[big region,fit=(IDWorld)(SiteID),inner sep=15pt,] (r0) {};
	\node[below right, inner sep=0pt, shift={(0.1,-0.1)}, ] at (r0.north west) {0};
	\node[big region,fit=(BoundaryWorld),inner sep=15pt,] (r1) {};
	\node[below right, inner sep=0pt, shift={(0.1,-0.1)}, ] at (r1.north west) {1};
	
	\node[above=0.3 of r1] (node215742) {node 215742};
	\draw[big edge] (SiteEssex)  to[out=90,in=270]  (node215742);
	\draw[big edge] (SiteCambridgeshire)  to[out=90,in=270]  (node215742);
\end{tikzpicture}
\centering
\caption{Cambridgeshire and the rest of the world. Sites 6 and 7 are placeholders for the rest of the nodes in the bigraph of Cambridgeshire.}
\label{fig:composed-bigraph}
\end{figure}

In order to nest the $Cambridgeshire$ bigraph into the $Context$ bigraph, the $\phi$ placing (Figure \ref{fig:connector}) is constructed using the elementary bigraphs $id$ and $symmetry$ introduced in \S \ref{elementary-bigraphs} to reorder the ID and Physical regions of $Cambridgshire$. Then, the nesting \[Context.Connector.(id || id || id || id || id || id || Cambridgeshire)\]  puts the bigraph of Cambridgeshire into its place in the contextual bigraph and results in the bigraph illustrated in Figure \ref{fig:composed-bigraph}.

\section{Modelling street connectivity using shared junctions} \label{section:shared-junctions}

\begin{figure}[h]
\begin{subfigure}{\linewidth}
\begin{tikzpicture}
	\node[draw, rounded corners=2] at (0,0) (Junction0) {Junction};

	\node[left=1.5 of Junction0, shift={(-0.5,-0.5)}, big site, inner sep=10pt,thick] (SiteHighCross) {};
	\node[below right, inner sep=0pt, shift={(0.1,-0.1)}, ] at (SiteHighCross.north west) {1};
	\node[draw, ellipse, fit=(Junction0)(SiteHighCross), rotate=10,inner sep=7pt,] (StreetHighCross){};
	\node[above=0.15, shift={(0.7,0)},inner sep=0pt, rotate=0] at (StreetHighCross.south west) {High Cross};

	\node[right=1.5 of Junction0, shift={(0.5,-0.5)}, big site, inner sep=10pt,thick] (SiteAda) {};
	\node[below right, inner sep=0pt, shift={(0.1,-0.1)}, ] at (SiteAda.north west) {2};
	\node[draw, ellipse, fit=(Junction0)(SiteAda), rotate=-10,inner sep=7pt,] (StreetAda){};
	\node[above=0.25, shift={(-1,0)}, inner sep=0pt, rotate=0] at (StreetAda.south east) {Ada Lovelace Road};

	\node[above=1 of Junction0, big site, inner sep=10pt,thick] (SiteCharles) {};
	\node[below right, inner sep=0pt, shift={(0.1,-0.1)}, ] at (SiteCharles.north west) {0};
	\node[above left=1 of SiteCharles.north, draw, rounded corners=2] (Junction1) {Junction};
	\node[big region,fit=(Junction1),inner sep=10pt,] (rJunction1) {};
	\node[below right, inner sep=0pt, shift={(0.1,-0.1)}, ] at (rJunction1.north west) {1};
	\node[above right=1 of SiteCharles.north, draw, rounded corners=2] (Junction2) {Junction};
	\node[big region,fit=(Junction2),inner sep=10pt,] (rJunction2) {};
	\node[below right, inner sep=0pt, shift={(0.1,-0.1)}, ] at (rJunction2.north west) {2};
	\node[draw, circle, fit=(Junction0)(Junction1)(Junction2),shift={(0,1)}] (StreetCharles){};
	\node[below=0.85, inner sep=0pt] at (StreetCharles.north) {Charles Babbage Road};

	\node[big region,fit=(StreetHighCross)(StreetAda)(StreetCharles),inner sep=10pt,] (r0) {};
	\node[below right, inner sep=0pt, shift={(0.1,-0.1)}, ] at (r0.north west) {0};
\end{tikzpicture}
\centering
\caption{Bigraph with sharing}
\end{subfigure}

\begin{subfigure}{\linewidth}
 \begin{forest}
for tree={edge = {-latex}}
[,phantom
[0, big region,name=r0
	[Charles Babbage Road
		[Junction,name=Junction1]
		[Junction,name=Junction2]
		[0,big site]
		[Junction, name=Junction0]
	]
	[High Cross, name=StreetHighCross
		[,phantom]
		[1,big site]
	]
	[Ada Lovelace Road, name=StreetAda
		[,phantom]
		[2,big site]
	]
]
]
\node[big region] at (Junction1|-,|-r0) (r1){1};
\node[big region] at (Junction2|-,|-r0) (r2){2};
\draw[-latex] (r1) -- (Junction1);
\draw[-latex] (r2) -- (Junction2);
\draw[-latex] (StreetHighCross) -- (Junction0);
\draw[-latex] (StreetAda) -- (Junction0);
\end{forest}
\centering
\caption{Place graph, which is a DAG}
\end{subfigure}
\centering
\caption{Bigraph with sharing representation of intersection between Charles Babbage Road, High Cross and Ada Lovelace Road.}
\label{fig:bigraph-connected-streets-sharing}
\end{figure}
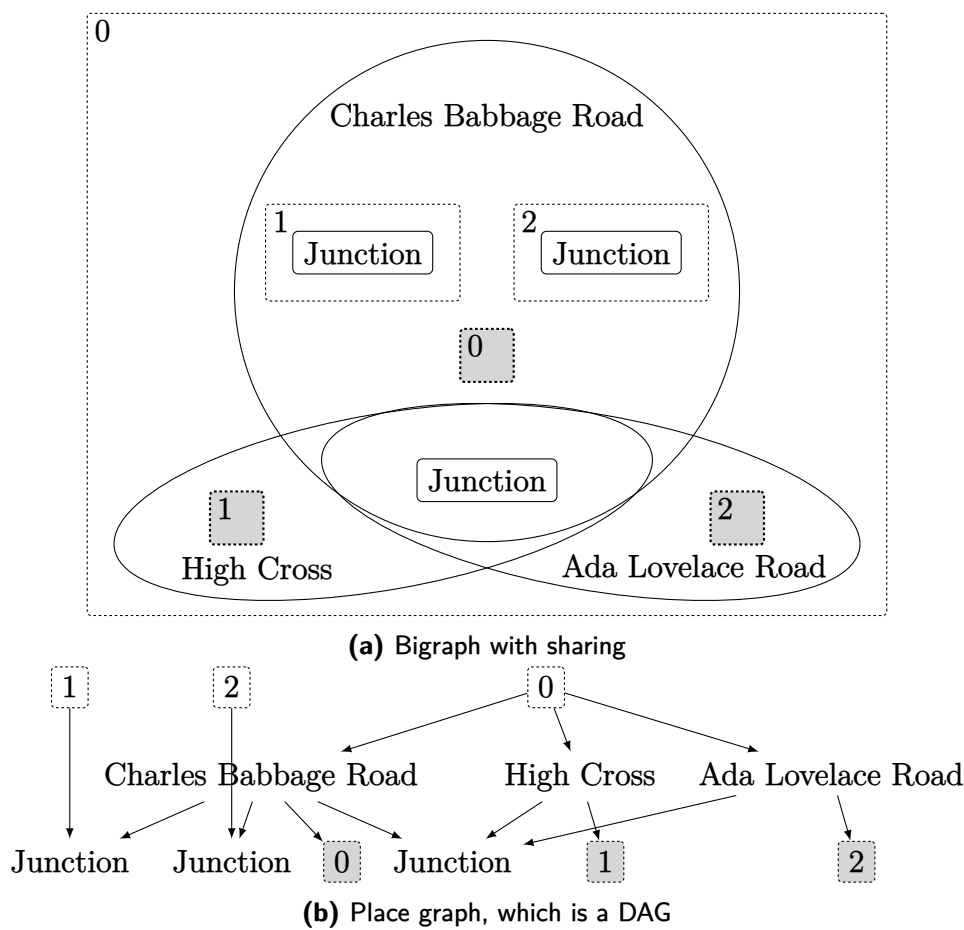

In \S \ref{street-connectivity-link}, the specification for building a bigraph of the real world described modelling connected streets by having the Street nodes corresponding to intersecting streets each contain a Junction node, which are all linked. Here, an alternative is considered: using bigraphs with sharing (\S\ref{section:sharing}), a single Junction node can be contained by two or more Street nodes, and this sharing represents the intersection between those streets. The street intersection previously modelled using multiple linked junctions in Figure \ref{fig:bigraph-connected-streets} is now represented with a single shared junction, as illustrated in Figure \ref{fig:bigraph-connected-streets-sharing}.

The shared junction approach has a number of advantages. Firstly, it is an abstraction that is closer to OSM's representation of streets: way elements of intersecting streets share a single node element. Secondly, it stays true to the spirit of bigraphs: placing and linking can be considered independently; meanwhile the linked junctions approach uses linking to represent locality of connected streets. Thirdly, it encodes the intersection of streets using less nodes: one Junction node per intersection instead of multiple. 

However, while it might be possible to express sharing when constructing the complete bigraph of the whole world at once, it becomes extremely difficult to construct subgraphs for different parts of the world separately and then combine them like in \S\ref{composition-world}. Consider the parallel product of the bigraphs of Cambridgeshire and Essex (Figure \ref{fig:cambridgeshire_essex}) but now modelling connected streets using sharing. One approach is for one bigraph to have a Junction node shared by a Street node and a region and the other bigraph to nest a site in a Street node that can join with the region containing the Junction. However, it is difficult to coordinate this when constructing the bigraphs separately. Another solution is for both to nest a site in their Street nodes in place of a Junction node; after the parallel product, a Junction node shared by two regions can be nested in. Unfortunately, this quickly becomes unwieldy as placings constructed using the elementary bigraphs $symmetry$ and $id$ (\S \ref{elementary-bigraphs}) are required to reorder the regions to link to the correct sites. The scale of bigraphs of the real world aggravates this: again, the complete bigraph of Cambridgeshire has 225 outer names (Figure \ref{fig:cambridgeshire-outer-names}). Nesting into a contextual bigraph of the whole world would be even more complex.

The key insight here is that the names given to open links make combining bigraphs extremely convenient: when forming parallel products or nestings, links with the same names are combined. Meanwhile, regions and sites do not have names; instead, the $i^{th}$ root is joined to the $i^{th}$ site in the nesting (\S\ref{bigraph-operations}), hence placings constructed using $symmetry$ and $id$ are required to reorder the regions. Since this dissertation requires that bigraphs of different parts of the world can be constructed independently then combined, the specification chooses to model connected streets using linked junctions.

\section{Unifying bigraphs that model virtual spaces}
\begin{figure}[h!]
\centering
\begin{subfigure}{0.5\textwidth}
\centering
\includegraphics[width=\textwidth]{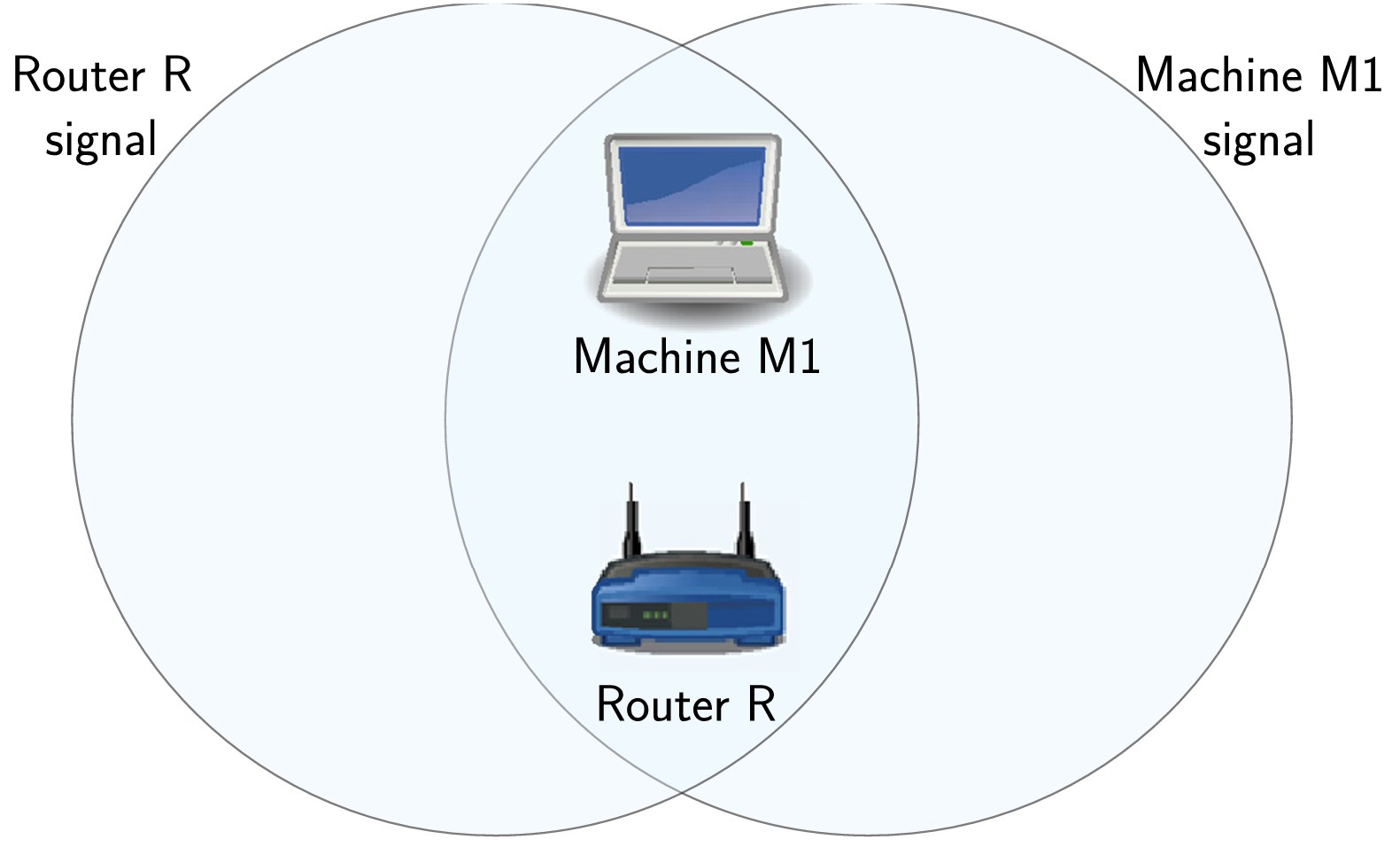}
\caption{WLAN with one machine and a router.}
\end{subfigure}
\hfill
\begin{subfigure}{0.44\textwidth}
\centering
\includegraphics[width=0.7\textwidth]{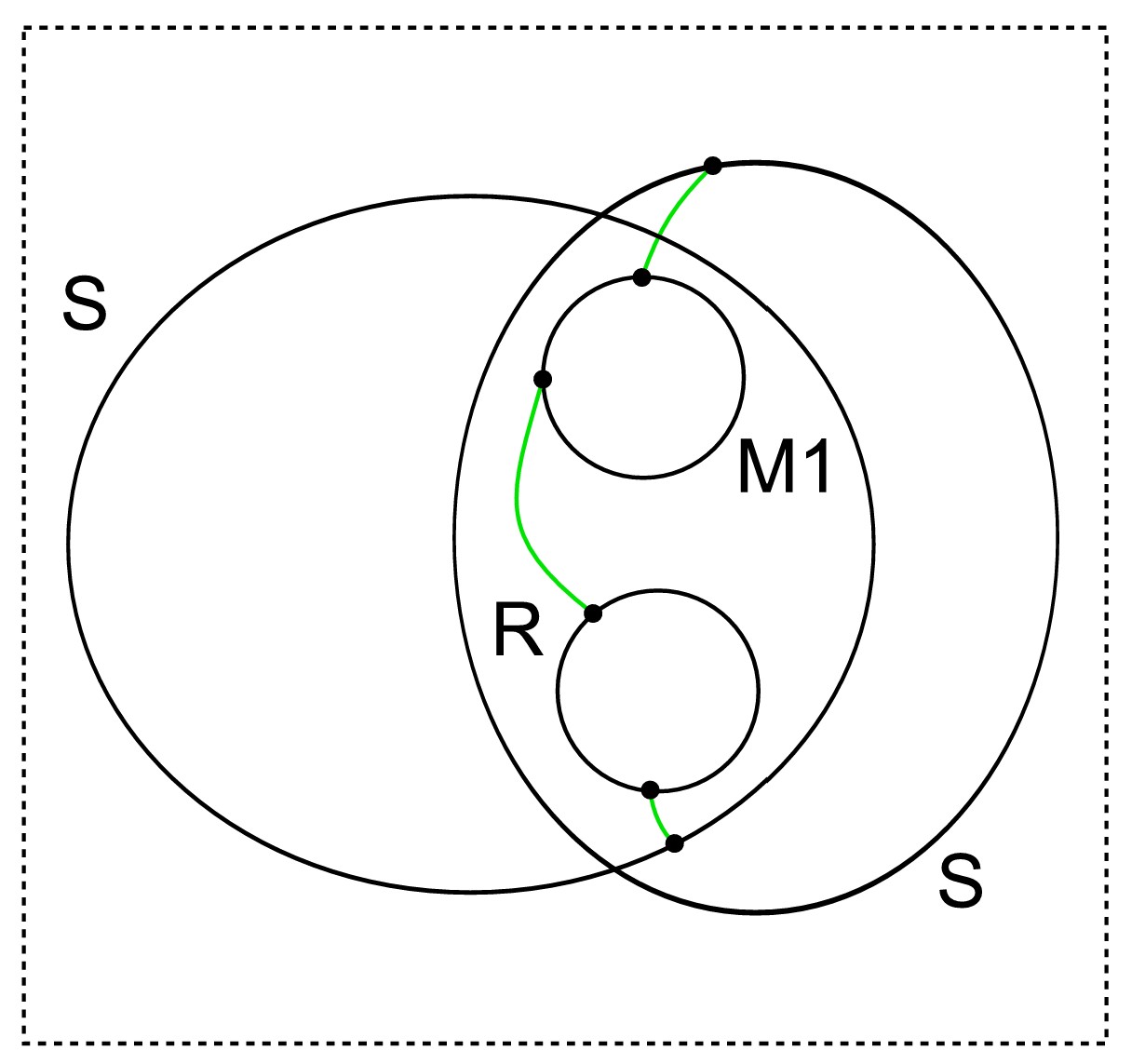}
\caption{Bigraph representation of  WLAN with one machine and a router.}
\label{fig:WLAN-bigraph}
\end{subfigure}
\caption{Bigraph model of example wireless network.}
\end{figure}

Even bigraphs modelling only virtual spaces can be composed into the bigraph of the real world. Wireless networks were modelled using bigraphs with sharing \cite{wireless-home-networks}. The range of signals were modelled by placing and, using bigraph with sharing, devices can be in multiple signal ranges (Figure \ref{fig:WLAN-bigraph}). This bigraph of a wireless network can be combined in a parallel product with a bigraph of the real world, as virtual and physical perspectives respectively. Additionally, the machine and router in the virtual perspective, which can move in and out of signal ranges, can be linked to physical entities located in the physical perspective, which are able to move around the world. Another approach to combining the bigraphs uses sharing (\S\ref{section:sharing}), so that the machine and router can simultaneously be inside virtual signal ranges and a physical location.

\section{Transition system of the BRS for motion}
\begin{figure}[h!]
\includegraphics[width=\textwidth]{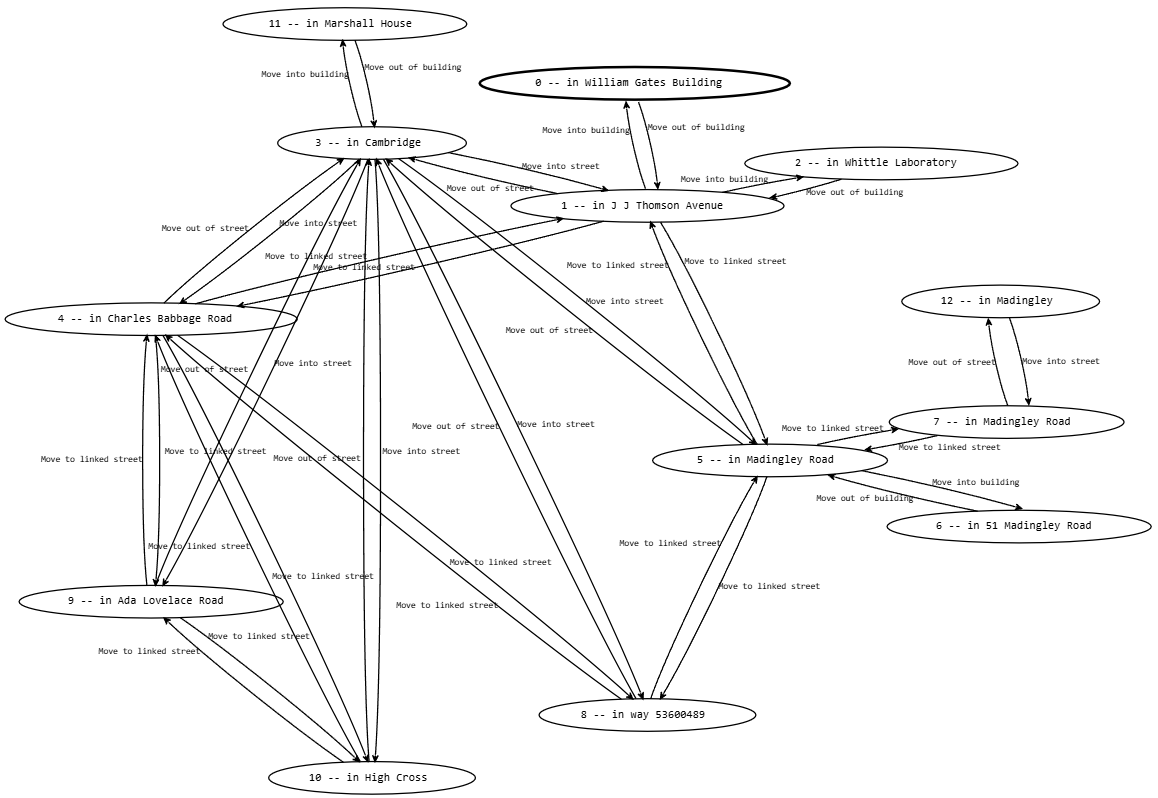}
\centering
\caption{Transition system from applying the reaction rules for motion on the bigraph of Cambridgeshire}
\label{fig:transition-system-motion}
\end{figure}

 The full transition system of bigraphs from applying the reaction rules for motion (\S\ref{section:react-motion}) on the simplified bigraph of Cambridgeshire (Figure \ref{fig:bigraph-Cambridgeshire}) is illustrated in Figure \ref{fig:transition-system-motion}.

\afterpage{%
    \clearpage
\addtocounter{section}{1}
\addcontentsline{toc}{section}{\protect\numberline{\thesection}Key statistics of bigraphs of selected regions}
	\small
\setlength{\tabcolsep}{2pt}
\renewcommand{\arraystretch}{0.9}
    \begin{landscape}%
\begin{longtable}{p{1.5cm}llllp{1.1cm}p{1.8cm}llll}
\toprule
admin\_level & id & name & Nodes & Edges & Outer names & Peak open links & Subdivisions & Streets & Buildings & Junctions \\
\midrule
4 & 62422 & Berlin & 277,289 & 135,449 & 208 & 822 & 114 & 15,213 & 96,373 & 53,887 \\
4 & 162069 & District of Columbia & 267,244 & 132,694 & 262 & 1,580 & 9 & 4,160 & 118,037 & 22,830 \\
2 & 536780 & Singapore & 212,140 & 104,497 & 4 & 6,086 & 12 & 11,917 & 70,981 & 46,318 \\
6 & 180837 & Cambridgeshire & 210,768 & 103,274 & 225 & 1,875 & 241 & 16,190 & 69,754 & 38,396 \\
6 & 62428 & Munich & 164,233 & 78,594 & 140 & 939 & 141 & 10,378 & 54,375 & 34,443 \\
8 & 118362 & Leeds & 161,213 & 79,732 & 132 & 14,654 & 38 & 13,715 & 47,115 & 39,475 \\
9 & 30353 & Wandsbek & 149,385 & 74,302 & 113 & 574 & 19 & 2,582 & 68,076 & 8,029 \\
8 & 146656 & Manchester & 96,802 & 48,052 & 235 & 10,189 & 1 & 7,140 & 30,599 & 21,320 \\
8 & 2305280 & Oceanside & 78,005 & 38,904 & 86 & 3,945 & 0 & 2,709 & 32,250 & 8,085 \\
8 & 111848 & Chula Vista & 77,630 & 38,686 & 129 & 4,959 & 0 & 3,186 & 30,541 & 10,174 \\
8 & 295353 & South Cambridgeshire & 75,961 & 37,326 & 178 & 688 & 104 & 4,663 & 27,711 & 11,003 \\
9 & 55734 & Steglitz-Zehlendorf & 74,559 & 36,943 & 102 & 522 & 11 & 1,829 & 32,366 & 6,145 \\
6 & 1906767 & Glasgow City & 70,524 & 34,790 & 194 & 11,685 & 0 & 7,427 & 15,678 & 24,312 \\
8 & 295355 & Cambridge & 62,838 & 31,355 & 73 & 1,875 & 0 & 1,612 & 27,868 & 3,876 \\
8 & 2305279 & Carlsbad & 60,972 & 30,425 & 63 & 3,361 & 0 & 2,407 & 24,657 & 6,842 \\
8 & 51781 & City of Westminster & 57,179 & 28,460 & 183 & 2,750 & 1 & 1,809 & 23,816 & 5,925 \\
8 & 7444 & Paris & 56,622 & 24,667 & 324 & 1,320 & 100 & 9,071 & 4,752 & 28,774 \\
8 & 172987 & Liverpool & 50,577 & 24,816 & 122 & 10,168 & 0 & 6,879 & 7,769 & 21,279 \\
6 & 1920901 & City of Edinburgh & 48,146 & 23,286 & 57 & 3,429 & 30 & 6,138 & 9,268 & 17,272 \\
8 & 8450265 & London Borough of Southwark & 42,145 & 21,031 & 193 & 2,931 & 0 & 2,038 & 16,062 & 5,943 \\
10 & 271110 & Amsterdam & 41,306 & 20,189 & 152 & 10,631 & 0 & 5,082 & 4,476 & 22,188 \\
8 & 295351 & Huntingdonshire & 37,597 & 18,234 & 102 & 679 & 81 & 4,899 & 7,931 & 11,773 \\
8 & 51805 & London Borough of Tower Hamlets & 32,888 & 16,367 & 92 & 2,469 & 0 & 1,885 & 12,013 & 5,090 \\
5 & 10264792 & Hong Kong Island & 21,676 & 10,647 & 6 & 759 & 4 & 1,621 & 6,555 & 5,314 \\
8 & 1544956 & Mountain View & 20,224 & 10,077 & 108 & 1,700 & 0 & 1,027 & 7,350 & 3,468 \\
8 & 295352 & East Cambridgeshire & 19,619 & 9,566 & 85 & 514 & 35 & 2,383 & 4,556 & 5,669 \\
8 & 71033 & Strasbourg & 15,819 & 7,327 & 77 & 462 & 15 & 2,641 & 1,141 & 8,223 \\
8 & 295349 & Fenland & 13,480 & 6,530 & 64 & 552 & 16 & 2,283 & 1,688 & 5,504 \\
6 & 51800 & City of London & 4,193 & 2,089 & 93 & 580 & 1 & 431 & 1,077 & 1,173 \\
10 & 2604777 & Dover & 2,184 & 1,069 & 25 & 520 & 0 & 391 & 158 & 1,084 \\
\bottomrule
\end{longtable}
\captionof{table}{Key statistics of bigraphs of selected regions.\label{table:bigraph-stats}}
  \end{landscape}
    \clearpage%
}

\chapter{Stack traces of reaction rules}

\begin{figure}[h]
\includegraphics[width=\textwidth]{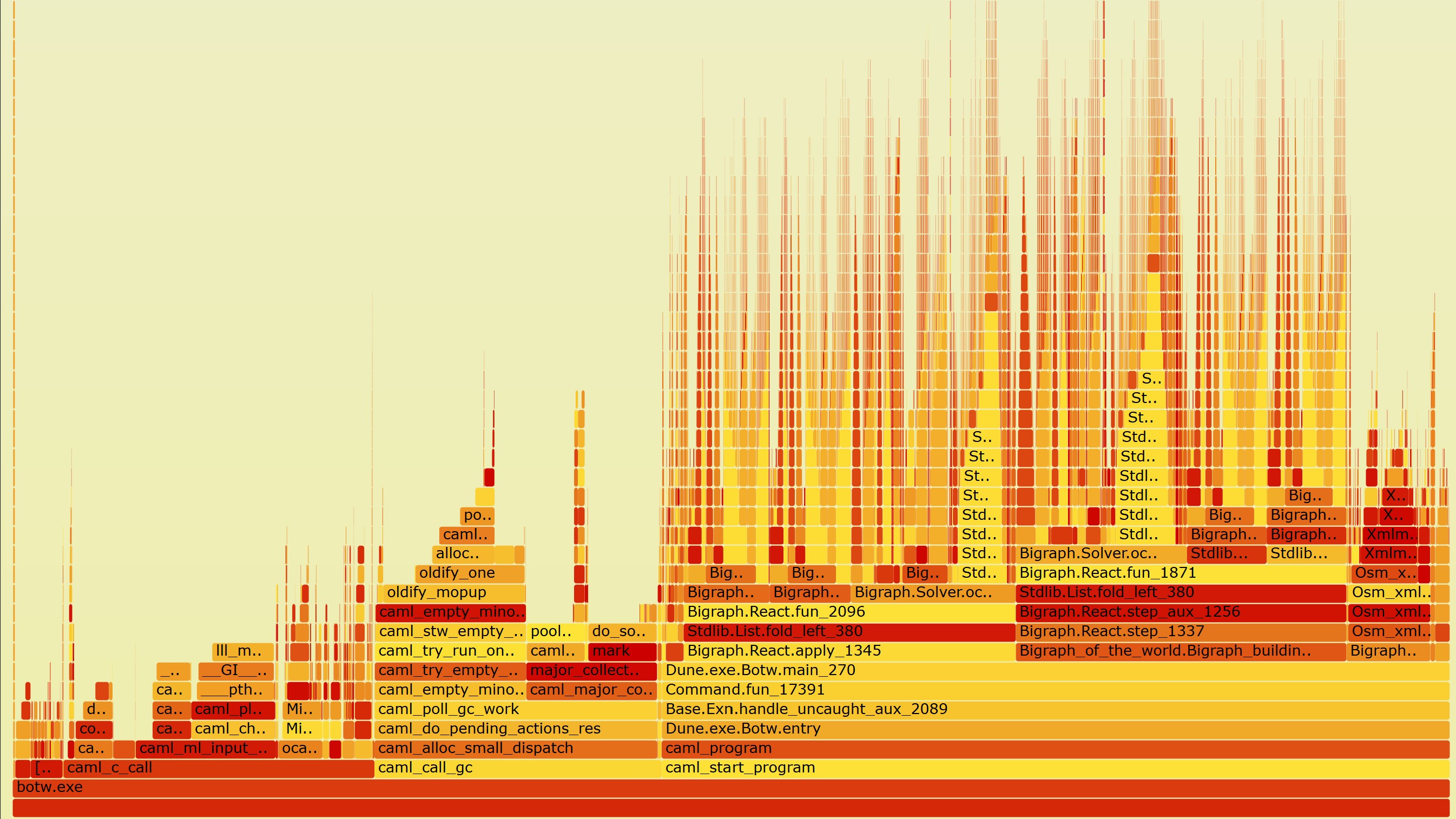}
\centering
\caption{Stack trace of applying a single reaction rule when using the $n\times DFS$ algorithm,  visualised as a flame graph.}
\label{fig:new-trans-flamegraph}
\end{figure}

\begin{figure}[h]
\includegraphics[width=\textwidth]{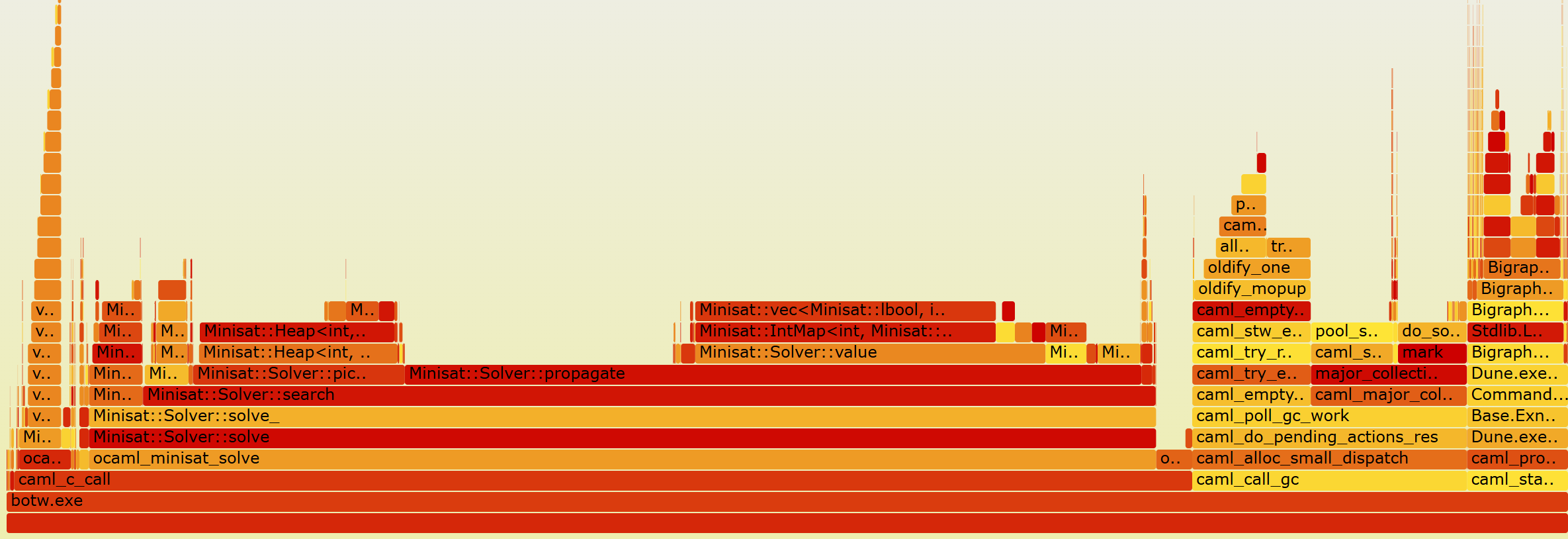}
\centering
\caption{Stack trace of applying the reaction rule \texttt{enter\_building}, visualised as a flame graph.}
\label{fig:sat-solver}
\end{figure}

\chapter{Project proposal}

\centerline{\Large \@title}
\vspace{2em}
\centerline{\large Candidate \candidatenumber}
\vspace{1em}
\centerline{\large Supervisors: Ryan Gibb and Anil Madhavapeddy}

\section*{Introduction}
Bigraphs were first introduced by Robin Milner as a universal model for representing the space in which agents move and interact. Bigraphs are represented by two constituents: a hierarchical forest of nodes called a place graph, and a hypergraph of edges called a link graph. They have since been applied in areas such as home network management and computer networking. 

Bigraphs were originally proposed as a model for the behaviour of ubiquitous systems since interaction between mobile devices is dependent on both placing (locality) and linking (connectivity). However, there has yet to be a bigraph that represents the complete physical world. Such a bigraph will enhance the computer's representation of its location from a simple latitude-longitude pair to a context more familiar to humans: the room it is in, the street the building is on, and the town the street is in. This will allow for location-aware applications and policies about connectivity of mobile devices to work based on the defined locality of buildings, streets and administrative regions.

The physical world has also long been represented by maps. OpenStreetMap (OSM) is an freely-licensed geographic database built by a community of volunteers through the annotation of data collected through surveys, aerial imagery and other free geodata sources. Boasting a user base of 10 million, OSM has labelled buildings, streets and regions with impressive detail comparable with commercial counterparts. The map elements are supplemented with key-value pairs called tags that describe characteristics of the element. Tagging conventions vary across countries, but there are standard practices such as the “addr” tag on buildings to describe its address.

This project will demonstrate modelling the physical world as a bigraph. Places marked on OSM will be hierarchically structured in a place graph, guided by administrative boundaries such as country, state, city etc. Then, a link graph will be built on top of the place graph to model the network of connected streets. The use of such bigraphs for ubiquitous systems will be demonstrated with the use case of device connectivity, using reaction rules that allow devices to move to a new place and form links with other devices in its proximity.

\section*{Goals}
This project will involve modelling the places contained in the county of Cambridgeshire, UK. There are approximately 273,869 buildings and 126,807 streets segments and paths in Cambridgeshire. This scope is chosen as we are most interested in how bigraphs model relationships between proximate locations. The bigraph for the county built in this project can be composed with a contextual bigraph, to connect to the rest of the United Kingdom and also the rest of the world.

First, a place graph modelling physical locations will be built. Places marked on OSM will be arranged in an hierarchical tree from County, City, (Suburb), Street, to Building. This structure will be built primarily by parsing OSM data in OCaml and referencing the tags: administrative boundaries such as cities are represented by relations with a “admin\_level” tag, streets and paths are represented by ways with a  “highway” tag, and buildings are tagged with “addr:street” that describe the street it is on. Since different parts of the world have different tagging conventions on OSM, different hierarchies are expected and any information missing from tags will be calculated by finding nearest neighbours or the containing boundary. After all places have been arranged in a hierarchy, they will be modelled with a static place graph using BigraphER, a tool for computing, validating and simulating bigraphs which provides an OCaml library with APIs to build bigraphs programmatically. This place graph will have Cambridgeshire as the biggest node that contains all other nodes, and the smallest unit nodes will be the individual buildings. In BigraphER, an ontology of controls will be defined for County, City, Street and Building, and places will be represented by nodes of one of the above controls together with an identifier stored as a string parameter.

\begin{figure}[h]
\includegraphics[width=\textwidth]{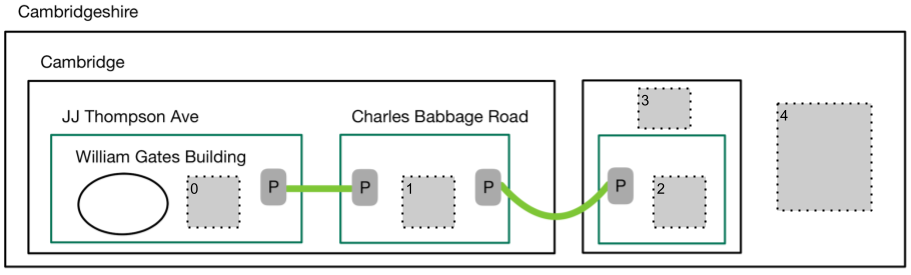}
\centering
\caption{Bigraphical representation of Cambridgeshire. The place graph is represented by the containment of nodes and the link graph is specified by green edges. Sites, where smaller bigraphs can be composited into, are represented by shaded rectangles.}
\end{figure}

Next, a link graph will be added to the place graph to model the connectivity of places. Buildings in the same street node are connected by that street. Additionally, streets are connected to other streets and this will be represented with links. However, different streets nodes have different degree but controls in a bigraph have fixed arity, so a Port control for each edge will be added to the Street control. Streets that are connected will therefore each have a Port that is linked to the other. The graph of connected streets will be constructed by comparing the streets’ OSM relation data: connected streets share a common OSM node. Together, the place graph and link graph constitute a static bigraph, which will be modelled using BigraphER. 

Finally, a use case of the bigraph will be demonstrated: modelling local peer-to-peer connectivity of devices. Devices will be added to the bigraph as atomic nodes that are contained in either a building or street. Reaction rules will be added to allow them to move between buildings and streets that are connected. For this use case, we will assume devices in the same building can connect to each other. This connectivity will be modelled using reaction rules: devices in the same building can form links with each other, but when a device leaves the building it must disconnect from other devices. The use case demonstrated models a short-ranged communication technology that is affected by attenuation of walls (such as Bluetooth and Wifi), as building-grained separation is provided by the topological representation of our bigraph. The dynamic bigraph will then be simulated with BigraphER to demonstrate how the bigraph mutates as devices move between buildings and streets, and form links with other devices according to the reaction rules.

\begin{figure}[h]
\includegraphics[width=0.75\textwidth]{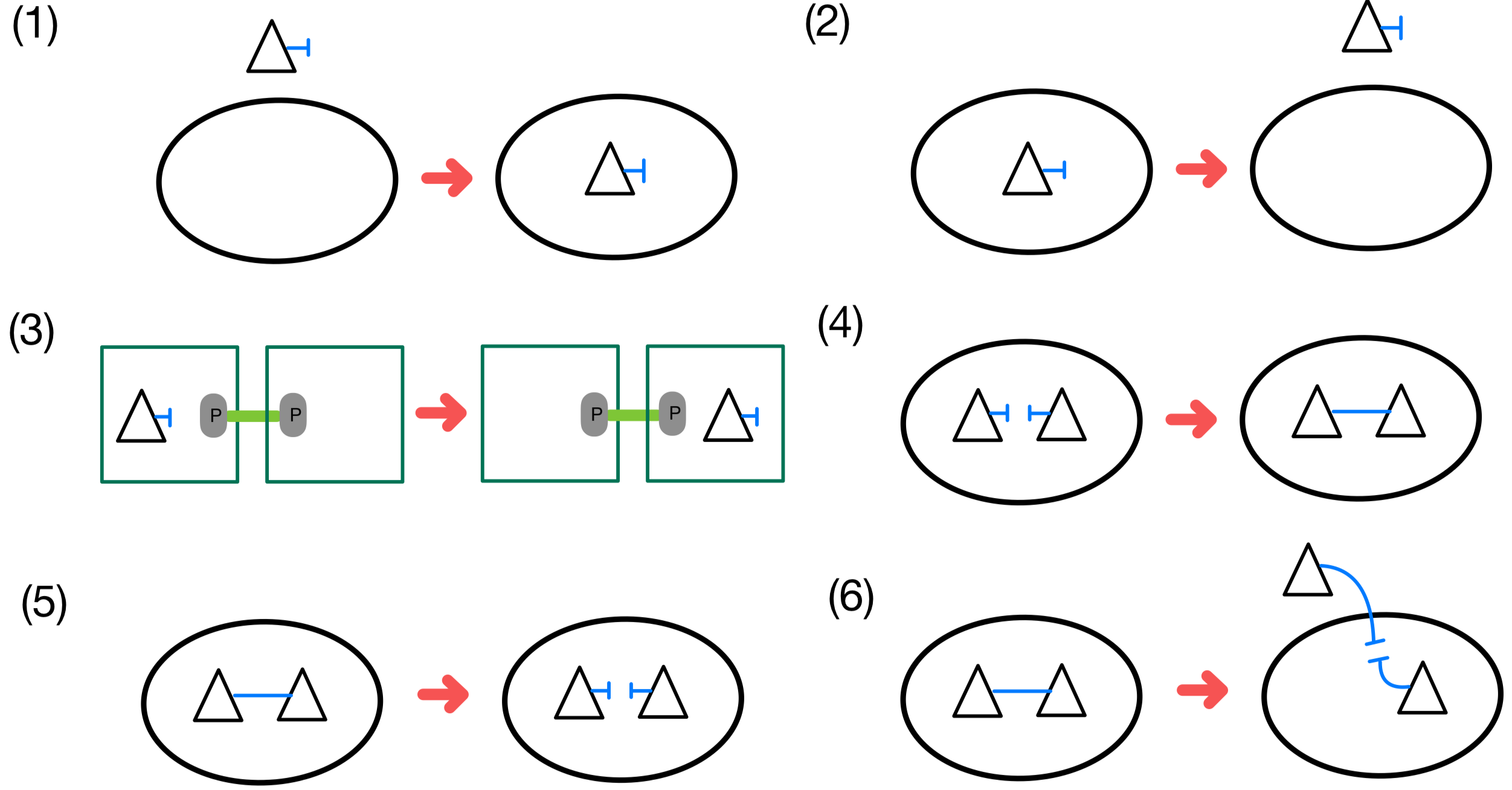}
\centering
\caption{Reaction rules. (1) and (2) allow devices to move from a street to a building, or vice versa. (3) allows a device to move to a connected street. (4) and (5) allow devices in the same building to connect or disconnect. (6) encodes that a device must disconnect from other devices when leaving a building.}
\end{figure}

\section*{Starting Point}

This project is focused on the use of bigraphs to model physical locations in OSM. As background preparation for the project, I have gone through a six-lecture series on Bigraphs by Milner, and experimented briefly with BigraphER. I have familiarised myself with the format of OSM geodata and tools to process OSM data, and done pre-reading on OSM’s tagging conventions.

I have also discussed in depth with my supervisor the spatial context of ubiquitous systems, especially regarding the Spatial Naming System, a proposed extension of the Domain Name System that promises faster discovery of nearby devices and enables connection using a variety of connectivity options such as Bluetooth.

\section*{Success Criteria}
This project is considered a success if:
\begin{itemize}
\item I arrange all the named buildings and streets in Cambridgeshire UK marked on OSM into a hierarchical tree.
\item I model the hierarchy of places with a place graph.
\item I build the graph of connected streets in Cambridgeshire UK, without loss of connectivity information from the original OSM data.
\item I model connected streets and places with a static bigraph.
\item I evaluate the static bigraph’s storage requirements in comparison with the size of the original OSM data, and the performance of the computation of the model.
\item I model local device connectivity by adding to the bigraph device nodes and correct reaction rules.
\item I evaluate the dynamic bigraph on the performance and storage requirements of computation and simulation.
\end{itemize}

\section*{Extensions}
\begin{itemize}
\item I additionally model another county/city in a different part of the world that has a different OSM tagging convention, to demonstrate that bigraphs are still a suitable metamodel despite the difference in hierarchies.
\item I include room nodes in the bigraph, which are contained in a building. It is uncommon that buildings are tagged to room-grained detail in OSM, but select buildings such as the Palace of Westminster and British Museum are. Room-grained detail in the bigraph will enable greater expressivity in reaction rules and therefore more use cases.
\item I implement bigraphs with sharing to represent buildings that are connected to two streets and streets that cross city lines. 
\item I demonstrate another use case of the bigraph: expressing access control policies for location aware applications based on the device’s position in the bigraph. A specific application could be access control of resource records for a Spatial Name System.
\end{itemize}

\section*{Timetable and Milestones}
\begin{enumerate}
\subsection*{Michaelmas Term: Implementation}

\item \textbf {17 Oct - 30 Oct (Week 2-3): Preparation}

{\itshape 21 Oct: Final proposal deadline}
\begin{itemize}
\item Read Robert Milner’s book “The Space and Motion of Communicating Agents”, specifically chapters on definition, structure and reactions of bigraphs as well as motivation for modelling ubiquitous systems.
\item Set up a GitHub repository and Google Drive for version control and backup.
\item Choose a software engineering methodology.
\item Export OSM data for places contained in Cambridgeshire, using Overpass API, a read-only API for selecting custom parts of the OSM map data.
\item Filter OSM data to keep only information relevant to bigraph development and the use case of device connectivity, using osmfilter, a command line tool used to filter OSM data files for specific tags.
\end{itemize}
Milestones:
\begin{itemize}
\item A short LaTeX write-up on the definition of bigraphs and Robert Milner’s original intention to model ubiquitous systems.
\item A filtered OSM data file containing only buildings, streets and boundaries in Cambridgeshire and only relevant tags.
\item A point-form write-up of filtering procedure.
\end{itemize}

\item \textbf {31 Oct - 13 Nov (Week 4-5): Data parsing in OCaml (I)}

\begin{itshape}
4 Nov: DSP module assignment 1 deadline

11 Nov: DSP module assignment 2 deadline
\end{itshape}
\begin{itemize}
\item Parse the filtered OSM data in OCaml, using objects/data structures to store buildings and streets in a hierarchical tree.
\end{itemize}
Milestones:
\begin{itemize}
\item OCaml code to build the hierarchy of places and data file export.
\item A point-form write-up on the methodology of parsing in OCaml
\end{itemize}

\item \textbf {14 Nov - 27 Nov (Week 6-7): Data parsing in OCaml (II)}

\begin{itshape}
18 Nov: DSP module assignment 3 deadline

22 Nov: DSP module written test
\end{itshape}
\begin{itemize}
\item Identify connected streets and build the graph of connected streets in OCaml.
\item Gain familiarity with BigraphER OCaml API.
\end{itemize}
Milestones:
\begin{itemize}
\item OCaml code to build the graph of connected streets and data file export.
\item A point-form write-up of the methodology of identifying connected streets.
\end{itemize}

\item \textbf {28 Nov - 4 Dec (Week 8): Building the static bigraph}
\begin{itemize}
\item Build a bigraph based on the hierarchy of places using BigraphER OCaml API.
\item Add the link graph based on the graph of connected streets using BigraphER OCaml API.
\end{itemize}
Milestones:
\begin{itemize}
\item Static bigraph of Cambridgeshire and its file export.
\end{itemize}

\subsection*{Winter Break: Implementation and Evaluation}
\item \textbf{5 Dec - 25 Dec (Week 1-3): Buffer}

\begin{itshape}
5 Dec - 9 Dec: Block leave
14 Dec - 21 Dec Block leave
\end{itshape}
\begin{itemize}
\item Flex period for unaccounted delays.
\end{itemize}

\item \textbf{26 Dec - 8 Jan (Week 4-5): Building the dynamic bigraph and Simulating}
\begin{itemize}
\item Add reaction rules to model device connectivity using BigraphER.
\item Run simulations on the dynamic bigraph, with different configurations (number and initial spread of devices etc.).
\end{itemize}
Milestones:
\begin{itemize}
\item Dynamic bigraph of Cambridgeshire including atomic devices and its file export. 
\item Simulation data output.
\item A point-form writeup on simulation configurations and aims.
\end{itemize}

\item \textbf{9 Jan - 22 Jan (Week 6-7): Evaluation of performance and simulation results}
\begin{itemize}
\item If necessary, any adjustments to bigraph and simulation.
\item Evaluate static and dynamic bigraphs computation and storage requirements, in comparison with filtered OSM data.
\end{itemize}
Milestones:
\begin{itemize}
\item A short LaTex write-up on the evaluation of bigraphs and simulation results
\end{itemize}

\subsection*{Lent Term: Extensions}
\item \textbf{23 Jan - 5 Feb (Week 1-2): Progress Report; Chapter 1 \& 2}

\begin{itshape}
7 Feb: Progress Report deadline

12 Feb - 19 Feb: Progress Report Presentation
\end{itshape}
Milestones:
\begin{itemize}
\item Progress report should include some visualisation of the bigraph of Cambridgeshire and examples of the bigraph mutating according to reaction rules.
\item LaTeX write-up of Introduction and Preparation chapters of dissertation.
\end{itemize}

\item \textbf{6 Feb - 19 Feb (Week 3-4): Extension (I)}
\begin{itemize}
\item Core success criteria should be achieved at this point.
\item Start work on one of the extension tasks, as described in Extensions.
\end{itemize}
Milestones:
\begin{itemize}
\item A point-form write-up of extension implementation and results.
\end{itemize}

\item \textbf{20 Feb - 5 Mar (Week 5-6): Extension (II)}

\begin{itshape}
CC module assignment 1 and 2 projected deadline
\end{itshape}
\begin{itemize}
\item Continue work on extension tasks.
\end{itemize}
Milestones:
\begin{itemize}
\item A point-form write-up of extension implementation and results.
\end{itemize}

\item \textbf{6 Mar - 19 Mar (Week 7-8): Buffer; Freeze all code}
\begin{itemize}
\item Flex period for unaccounted delays.
\end{itemize}
Milestones:
\begin{itemize}
\item Complete and freeze code related to extensions. After this, no more development work is expected.
\end{itemize}

\subsection*{Easter Break: Dissertation}

\item \textbf{20 Mar - 2 Apr (Week 1-2): Chapter 3 \& 4}
\begin{itemize}
\item Possible week-long block leave
\end{itemize}
Milestones:
\begin{itemize}
\item LaTeX write-up of Chapters 3 and 4, for vetting by supervisor.
\end{itemize}

\item \textbf{3 Apr - 16 Apr (Week 3-4): Chapter 5; Buffer}
\begin{itemize}
\item Flex period for unaccounted delays.
\end{itemize}
Milestones:
\begin{itemize}
\item LaTeX write-up of Chapter 5, for vetting by supervisor, UTO marker and DoS.
\end{itemize}

\item \textbf{17 Apr - 30 Apr (Week 5-6): Refine dissertation}
\begin{itemize}
\item Continue refining dissertation based on feedback.
\item Clean up code, package for submission.
\end{itemize}

subsection*{Easter Term: Finalisation}

\item \textbf{1 May - 14 May (Week 1-2): Final touches}
\begin{itemize}
\item Final edits to dissertation.
\end{itemize}
Milestones:
\begin{itemize}
\item All deliverables to be ready.
\end{itemize}
\begin{itshape}
16 May: Dissertation Deadline
\end{itshape}

\end{enumerate}

\section*{Resource Declaration}
The non-standard resources used in this project are:
\begin{itemize}
\item My own machine (AMD Ryzen 5 7640U with 16GB of RAM)

I will use my own laptop for development, employing the use of GitHub for version control and making frequent backups of code and data to Google Drive. Should my laptop encounter an unexpected failure, I will either continue work using the machines in my college computer room or the Intel Lab, while I try to replace my laptop as soon as possible. I accept full responsibility for this machine and I have made contingency plans to protect myself against hardware and/or software failure.
\item OSM data

This project will make use of open-source data from OSM. In the unlikely event that OSM is taken offline or experiences failure, exported data is still available from other sources such as Internet Archive. There are also other geographic databases such as Who’s On First and Yahoo’s GeoPlanet that can serve as replacements for OSM.
\item BigraphER tool

This project will make use of the open-source tool BigraphER. The lead developer Michele Sevegnani has expressed that she is able to provide assistance should it be needed. Alternatively, the Bigraph Toolkit Suite, an open-source Java-based framework by Dominik Grzelak from the Technische Universität Dresden, can also be used to build and simulate bigraphs.
\end{itemize}

\section*{References}
\begin{enumerate}
\item Milner, Robert. Lecture notes on Bigraphs: a Model for Mobile Agents. November 2008. Bigraph Notes

\url{https://www.cl.cam.ac.uk/archive/rm135/Bigraphs-Notes.pdf. }
\item Sevegnani, M., Calder, M. (2016). BigraphER: Rewriting and Analysis Engine for Bigraphs. In: Chaudhuri, S., Farzan, A. (eds) Computer Aided Verification. CAV 2016. Lecture Notes in Computer Science(), vol 9780. Springer, Cham. 

\url{https://doi.org/10.1007/978-3-319-41540-6\_27}
\item Overpass API \url{https://overpass-api.de/}
\item Osmfilter \url{https://wiki.openstreetmap.org/wiki/Osmfilter}
\item Grzelak, D. Bigraph Ecore Metamodel (BEM): An Emof-compliant Specification for Bigraphs. Zenodo, 26 Oct. 2023, doi:10.5281/zenodo.10043063. 

\url{https://zenodo.org/records/10043063}
\end{enumerate}

\label{lastpage}
\end{document}